\documentclass{aa}  
\pdfoutput=1
\usepackage{graphicx}
\usepackage{color}
\usepackage{rotating}
\usepackage{natbib}
\usepackage{longtable,lscape}
\usepackage[varg]{txfonts}

\newcommand{\mcol}[3]{\multicolumn{#1}{#2}{#3} }
\newcommand{\teff}{$T_{\mathrm{eff}}$~}
\newcommand{\logg}{$\mathrm{log}\,g$~} 
\newcommand{\vsini}{$v\,\mathrm{sin}\,i$~} 
\newcommand{\teffi}{$T_{\mathrm{eff}}$}
\newcommand{\loggi}{$\mathrm{log}\,g$} 
\newcommand{\vsinii}{$v\,\mathrm{sin}\,i$} 
\newcommand{\ms}{m\,s$^{-1}$}
\newcommand{\kms}{km\,s$^{-1}$~}
\newcommand{\kmsi}{km\,s$^{-1}$}

\newcommand{\struutup}{\rule{0ex}{3.2ex}}
\newcommand{\struutdown}{\rule[-2ex]{0ex}{2ex}}

\bibpunct{(}{)}{;}{a}{}{,}

\begin{document} 

   \title{Multi-technique investigation of the binary fraction among A-F type candidate hybrid variable stars discovered by 
   \textit{Kepler}\thanks{Based on observations obtained with the \textit{Hermes} spectrograph, which is supported by 
   the Research Foundation - Flanders (FWO), Belgium, the Research Council of KU Leuven, Belgium, the Fonds National de la 
   Recherche Scientifique (F.R.S.-FNRS), Belgium, the Royal Observatory of Belgium, the Observatoire de Genève, Switzerland 
   and the Th\"uringer Landessternwarte Tautenburg, Germany. }}
   \authorrunning{P. Lampens et al.}
   \titlerunning{A spectroscopic investigation of \textit{Kepler} A-F type candidate hybrid stars}
   \author{
           P. Lampens\inst{1}
          \and
           Y. Fr\'emat\inst{1} 
          \and
           L. Vermeylen\inst{1}
          \and
           \'A. S\'odor\inst{2}
          \and
           M. Skarka\inst{2}
          \and
           P. De Cat\inst{1}
          \and
           Zs. Bogn\'ar\inst{2}
          \and
           R. De Nutte\inst{3}
          \and
           L. Dumortier\inst{1}
         \and
           A. Escorza\inst{3,4}
          \and
           G. M. Oomen\inst{3}
	  \and
           G. Van de Steene\inst{1}
          \and
           D. Kamath\inst{3}
          \and
           M. Laverick\inst{3}
          \and
           A. Samadi\inst{3}
          \and
           S. Triana\inst{1}
          \and
           H. Lehmann\inst{5}
          }
   \institute{
   Koninklijke Sterrenwacht van Belgi\"e, Ringlaan 3, B-1180 Brussel, Belgium
   \email{patricia.lampens@oma.be}
   \and
   Konkoly Observatory, MTA CSFK, Konkoly Thege M. u. 15-17, H-1121 Budapest, Hungary
   \and
   Instituut voor Sterrenkunde (IvS), Katholieke Universiteit Leuven, Leuven, Belgium
   \and
   Institut d'Astronomie \& Astrophysique, Universit\'e Libre de Bruxelles, Brussels, Belgium
   \and
   Th\"uringer Landessternwarte, Tautenburg, Germany
          }

   \date{Received  ; accepted }

  \abstract
   {Hundreds of candidate hybrid pulsators of intermediate type A--F were revealed by the recent space missions.
   Hybrid pulsators allow to study the full stellar interiors, where both low-order \textit{p}- and 
   high-order \textit{g}-modes are simultaneously excited. The true hybrid stars must be identified since 
   other processes, related to stellar multiplicity or rotation, might explain the presence of (some) 
   low frequencies observed in their periodograms. } 
   {We measured the radial velocities of 50 candidate $\delta$ Scuti - $\gamma$ Doradus hybrid stars from the \textit{Kepler} mission 
    with the \textsc{Hermes} 
    and \textsc{Ace} spectrographs over a time span of months to years. 
   We aim to derive the fraction of binary and multiple systems and to provide an independent and homogeneous determination of the atmospheric 
   properties and \vsini for all targets. 
   The long(er)-term objective is to identify the (probable) physical cause of the low frequencies. }
   {We computed 1-D cross-correlation functions (CCFs) in order to find the best set of parameters in terms of the number 
   of components, spectral type(s) and \vsini for each target. Radial velocities were measured from spectrum synthesis 
   and by using a 2-D cross-correlation technique in the case of double- and triple-lined systems. Fundamental parameters 
   were determined by fitting (composite) synthetic spectra to the normalised median spectra corrected for the appropriate 
   Doppler shifts. } 
   {We report on the analysis of 478 high-resolution \textsc{Hermes} and 
   41 \textsc{Ace} spectra of A/F-type candidate hybrid pulsators from the \textit{Kepler} field. We determined their radial 
   velocities, projected rotational velocities, atmospheric properties 
   and classified our targets based on the shape of the CCFs and the temporal behaviour of the radial velocities. We 
   derived orbital solutions for seven new systems. Three long-period
   preliminary orbital solutions are confirmed by a photometric time-delay analysis.
   Finally, we determined a global multiplicity fraction of 27\% in our sample of candidate hybrid stars. }
   {}


   \maketitle
%

\section{Introduction}
\label{sect:intro}

Many different physical processes compete in the outer atmospheres of A- and F-type main-sequence (MS) stars and their slightly 
more evolved cousins. 
In the corresponding region of the H-R diagram, we find the following important transitions:\\ 
- 1. the transition from average "slow" to average "fast" rotation. The distribution of \vsini 
as a function of spectral type shows that stars cooler than F5 have small \vsini (typically 
< 10 \kmsi), whereas \vsini can reach several hundreds of \kms for hotter stars 
(\citealt{Royer2009LNP...765..207R}, cf. Fig.~2 from \citealt{Royer2014psce.conf..265R});\\
- 2. the transition { from deep to shallow convection, i.e. from convective to radiative envelopes}. Theoretical models 
indicate a dramatic change in the structure of the outer envelopes near the red edge (RE) of the $\delta$ Sct instability 
strip, i.e. near \teff = 7000~K \citep{ChristensenDalsgaard2000ASPC..210..187C}. This transition 
(where the sudden onset of convection in the stellar envelope starts, see \citealt{DAntona2002A&A...395...69D}) has 
also been described as a second "B\"ohm-Vitense gap";\\
- 3. the transition from mode driving by convective blocking near the base of the convective envelope (exciting gravity 
modes of type $\gamma$ Dor) 
\citep{Dupret2004A&A...414L..17D,Dupret2005A&A...435..927D} to mode driving by the opacity mechanism (exciting acoustic 
modes of type $\delta$ Sct). Both instability strips are largely overlapping, which suggests that 
two regimes of modes could be simultaneously excited in some stars (i.~e. the so-called ``hybrid'' stars). A recent study of 
a large sample of candidate $\gamma$ Dor stars suggests that the latter are confined to the \teffi-range from 6900~to 7400~K 
on the MS, which corresponds to the region in the Hertzsprung-Russell diagram where $\gamma$ Dor pulsations are theoretically
predicted \citep{Tkachenko2013A&A...556A..52T};\\
- 4. the transition from significant chromospheric activity and 
(coronal) X-ray emission to (almost) null emission. Such X-ray emission is expected for cool stars (type A7 and later) 
due to magnetic activity as well as for hot stars (type B2 and earlier) due to wind shocks. Intermediate spectral types are 
virtually X-ray dark \citep{Schroder2007A&A...475..677S,Robrade2009A&A...497..511R}. 
The { former} transition occurs abruptly, over a temperature interval no larger than 100~K in width, i.e. at approximately 
\teff = 8250~$\pm$~50~K \citep{Simon2002ApJ...579..800S}.
Within this narrow temperature range, chromospheric emission abruptly drops from solar brightness levels to 
more than one order of magnitude less. These phenomena are linked to the presence of surface convection zones and 
stellar magnetic fields, ultimately held responsible for the observed activity \citep{Schroder2007A&A...475..677S}.
Exceptions are the young Herbig Ae/Be stars and the peculiar Ap/Bp stars, where fossil magnetic fields are 
thought to play a role in the production of X-rays.\\
The origin of the stellar magnetic fields should also be considered. Dynamo processes generate 
fields in most of the low-mass stars of the MS, whereas the origin of the large-scale magnetic fields 
in massive stars is not yet understood \citep[]{Fossati2015A&A...574A..20F,Morel2015IAUS..307..342M}. 
On the other hand, magnetic fields are rare among the intermediate-mass stars. There is indeed no normal A star 
known with a fossil magnetic field of average strength \citep{Auriere2007A&A...475.1053A,Lignieres2014IAUS..302..338L} : 
either the fields are (very) strong (B$_{long}$ > 100 G) as in Ap stars \citep[e.g.][]{Mathys2001ASPC..248..267M}, 
or ultra weak (B$_{long}$ < 1 G) as in Am stars \citep{Blazere2014sf2a.conf..463B}. 
Most chemically peculiar stars, however, appear to possess a magnetic field which brakes the rotation and stabilizes 
the atmosphere, enabling processes of atomic diffusion \citep{2015ads..book.....M}.\\ 
 
In this complex region \citep[see also][]{Antoci2014IAUS..301..333A}, a new group of pulsators has been revealed 
on the basis of the analysis of the {\it Kepler} light curves: 
they are called the $\delta$\,Sct -- $\gamma$\,Dor or $\gamma$\,Dor -- $\delta$\,Sct {\it candidate hybrid} stars 
\citep{Grigahcene2010ApJ...713L.192G,Uytterhoeven2011A&A...534A.125U}. 
These candidate hybrid stars are A/F-type pulsating stars located across the instability strips of the $\delta$\,Sct 
and the $\gamma$\,Dor pulsators. Their light curves exhibit frequencies in both regimes.
A first suspicion of the co-existence of both \textit{p}- and \textit{g}- modes already arose from a search for multiperiodicity 
in early A-type stars based on the Hipparcos epoch photometry \citep[e.~g.][]{Koen2001MNRAS.321...44K}. 
Because the hotter ones are unexplained by theoretical models \citep{Grigahcene2010ApJ...713L.192G,Balona2015MNRAS.452.3073B}, 
it is important to unravel the physical origin of their low frequencies and to confirm the cases of {\it genuine} hybrid 
pulsation. In the case of KIC~9533489, an object with \teff of 7500~K, the authors concluded in favour of 
a true hybrid character of the oscillations \citep{Bognar2015A&A...581A..77B}. 
It is furthermore critical to look for binarity in such stars. One reason is that some of the low frequencies should not be 
attributed to pulsation but perhaps to ellipsoidality and reflection or even something more exotic (like shallow eclipses or a 
heartbeat phenomenon). Another reason is that tidal forces in close companions may excite a number of \textit{g}-modes (as harmonics 
of the orbital frequency) which would not be excited if the star was single. { This could be the case of a pulsating star in 
an eccentric, close binary, where the dynamical tidal forces excite g-mode pulsations \citep{2002A&A...384..441W}.} 
Moreover, a normal $\gamma$\,Dor star coupled to a normal $\delta$\,Sct star will also mimic a $\gamma$\,Dor -- $\delta$\,Sct hybrid.\\

The confirmation of genuine cases of hybrid pulsation thus represents an unavoidable step in the study of the A/F-type hybrid 
phenomenon. In this work, we aim to estimate the fraction of short-period (i.e. with orbital periods between about 1 and 50\,d) 
spectroscopic systems among a sample of brighter A/F-type candidate hybrid stars discovered by the \textit{Kepler} satellite. 
Some preliminary results have been reported by \citet{Lampens2015EPJWC.10106043L}. This survey is based on the 171 candidate 
hybrid \textit{Kepler} stars first studied by \citet{Uytterhoeven2011A&A...534A.125U}. 
The following definition of a hybrid star was used in their study:
\begin{itemize}
\item frequencies detected in the $\gamma$\,Dor (i.e. < 5 d$^{-1}$) and $\delta$\,Sct (i.e. > 5 d$^{-1}$) domains;
\item the amplitudes in both domains are either comparable, or they do not differ by more than a factor of 5--7;
\item at least two independent frequencies detected in both regimes with amplitudes higher than 100 ppm.\\
\end{itemize} 

Section~\ref{sect:obs} describes the target selection, the observational strategy, the campaigns and the observations. 
Section~\ref{sect:spec} deals with the data processing. In Section~\ref{sect:anal}, we explain the methodology and 
the data analysis. In Sections~\ref{sect:clas} and~\ref{sect:orb}, the results of the classification and the orbital 
solutions of the newly discovered systems, respectively, are presented. The extraction of the physical parameters 
is discussed in Section~\ref{sect:phys}. In Sect.~\ref{sect:timeDelay}, we study the periodograms based on the \textit{Kepler} 
data and we present an observational H-R diagram in Sect.~\ref{sect:HRD}. A discussion and conclusions from this work 
are provided in Section~\ref{sect:con}. 

\section[]{Sample and observational strategy}
\label{sect:obs}

We selected 50 of the brightest objects among the A/F-type candidate hybrid stars classified by 
\citet[][cf.~Table 3]{Uytterhoeven2011A&A...534A.125U}. The observations were performed in the observing seasons from Aug. 2013 until 
Dec. 2016 with the high-resolution fibre-fed \'echelle spectrograph \textsc{Hermes} (High Efficiency and Resolution Mercator Echelle 
Spectrograph, \citet{Raskin2011A&A...526A..69R}) mounted at the focus of the \mbox{1.2-m} Mercator telescope located at the international 
observatory {\rm Roque de los Muchachos} (ORM, La Palma, Spain). The instrument is operated by the University of Leuven under the supervision 
of the \textsc{Hermes} Consortium. It records the optical spectrum in the range $\lambda$ = 377\,- 900\,nm across 55 spectral orders in a 
single exposure. The resolving power in the high-resolution mode is R = 85\,000. Advantages of the instrument are its broad spectral coverage, 
high stability (the velocity stability of a radial velocity standard star is about 50 \ms, priv. commun., IvS, Leuven) and excellent throughput  
\citep{Raskin2011A&A...526A..69R}. Table~\ref{tabl:logobs} lists the journal of the spectroscopic campaigns.\\

More spectra were obtained with the new {\sc Ace} fibre-fed \'echelle spectrograph attached to the 1-m RCC telescope of the Konkoly Observatory 
at Piszk\'es-tet\H o, Hungary. The  {\sc Ace} spectrograph covers the 415\,-\,915\,nm wavelength range 
with a resolution R\,=\,20\,000. A total of 41 {\sc Ace} spectra was acquired for 11 targets during the winter of 2014-2015 (Table~\ref{tabl:logobs}).\\

\begin{table}
\caption{\label{tabl:logobs} Log of spectroscopic observations. 
}
\centering
\begin{tabular}{lc@{}c}
\hline\hline
\multicolumn{1}{c}{Range of JDs} & \multicolumn{1}{c}{Range of dates} & \multicolumn{1}{c}{Nr of spectra}\\
\multicolumn{1}{c}{From - to} & \multicolumn{1}{c}{From - to} &\\
\hline
$24[55198 - 56293]$ & Jan, 2010 - Dec, 2012 & 62 \\ 
$24[56513 - 56607]$ & Aug 08 - Nov 10, 2013 & 64 \\ 
$24[56772 - 56974]$ & Apr 24 - Nov 12, 2014 & 188 \\ 
$24[56932 - 57097]$ & Oct 01 - Mar 15, 2014-15 & 41\tablefootmark{n}\\
$24[57150 - 57324]$ & May 07 - Oct 28, 2015 & 105 \\
$24[57466 - 57739]$ & Mar 18 - Dec 16, 2016 & 59 \\
\hline
\end{tabular}
\tablefoottext{n}{These spectra were collected with the \textsc{Ace} spectrograph.}
\end{table}

\begin{figure}
\begin{center}
\includegraphics[width=8.9cm]{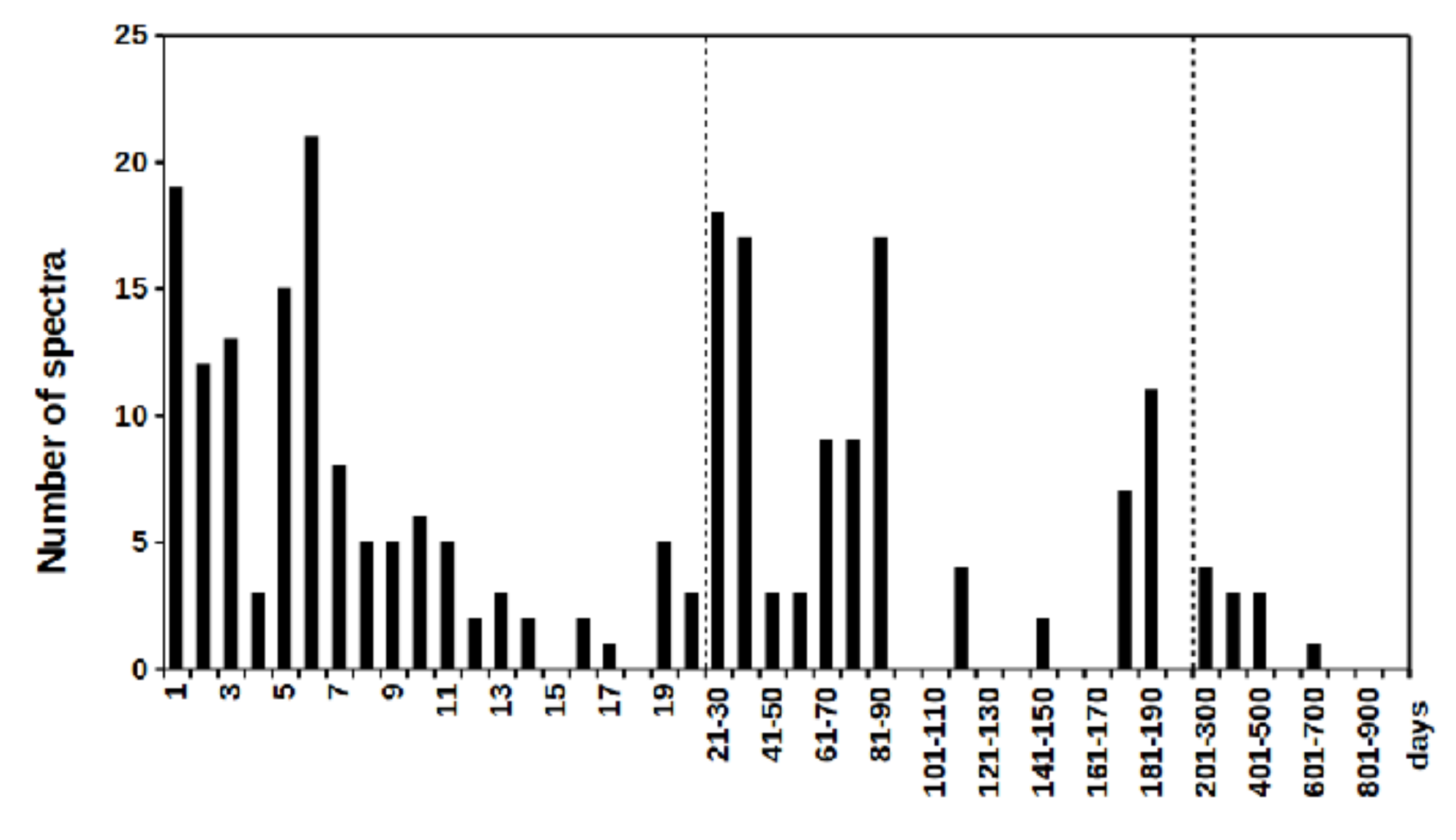}
\end{center}
\caption{Distribution of time intervals between successive \textsc{Hermes} spectra for the sample stars showing peaks of 1 and 6 days  
as well as $\sim$1 and $\sim$3 months. Note the rapidly increasing bin size along the X-axis.}
\label{fig:hist}
\end{figure}

The KIC magnitudes of our targets are brighter than or equal to 10.3 mag. For a target brightness of 9.5 mag, an exposure 
time of 10 min was usually sufficient to reach a signal-to-noise ratio (S/N) of at least 50 per bin (at $\lambda = 650$\,nm). 
The exposure times ranged between 5 and 30~min mostly, leading to a S/N of about 200 in the CCF (with over 100 useful lines in 
the spectrum). We expected to achieve a precision of 1\,\kms in the measurement of the radial velocities in most cases.  
We acquired a {\it minimum} of four high-resolution spectra for all our targets. Sampling was done irregularly over a total time base of 
four years. We planned at observing each target during two successive nights in the first week, once after at least a full week, and 
once more after a time lapse of at least one month. This intended scheme could, however, not always be maintained due to practical 
circumstances. On average, we dispose of 5-10 \textsc{Hermes} spectra per target. Our spectra cover time scales of a few days, weeks and 
months up to a few years per target. The objective is to be able to (spectroscopically) detect binarity for orbital periods ranging 
from a day up to a few months. The actual temporal distribution of most spectra is illustrated in Fig.~\ref{fig:hist}.  \\ 

An additional goal is to determine improved atmospheric stellar parameters and \vsini for our targets, in particular for the 
single-lined cases, in order to pinpoint their position in the H-R diagram. The combination of (at least) four spectra per 
target allows us to reach a reasonable precision on the effective temperature (\teffi) and the surface gravity (\loggi). \\

\section[]{Data reduction and processing}
\label{sect:spec}

The spectra were reduced with the dedicated reduction pipeline elaborated for \textsc{Hermes} spectra which 
includes the subtraction of the bias and stray light, order-by-order extraction, correction for the flat-field, 
and the scaling in wavelength using calibration frames (obtained with Thorium--Argon lamps) followed by cosmic rays 
removal and order merging \citep{Raskin2011A&A...526A..69R}.  
The procedure also provides the S/N, which in our case is in the range of 60-70 (at $\lambda = 650$\,nm) 
on average. For all the spectra, the normalisation to the local continuum was done by fitting a low-order polynomial 
through the continuum parts in wavelength bins of about 50\,nm long using the IRAF\footnote{IRAF is distributed by the 
National Optical Astronomy Observatories, which are operated by the Association of Universities for Research in Astronomy, 
Inc., under cooperative agreement with the National Science Foundation.} task 'continuum'. The average S/N of 
the median spectra is of the order of 150-200.\\ 

The \textsc{Ace} spectra were reduced using IRAF standard tasks including bias, aperture extraction, dark and flat-field 
corrections, and wavelength calibration using Thorium--Argon exposures. The normalisation, 
cosmic rays filtering, order merging (and cross-correlation, see below) was performed by in-house programmes (developed by \'AS). 
The systematic errors introduced by the data processing and the stability of the wavelength calibration system of 
the \textsc{Ace} instrument was found to be better than 0.36\,\kmsi, based on observations of radial velocity
standards \citep{Derekas2017MNRAS.464.1553D}. All the spectra were systematically corrected for barycentric motion.\\

\section[]{Data analysis}
\label{sect:anal}

As a first-look analysis, we computed a series of one-dimensional cross-correlation functions (CCFs) in the wavelength 
intervals $415-445, 445-510, 510-570$\,nm using different pre-selected masks. The correlation masks have been built 
from line lists computed with the code \textsc{synspec} \citep{Hubeny1995ApJ...439..875H}. 
In the case of a double- or multiple-lined system, these one-dimensional CCFs may show striking features such as a variable 
(strong) asymmetric broadening (Fig.~\ref{fig:CCF1}) or multiple minima (Fig.~\ref{fig:CCF2}). Occasionally, a 
composite profile consisting of a narrow profile embedded in a broader one was also detected (e.g. KIC~11572666, 
Fig.~\ref{fig:KIC115_ccfcomb}). In the complex situations, using various masks, we were able to estimate the number 
of components in the system as well as to derive preliminary values for the component's spectral types and projected 
rotational velocities. \\

\subsection[]{Radial velocities}
\subsubsection[]{Single stars and single-lined systems}
\label{sect:spec_single}

In the following step, we searched for the most adequate set of fundamental parameters by fitting synthetic models 
(templates) to the spectrum on the basis of a 'minimum distance' method (i.e. by determining the smallest $\chi^2$). 
Thus, we explored the parameter space in terms of spectral type and \vsini and confronted each observed spectrum 
to a variety of synthetic spectra in four distinct spectral bins chosen in the wavelength interval $415 - 570$\,nm. 
The synthetic spectra, suitably broadened to the estimated projected rotational velocity, were computed using the code 
\textsc{Synspec} \citep{Hubeny1995ApJ...439..875H} together with the \textsc{atlas9} atmosphere models for \teff < 15000~K  
and \logg = 4 \citep{Castelli2003IAUS..210P.A20C}. Except for Section~\ref{Sect:multifit}, the same method 
will be used for all synthetic spectra discussed from hereon. 
Upon finding the closest match in parameter space, we adopted this best-fit model and computed the radial velocity 
by Doppler shifting the synthetic spectrum until it matched the observed spectrum. 
This procedure was repeated for (typically 10) smaller wavelength bins, and the mean and scatter of the radial velocity 
measurement (RV) were derived from the values for the different bins, after the outliers were omitted.\\

\subsubsection[]{Double- or multiple-lined systems}
\label{sect:spec_double}

In the case of non-single or composite objects, we applied an in-house programme (developed by YF) which uses
the algorithm of \textsc{todcor} \citep{Mazeh1994Ap&SS.212..349M}. 
This code also makes use of \textsc{Synspec} \citep{Hubeny1995ApJ...439..875H} as well as the \textsc{atlas9} models 
\citep{Castelli2003IAUS..210P.A20C} with the estimated input parameters and \logg = 4 to build a suitably 
broadened synthetic spectrum {\it for each component}, 
and computes the two-dimensional cross correlation and the corresponding radial velocities. It allows us to 
extract the velocities even when the components are not fully resolved (i.e. the case of blended profiles). 
We tested different values of spectral type and \vsini and adopted the best-fitting parameters 
based on the same criterion of 'minimum distance' between the observed and the synthetic {\it composite} spectrum.
Since the ratio of the light contributions is another free parameter obtained during the fitting, we also derived its 
most probable value. For illustrations of a triple- and a double-lined system, we refer to Figs.~\ref{fig:KIC44} 
and~\ref{fig:KIC89}, respectively. 
Finally, we computed two-dimensional CCFs using this best-fit model for a large number of wavelength bins (usually 10 
bins of 6-20\,nm long) and derived the mean and scatter of the radial velocities (RVs) from the individual values, 
after the outliers were omitted. \\

In the case of a triple-lined system, we repeated this procedure in a two-step sequence: first, with components 
A and B to search for the best-fitting two-component model and the light ratio $l_{A,B}$ = $l_{B} / l_{A}$, and secondly, 
adding another component (C) to search for the best-fitting three-component model as well as the additional 
light ratio $l_{AB,C}$ = $l_{C} / l_{A+B}$.\\

\begin{figure}
\begin{center}
\includegraphics[width=8.9cm]{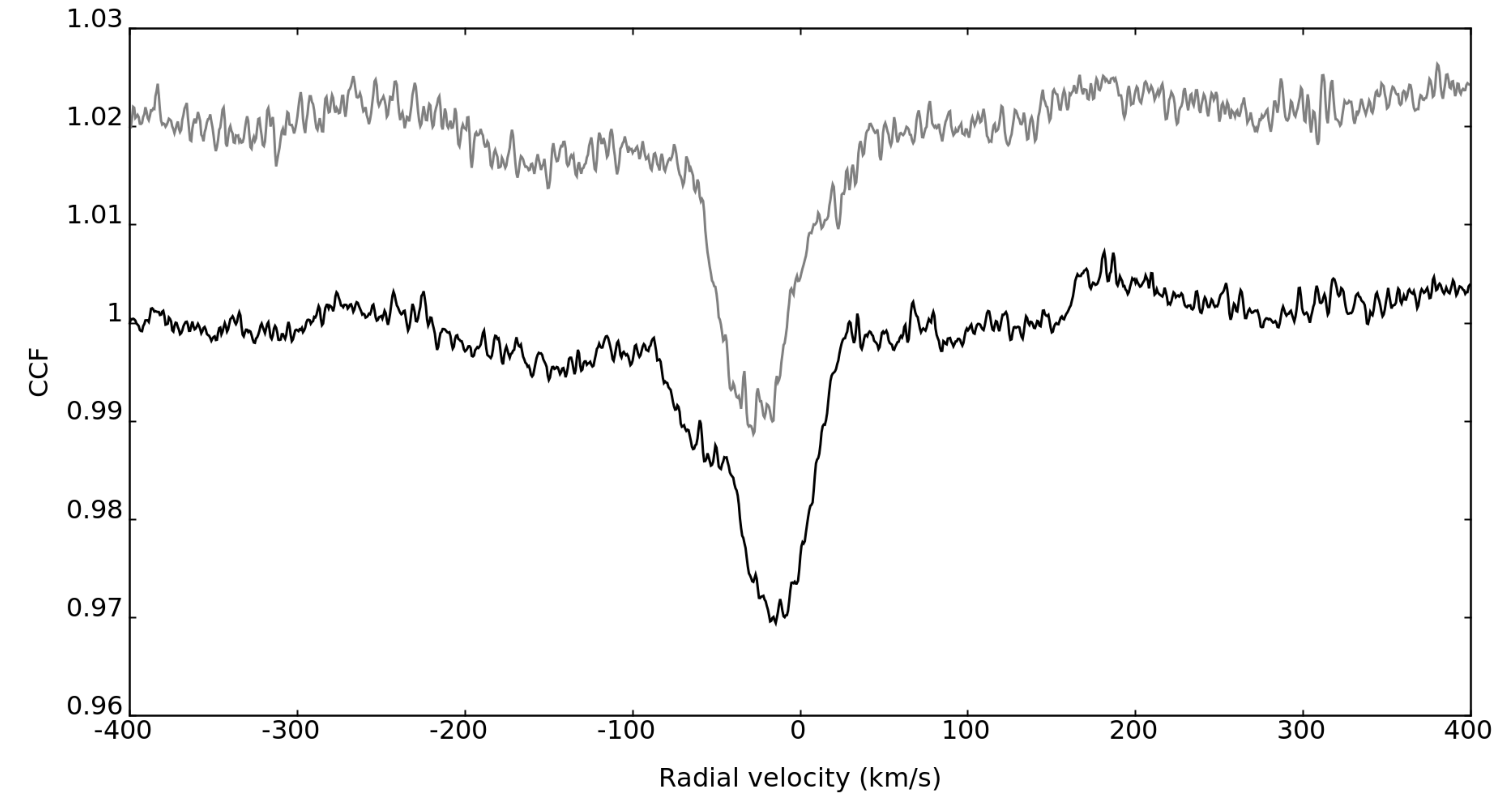}
\end{center}
\caption{Cross-correlation functions in the case of an unresolved double-lined spectroscopic binary system showing obvious 
asymmetric profiles as well as velocity shifts e.~g.~KIC~7756853.}
\label{fig:CCF1}
\end{figure}

\begin{figure}
\begin{center}
\includegraphics[width=8.9cm]{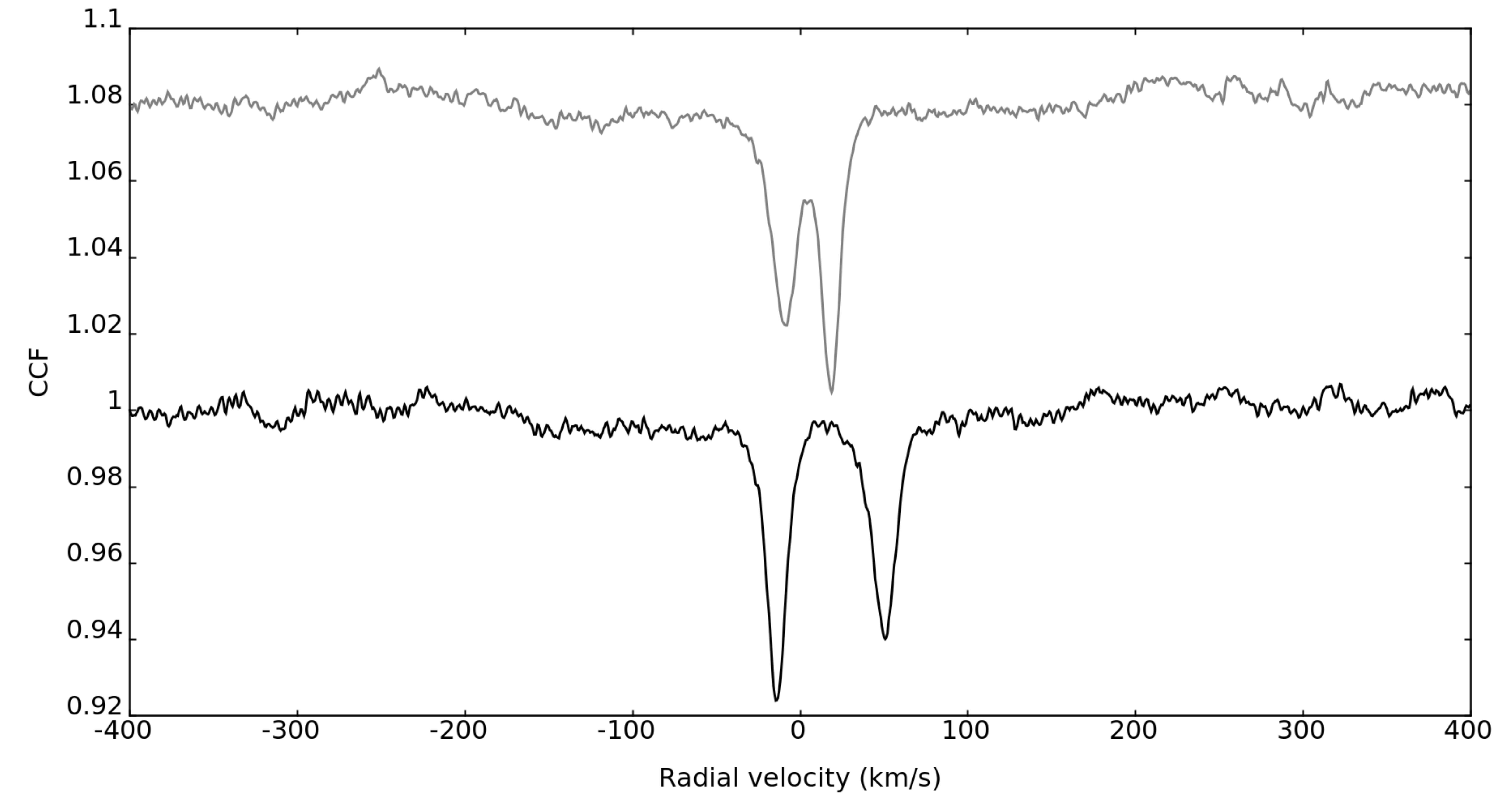}
\end{center}
\caption{Cross-correlation functions in the case of a well-resolved double-lined spectroscopic binary showing two distinct 
minima and velocity shifts in its profile e.~g.~KIC~5219533.}
\label{fig:CCF2}
\end{figure}

\begin{figure}
\begin{center}
\includegraphics[width=8.9cm]{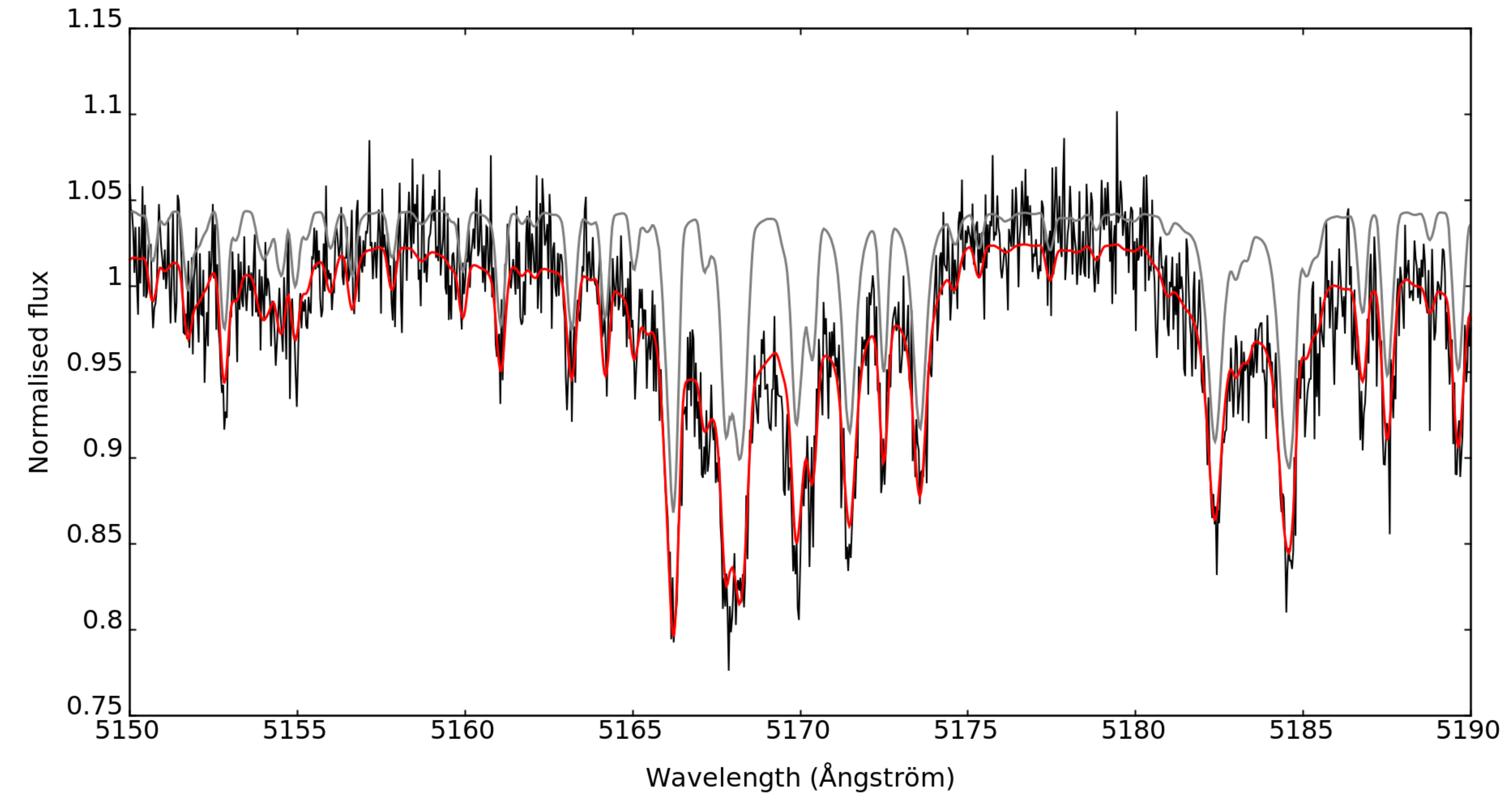}
\end{center}
\caption{Case of a triple system: part of the spectrum (in black) and composite model (in red) for KIC~4480321 using
 the synthetic spectra for three components, respectively of type (F0, \vsini=10), (F0, \vsini=10) and (A5, \vsini=160). 
 The thin line (in grey) represents the scaled contribution of the two (F0, \vsini=10) components.}
\label{fig:KIC44}
\end{figure}

\begin{figure}
\begin{center}
\includegraphics[width=8.9cm]{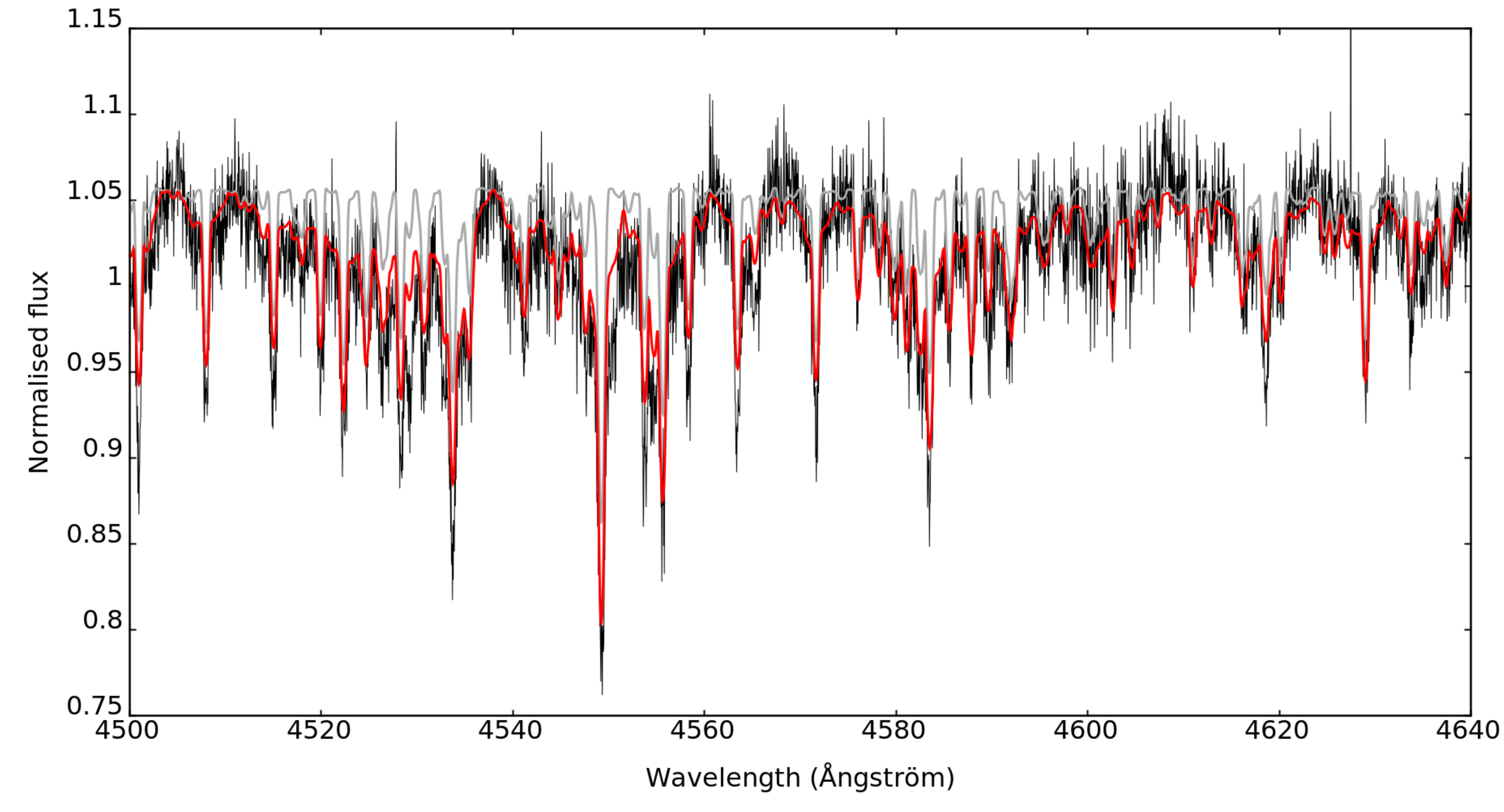}
\end{center}
\caption{Case of a binary system: part of the spectrum (in black) and composite model (in red) for KIC~8975515 using
 the synthetic spectra (A7, \vsini=170) and (A7, \vsini=30). The thin line (in grey) represents the scaled contribution 
 of the (A7, \vsini=30) component.}
\label{fig:KIC89}
\end{figure}

\subsection[]{Atmospheric properties}\label{sect:aps}

An additional purpose of the newly acquired spectra is the determination of atmospheric stellar properties. For a reliable
characterization and unambiguous location in the Hertzsprung-Russell diagram, the effective temperature and the surface 
gravity of the selected targets should be known as reliably and accurately as possible. This requires a comparison between 
observed and synthetic spectra. For single stars and single-lined systems, we fitted several regions of the normalised \textit{median} 
spectrum. For non-single objects, we fitted several regions of the normalised \textit{individual} spectra. 

\subsubsection[]{Single stars and single-lined systems}
\label{Sect:1D-atmos}

From each spectrum, three regions of width of 45-70\,nm centered on the H$\alpha$, H$\beta$, and H$\gamma$ lines, 
were extracted. Continuum normalisation was performed in two steps: first, by dividing each region of interest 
by the best-fitting model as defined in Sect.~\ref{sect:spec_single} and secondly, by dividing by a polynomial 
representation of the local continuum. This low-degree polynomial was computed independently also using sigma clipping. 
We next combined all the normalised spectra of each region into a normalised single median spectrum of higher quality. 
These median spectra were subsequently confronted to synthetic spectra using grids computed in slightly different ways.\\ 

\paragraph{Multi-region fits}
\label{Sect:multifit}

Firstly, we performed fits of the three regions of interest containing the lines H$\alpha$, H$\beta$, and H$\gamma$, 
respectively. An extended grid of high-resolution synthetic spectra computed with plane-parallel model atmospheres was 
retrieved from the \textsc{Pollux} database \citep{Palacios2010A&A...516A..13P}. We selected the models with a 
microturbulence of 2\,\kms (a typical value for A/F-type MS stars, \citet{Gebran2014psce.conf..193G}) and a solar metallicity, 
for which \teff ranged from 7000 to 9000~K (with a step of 200~K) and \logg ranged from 3.5 to 4.9 dex (with a step of 0.2 dex).
Only \teff and \logg were varied since the extracted regions include the Balmer lines to a large extent even though 
they may contain a few shallow metal lines. With this method, the synthetic spectra were selected at the nearest node of 
the atmospheric values. In general, we adopted the previous estimate for the parameter \vsini since it was found to be 
precisely determined (cf. Sect.~\ref{sect:spec_single}). 
This procedure provided an averaged value of \teff and a range of probable values for \logg which are listed in 
Table~\ref{tab:param1} (cols.~11-12).\\ 

\paragraph{Single-region fit}
\label{sect:girfit_single}

\begin{figure}
\begin{center}
\includegraphics[width=8.9cm]{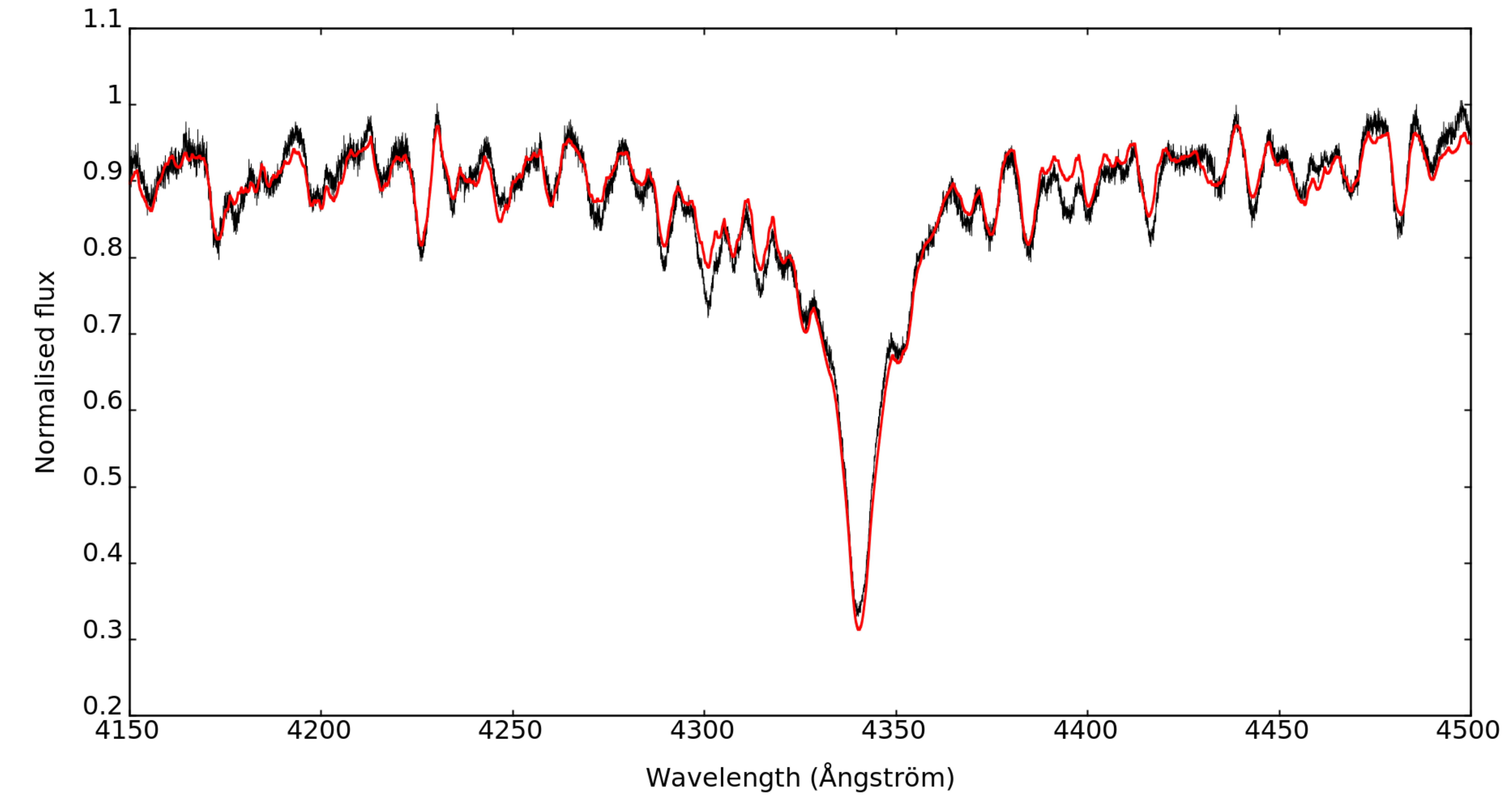}
\end{center}
\caption{Part of the observed spectrum (in black) for KIC~6432054 and model (in red) using the best-fitting set of atmospheric 
stellar properties \teff = 7542~K, \logg = 4.44 and \vsini = 184~\kms derived with \textsc{Girfit}.}
\label{fig:KIC64}
\end{figure}

As an independent, final step, we {\it simultaneously} (re)derived the three parameters \teffi, \logg and \vsini by 
fitting the range [415 - 450]\,nm (including H$\gamma$ cf. Fig.~\ref{fig:KIC64}) using the code \textsc{Girfit} 
\citep{Fremat2006A&A...451.1053F}. 
With this method, the model spectra  (cf. description above) are interpolated in a grid of (\teffi,~\loggi)-values 
instead of being selected at the nearest node of the atmospheric values.
These model spectra were next convolved with the rotational profile as well as a Gaussian instrument profile to account for the 
spectrograph's resolution.\\

The results of this spectrum synthesis method are displayed in Table~\ref{tab:param1} (cols.~13-15). At this point, the probable
errors are assumed to be of the order of the grid steps (i.e. $\epsilon_{T_{\mathrm{eff}}} = 250$~K for \teff and $\epsilon_{\mathrm{log}\,g}
= 0.5$ dex for \loggi), since the procedure uses interpolation. In Sect.~\ref{sect:phys}, we will be able to confirm that such assumptions 
are also realistic. An example of the adjustment using \textsc{Girfit} for a single object is presented in Fig.~\ref{fig:KIC64}. Note 
also the excellent agreement between both determinations (cols.~11-15 in Table~\ref{tab:param1}). A direct comparison between the results of 
both methods furthermore provides a reliable way to estimate the uncertainties involved. \\

\subsubsection[]{Double- or multiple-lined systems}
\label{sect:girfit_double}

The programme \textsc{Girfit} was next modified in order to extend the same analysis to spectra with {\it n} components
(in practice, the spectra of double- and triple-lined systems). The modified programme interpolates the 
component spectra in the usual grid, and combines them into a composite spectrum 
to find the solution which minimizes the residuals. The minimisation is performed using the Simplex algorithm. In 
this extended form of \textsc{Girfit} (developed by YF), the radial velocities and the light ratios can either 
be kept fixed or be considered as additional free parameters (our choice). The radial velocities served as a consistency 
check of the solutions, while the light ratios were considered as added value. For each system, we fitted all the individual 
normalised spectra using various sets of atmospheric input parameters and up to three spectral regions in order to verify 
the reliability of our results. 

\section{Classification}
\label{sect:clas}

Based on the shape of the one-dimensional CCFs and the evolution of the radial velocity measurements 
with time, we classified each target according to the following categories: S (for stable), VAR (for 
currently unexplained, possible long-term RV variations), SB (for a spectroscopic binary or triple
system), P (for a pulsating star with line-profile variability and/or rotating with the presence of 
structures on the surface such as chemical spots or temperature gradients) and CMP (i.e. composite, 
for stars with a narrow, shallow and almost central absorption feature in their -- 
usually -- broad profiles). The P-class may contain A/F-type pulsators of the $\delta$ Scuti type where 
chemical peculiarities appear on the stellar surface due to the 
presence of a (weak) magnetic field \citep[e.g. KIC~5988140 = HD~188774,][]{Neiner2015MNRAS.454L..86N} or where 
stellar rotation is so fast that it deforms the stellar surface and perturbs the homogeneity of the surface 
intensity distribution \citep[e.g.][]{Bohm2015A&A...577A..64B}.  
Our determination of \vsini alone does not suffice to distinguish between both scenarios. We provide a class for 
each target of the sample based on the currently available spectroscopic information in Table~\ref{tab:param1} (col.~9). 
To justify the adopted classification, we provide illustrations for all objects, except for the ones presented 
in the text, in the form of one-dimensional CCFs (computed with a mask of type F0 in the wavelength range 510-570\,nm 
for all but for KIC~9775454 where we used a K0-mask) as well as radial velocity plots in Appendices~\ref{sect:CCFs} 
and~\ref{sect:RVs}, respectively. We present the list of all radial velocity measurements in 
Tables~\ref{tab:radvel1}\footnotemark[2]~and~\ref{tab:radvel7}\footnotemark[2].\\ 

\footnotetext[2]{An electronic version is available at the CDS.}

Hereafter, we shortly comment on the classification of various particular targets, including the double- and triple-lined 
systems. For a detailed description of the orbital solutions, however, we refer to Sect.~\ref{sect:orb}. Further results 
concerning the classification of the selected targets will come from our analysis of the photometric data of the {\it Kepler} 
mission and their periodograms (cf. Sect.~\ref{sect:timeDelay}). \\

\section[]{Characterization of stellar atmospheres} 
\label{sect:phys}

Table~\ref{tab:param1}\footnotemark[2] summarises the physical information derived from the study of the CCFs and the spectral analyses. 
This includes the atmospheric stellar properties (cf. sect.~\ref{sect:aps}), as well as the classification of the target into one
of the subsequent categories: S (stable), VAR (RV variable), SB(1/2/3) (resp. single-, double-, or triple-lined spectroscopic system), 
P (pulsating or possibly rotating) or CMP (composite spectrum). We list the following information: the identifier (col.\,1), period (col.\,2), 
spectral type (col.\,3), \teff (col.\,4), \logg (col.\,5), and magnitude (col.\,6) all from the Kepler Input Catalogue (KIC) 
except for the spectral type which comes from \citet[][\, Table~1 and references therein]{Uytterhoeven2011A&A...534A.125U}, 
the number of \textsc{Hermes} and \textsc{Ace} spectra collected (col.\,7), a comment (col.\,8), the classification (col.\,9), and the model 
parameters adopted for reconstructing the (in casu composite) spectrum (col.\,10), the mean \teff (col.\,11), the mean \logg 
(col.\,12) based on three regions of interest ([620 - 686]~nm including H$\alpha$, [450 - 520]~nm including H$\beta$, [415 - 460]~nm 
including H$\gamma$), allowing to check for consistency, and, lastly, the \teff (col.\,13), \logg (col.\,14), and \vsini (col.\,15) 
as derived with \textsc{Girfit} in the wavelength range [415 - 450]~nm.\\ 

\begin{figure}
\begin{center}
\includegraphics[width=8.9cm,viewport = 5 27 365 270,clip=]{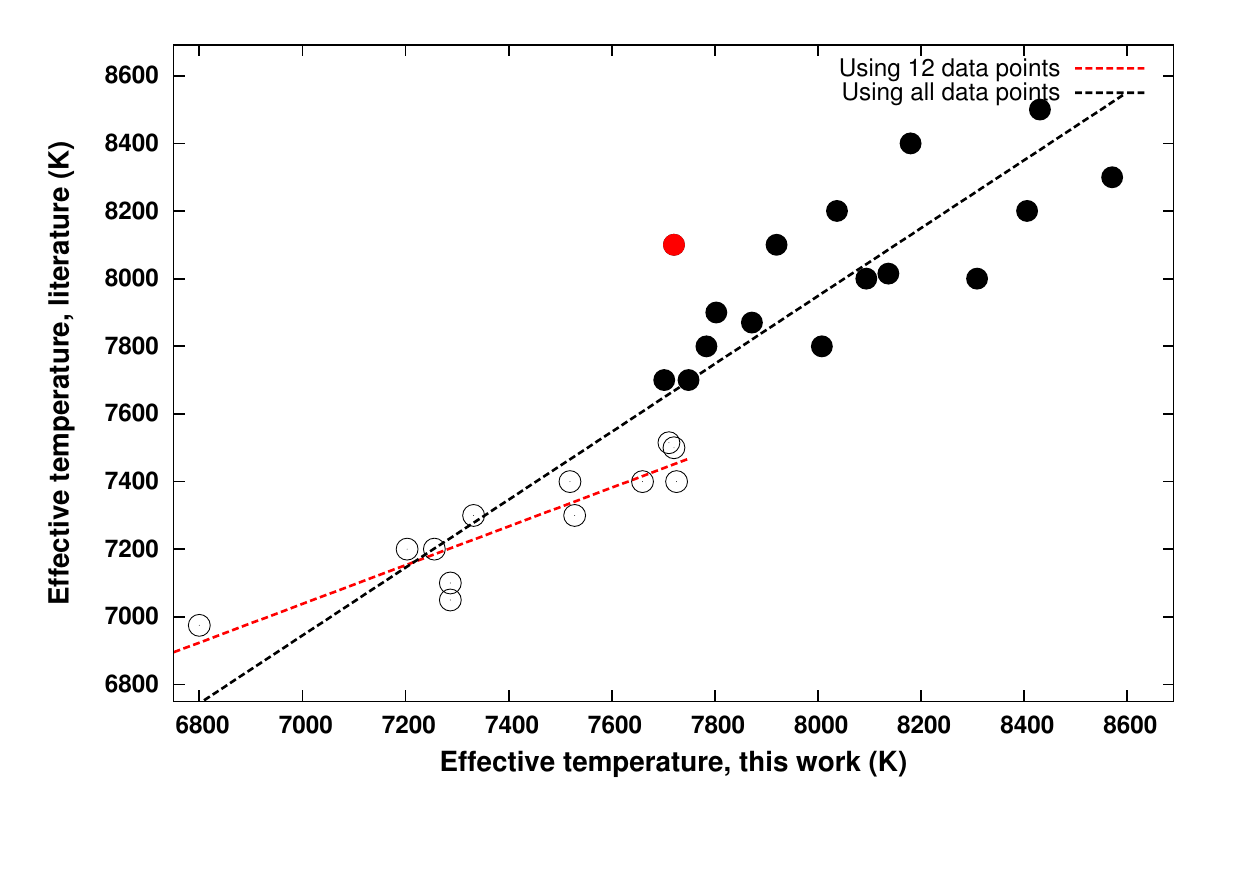}
\end{center}
\caption{Comparing the various determinations of \teffi. A linear fit using the literature values less than 7600~K (open symbols) 
shows an obvious discrepancy in the slope, illustrating the fact that our values for \teff less than 7700~K are systematically a little
larger than the literature values. The outlier (red symbol) is KIC~7119530 for which two very different literature values were found.} 
\label{fig:compT}
\end{figure}

\begin{figure}
\begin{center}
\includegraphics[width=8.9cm,viewport = 5 27 365 270,clip=]{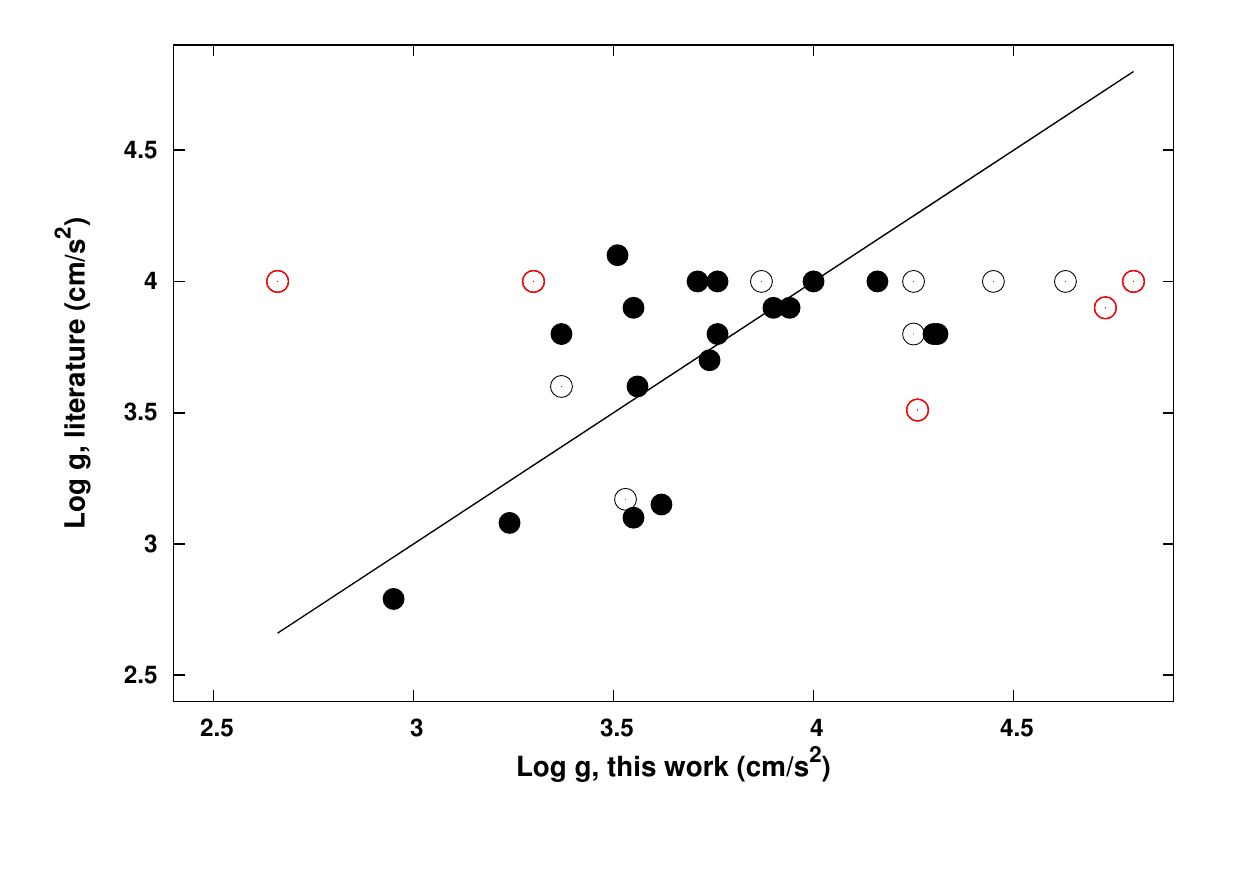}
\end{center}
\caption{Comparing the various determinations of \loggi. Same symbols as in Fig.~\ref{fig:compT}. The largest differences (red symbols) are 
found for KIC~3429637 (A9m),~5965837 (F2p) ,~9509296,~9764965 (Am), and~10537907.}
\label{fig:compG}
\end{figure}

\begin{figure}
\begin{center}
\includegraphics[width=8.9cm,viewport = 5 27 365 270,clip=]{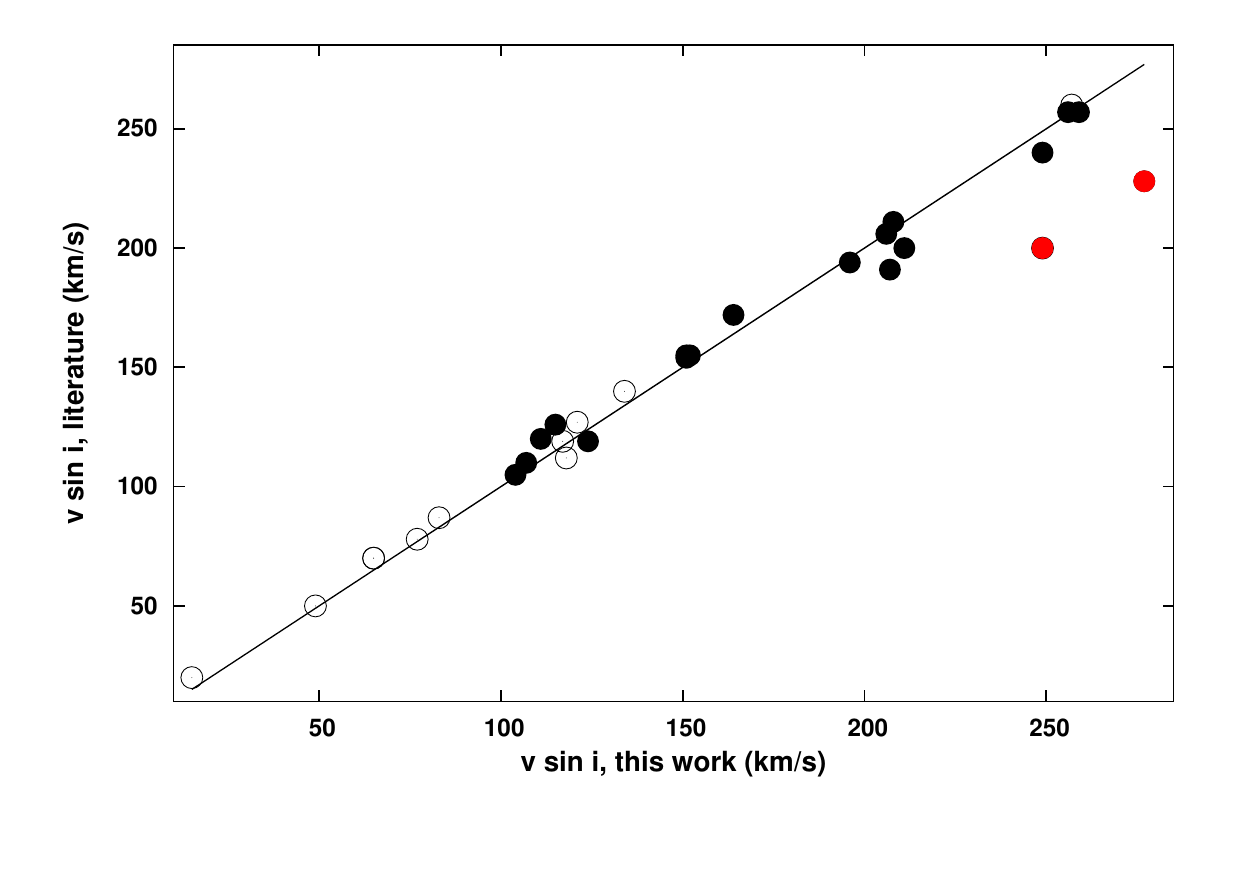}
\end{center}
\caption{Comparing the various determinations of \vsinii. Same symbols as in Fig.~\ref{fig:compT}. The largest differences (red symbols) are 
found for KIC~7119530 and~7827131.}
\label{fig:compV}
\end{figure}

We have 28 targets in common with other studies. This allowed us to perform a useful comparison. Table~\ref{tab:complit}\footnotemark[2] 
summarises the atmospheric stellar properties derived from this work, together with some recently published values. The referred studies 
are mentioned in the footnotes (col. 16 from Table~\ref{tab:complit}). In Figures~\ref{fig:compT},~\ref{fig:compG}, and~\ref{fig:compV} 
we illustrate the overall agreement between the various determinations. In particular, the agreement is excellent in \teff and 
\vsinii. The scatter in \teff matches well the estimated uncertainty of $\pm$ 250~K (Fig.~\ref{fig:compT}). While there is clearly no systematic 
trend in \vsinii, there seems to exist a small systematic offset in \teff in the temperature range [7300-7700]~K. Our temperature values appear 
to be slightly overestimated in this range with respect to the published values (or vice versa). This effect reminds us of the (known) fact that 
the KIC-values for \teff less than 7700~K are believed to be systematically underestimated. The scatter in \vsini reflects well the 
estimated uncertainty of $\pm$ 5-10~\kmsi\, but this depends a little on the absolute value (Fig.~\ref{fig:compV}). Two extremely fast rotating stars 
deviate from the 1:1 ratio line. The determination of \loggi, on the other hand, shows a huge spread which reflects its large uncertainty. It appears 
from Fig.~\ref{fig:compG} that some of the literature values were arbritarily set (or limited) to 4. This is often the case for targets cooler 
than 7700~K (open symbols). We also should mention that a comparison between previously published values (when possible) shows a similar degree 
of inconsistency, meaning that the errors are larger than thought. In five cases (e.g. KIC~3429637~(Am), KIC~5965837, KIC~9509296, KIC~9764965~(Am), 
and KIC~10537907), our determination of \logg is very different from that of \citet{Catanzaro2011MNRAS.411.1167C} or \citet{Niemczura2015MNRAS.450.2764N}.  
Two such cases (KIC~3429637 and KIC~5965837) are treated in detail in the discussion of the 22 individual targets hereafter. The remaining 
lack of agreement might also be due to the choice of the spectral intervals which may not be very sensitive to \logg for stars in the temperature 
range below 8000~K. Still, we conclude that there is no sign of any systematic trend in this plot and that other data points tend to follow 
the 1:1 ratio line, illustrating the existence of a rough agreement. \\  

\footnotetext[2]{An electronic version is available at the CDS.}

Searching for common targets also based on lower resolution spectra, we identified 22 targets (26 measurements) in common with 
the {\sc Lamost} project \citep{2015ApJS..220...19D}. In Table~\ref{tab:rotfit}\footnotemark[2], we compare the atmospheric stellar properties 
derived from this work to those derived by the {\sc Lamost} teams using the procedure {\sc Rotfit} \citep{2003A&A...405..149F,2016A&A...594A..39F}. 
In four cases, different values were derived from the {\sc Lamost} spectra (with a resolution of R $\sim$ 1800). Large differences, in \logg mostly, 
occur for the two fast rotators KIC~3453494 and KIC~9650390, as well as in the case of KIC~7748238 (because of our large value of \logg). In two 
cases (KIC~7668791 and KIC~9650390), our value of \teff differs from that of the {\sc Lamost} team, though our determinations of both \teff and 
\logg agree very well with the KIC values. This comparison illustrates the fact that the largest differences occur more often in the fast rotating 
A/F-type stars (for which the estimated errors are also large). \\

\subsection{KIC~3429637 = HD~178875}
HD~178875 is {\it not} a candidate A/F-type hybrid star from \citet{Uytterhoeven2011A&A...534A.125U}. This object was classified 
as a normal $\delta$ Scuti star while being suspected of binarity \citep{Catanzaro2011MNRAS.411.1167C}. For this reason, we 
included it in our programme. 
The RV data show variability with an amplitude smaller than $\sim$ 5~\kms on a time scale of several months, but no day-to-day 
variations (cf. Appendix~B). We found no direct trace of spectroscopic duplicity. Since we also detected line profile 
variations in the CCFs, we classified it as a pulsator accompanied by a long-term RV variability (``P+VAR''). We obtained 
the following stellar properties: \teff = 7266~K, \logg = 2.66, and \vsini = 49~\kmsi. While there is a good agreement in 
\teff and \vsini with \citet{Catanzaro2011MNRAS.411.1167C}, we however do not confirm their value of \logg = 4. 
In Fig.~\ref{fig:compG}, we notice a corresponding large discrepancy. We confirm its status of evolved Am star 
\citep[cf.][]{Murphy2012MNRAS.427.1418M}. Since this star is known to have a peculiar spectral type (kF2hA9mF3), its 
non-standard surface composition may interfere with the gravity determination.\\

\subsection{KIC~3453494 }
The CCF profiles of this fast rotating star (estimated \vsini of 220~\kmsi) are extremely broad and noisy which makes it difficult 
to detect any kind of variability. Nothwithstanding the noise and given the limited accuracy, we conclude that the radial velocity 
and the profiles are stable (``S''). We encountered several other cases of extreme stellar rotation, namely KIC~6756481,~7119530,~7827131,
~9650390 and~11602449, as well as KIC~4989900 and~6670742 (the latter two classified as ``S/P?'', with some ambiguity in their 
classification) (cf. Sect.~\ref{sect:con}).\\

\subsection{KIC~4480321 }\label{spec_KIC4480321}
This object is a new spectroscopic triple system (SB3). The spectrum shows the presence of a twin-like inner binary orbiting a 
slightly more luminous and fast rotating third component (cf. Fig.~\ref{fig:KIC44} and the CCF profiles in Appendix~A). We have 
excellent RV coverage for the close binary system and estimate the uncertainty of the component RV measurements to be significantly 
smaller than 1\,\kmsi. From two campaigns of 11 nights each, we derived an initial value near 9 days for the orbital period of the close 
binary. An orbital solution for the inner binary is presented in Sect.~\ref{sect:orb}. A tentative orbital solution for the outer system 
is also provided, though the period is probably longer than the available time span.\\ 

Using a composite model consisting of three synthetic spectra (A5, \vsini = 160), (F0, \vsini = 10) and (F0, \vsini = 10), we were 
able to reconstruct parts of the observed spectra very well (see Fig.~\ref{fig:KIC44}). We next applied the extended 3-D version 
of \textsc{Girfit} in three different regions of the normalised spectra using a composite model to find the solution which minimizes the 
residuals between the model and the observations. We fixed the \logg values to 4 (for MS phase) and the projected rotational 
velocities to our previous estimations (of resp. 160~\kmsi, 10~\kms and 10~\kmsi), and varied the initial values for the effective 
temperatures and the two luminosity ratio's (the only free parameters left). We found the component effective temperatures of \teffi$_{1}$ 
= 7900~$\pm$~100~K with both \teffi$_{2}$ and \teffi$_{3}$ in the range between 6300 and 6900~K (probably of similar temperature) in accordance 
with some of the observed spectra. However, no unique solution was found which would fit all our spectra in any of the inspected wavelength 
regions [415 - 450], [500 - 520] or [630 - 680]\,nm. The component properties are thus not well-known. We conclude that high-S/N spectra 
would be crucial for a more precise component characterization of this triple system.\\  

\subsection{KIC 5219533 }\label{spec_KIC5219533}
\begin{figure}
\begin{center}
\includegraphics[width=8.9cm]{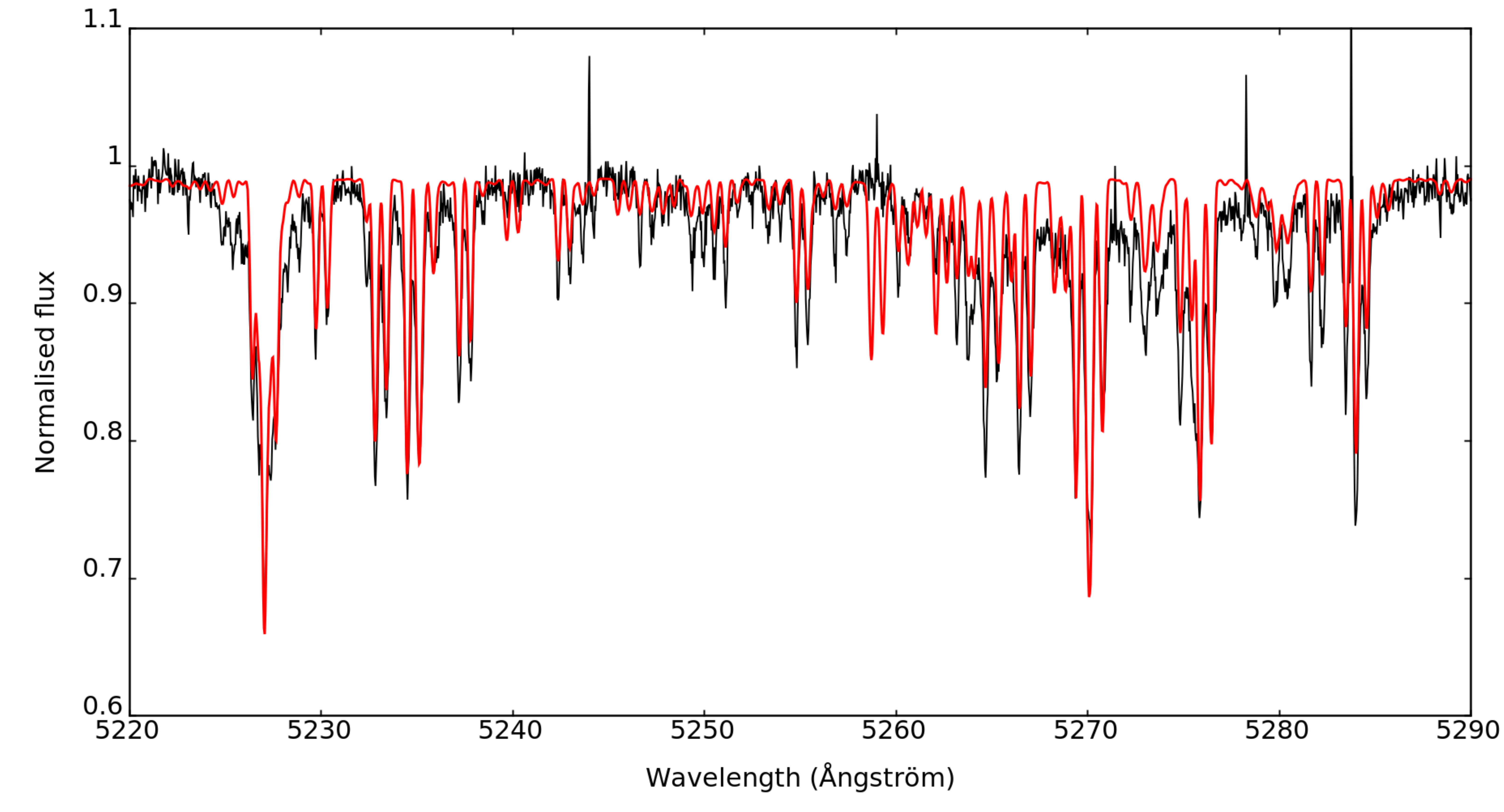}
\end{center}
\caption{Observed spectrum of KIC~5219533 (in black) vs. model spectra consisting of two narrow-lined components (in red). An extra 
broad-lined component of a similar spectral type could explain the additional wings and parts with missing continuum. } 
\label{fig:KIC52_sp1b}
\end{figure}

This target forms a visual double system with HD~189178 (5.5 mag, component~A) at an angular separation of 65\arcsec.  
It is a well-resolved, double-lined spectroscopic system (cf. Fig.~\ref{fig:CCF2}), with a primary component 
of type Am \citep[A2-A8,][]{Renson2009A&A...498..961R}.  
We furthermore deduce the presence of a more rapidly rotating component of a slightly cooler spectral type based on
two arguments: (a) from the unusual aspect of the (component and systemic) radial velocities (cf. Appendix~B), and (b) 
from a comparison of the observed spectra with models. For example, we found that a two-component synthetic spectrum 
is unable to reproduce the broad features found in various parts of the spectrum (cf. Fig.~\ref{fig:KIC52_sp1b}). We 
therefore conclude that this object is a new triple-lined spectroscopic system (SB3). An orbital solution for the inner 
binary is presented in Sect.~\ref{sect:orb}.\\

In order to compute reliable values for the component's atmospheric parameters and because the third component is very diluted, 
we applied the 2-D version of \textsc{Girfit} onto different regions of the normalised observed spectra in search of the 2-component 
model which best minimizes the residuals. Due to the complexity of this spectrum, we fixed the \logg values to 4 and kept the 
projected rotational velocities constant by adopting our previous estimations of \vsini (i.~e. 10~\kms for both components). 
The component effective temperatures and the luminosity ratio were the only free parameters. Allowing for more free parameters did 
not allow us to find any reliable solution. We chose various initial values and spectral regions, which gave us estimates for 
the uncertainties on the free parameters. Though both components are usually well-resolved, we were unable to find a unique 
solution for all our spectra. The most promising results were obtained in the interval [500 - 520]\,nm, where a consistent 
solution was found for most spectra (7 out of 12). In this case, the mean effective temperatures of \teffi$_1$ = 8300 $\pm$ 
100~K and \teffi$_2$ = 8200 $\pm$ 100~K coupled to a light ratio l$_1$ equal to 0.53 $\pm$ 0.02 were obtained. This indicates that 
the pair might consist of nearly identical ``twin'' stars of spectral type A5. In the ranges [415 - 450]\,nm and [640 - 670]\,nm, 
however, a more pronounced temperature difference between the components was derived but the solutions from the individual 
spectra showed less consistency. Since our models assume a standard solar composition, possible explanations for this behaviour 
might be the non-standard metallicity of the component(s) or the influence of the diluted third companion. A few spectra of 
high(er) S/N will be needed to improve the characterization of this new triple system.\\ 

\subsection{KIC~5724440 }
The CCF presents an extremely broad profile together with some additional features which seem to be variable. The broad profile 
appears to be stable in RV. We classified this target as ``P?''. The fast rotation could be the reason why previous determinations 
of \teff showed a large scatter \citep{Catanzaro2011MNRAS.411.1167C}, but our value matches perfectly both the 
spectroscopic determination by \citet{Niemczura2015MNRAS.450.2764N} and the photometric one by \citet{Masana2006A&A...450..735M}.
This object closely resembles KIC~6432054 which was also classified as a pulsator.\\

\subsection{KIC~5965837 }
\begin{figure}
\begin{center}
\includegraphics[width=8.9cm]{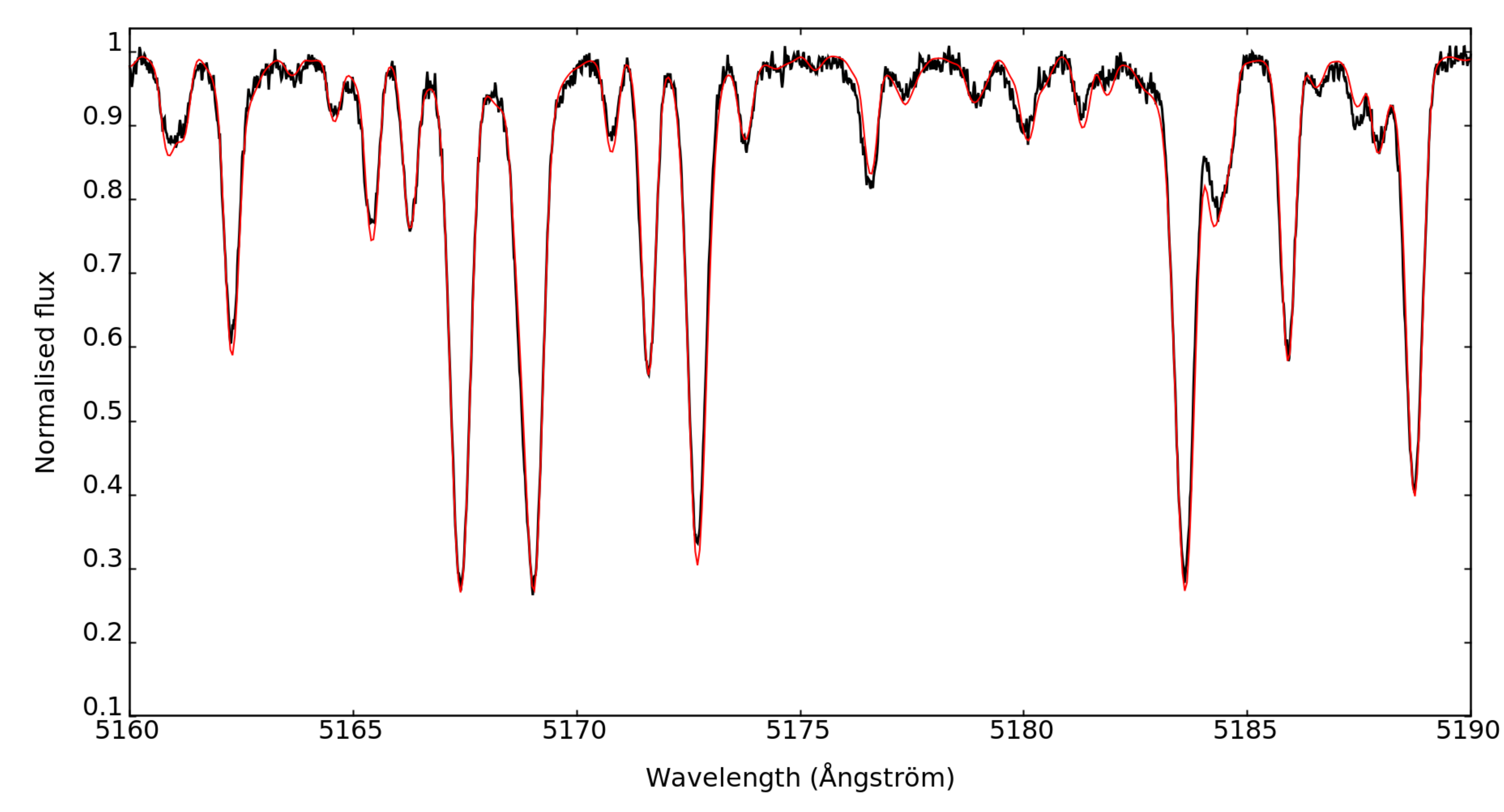}
\end{center}
\caption{Part of the observed spectrum for KIC~5965837 (in black) and corresponding part of the \textsc{elodie} 
spectrum of 20~CVn artificially broadened to the same rotation velocity (\vsini = 15~\kmsi) (in red).} 
\label{fig:KIC59}
\end{figure}

The CCF presents a narrow profile which is reflected by the extremely low \vsinii. The changes in the profile core are 
obviously due the presence of non-radial pulsations (classified as ``P'', cf. Appendix~A). The radial velocities show
a distinct variable pattern with a very low amplitude, a possible indication of radial pulsation. It is the coolest 
candidate hybrid star in our sample. We derived the following stellar properties: \teff = 6800~K, 
\logg = 3.3 and \vsini = 15 \kmsi. While there is an excellent agreement in \teff and \vsinii, 
we cannot confirm the value of \logg = 4 \citep{Catanzaro2011MNRAS.411.1167C}.
Instead, we derived \logg = 3.3 based on the \ion{Mg}{ii}~triplet using the spectral range [510-520]\,nm. Furthermore, we 
discovered that its chemical composition is non-standard, and heavily enhanced in metals. In fact, this star's spectrum 
almost perfectly matches that of 20~CVn, a well-known $\rho$ Puppis star (Fig.~\ref{fig:KIC59}). In conclusion, 
KIC~5965837 is a newly discovered $\rho$ Puppis star, i.e. an evolved and cool giant star with a metal-rich surface 
composition also showing pulsations. This is an interesting \textit{Kepler} star for a detailed study in view of the 
co-existence of pulsation and metallicism \citep{Kurtz1995MNRAS.276..199K}. \\ 

\subsection{KIC~6381306 }\label{spec_KIC6381306}
This object is another new spectroscopic triple system (SB3). It consists of a ``twin''-like binary and a slightly more 
luminous, more rapidly rotating primary component. For the inner binary system, we estimated an orbital period of 
$\sim$ 4~days. For the outer system, a periodicity of between about 100 and 200 days is expected. Both orbital solutions 
are presented in Sect.~\ref{sect:orb}. \\ 

We applied the extended 3-D version of \textsc{Girfit} in different regions of each spectrum to find the solution which 
best minimizes the residuals. The model spectrum is a composite of three components. We fixed the \logg values to 4 
and the projected rotational velocities were set close to our previous estimations (resp. 90-100~\kmsi, 5-10~\kms and 5-10~\kmsi). 
The component effective temperatures and two luminosity ratio's were the only free parameters left. We chose various 
initial values for these input parameters. Very similar results were found across both intervals [415 - 450] and [500 - 520]\,nm, 
though a consistent solution for only half out of 13 spectra was obtained at best. From these computations, we obtained the 
mean effective temperatures of \teffi$_1$ =  9000 $\pm$ 100 K, \teffi$_2$ = 7400 $\pm$ 100~K, and \teffi$_3$ = 7200 $\pm$ 50~K 
coupled to light factors of 0.79, 0.11 and 0.10 (for l$_1$, l$_2$ and l$_3$, respectively). It is moreover evident that 
such a complex model requires spectra of high(er) S/N ratio for a better characterization of this system. \\

\subsection{KIC~6756386 and KIC~6951642 }
Both objects have been classified as``P+VAR''. Their CCF profiles show clear line-profile variations which are due to 
the presence of stellar pulsations (cf. Appendix~A). In addition, the temporal evolution of both RV data sets indicates a 
long-term variability with a low amplitude, whose origin cannot yet be determined (cf. Appendix~B). 

\subsection{KIC~6756481 and KIC~7119530 }
\begin{figure}
\begin{center}
\includegraphics[width=8.9cm]{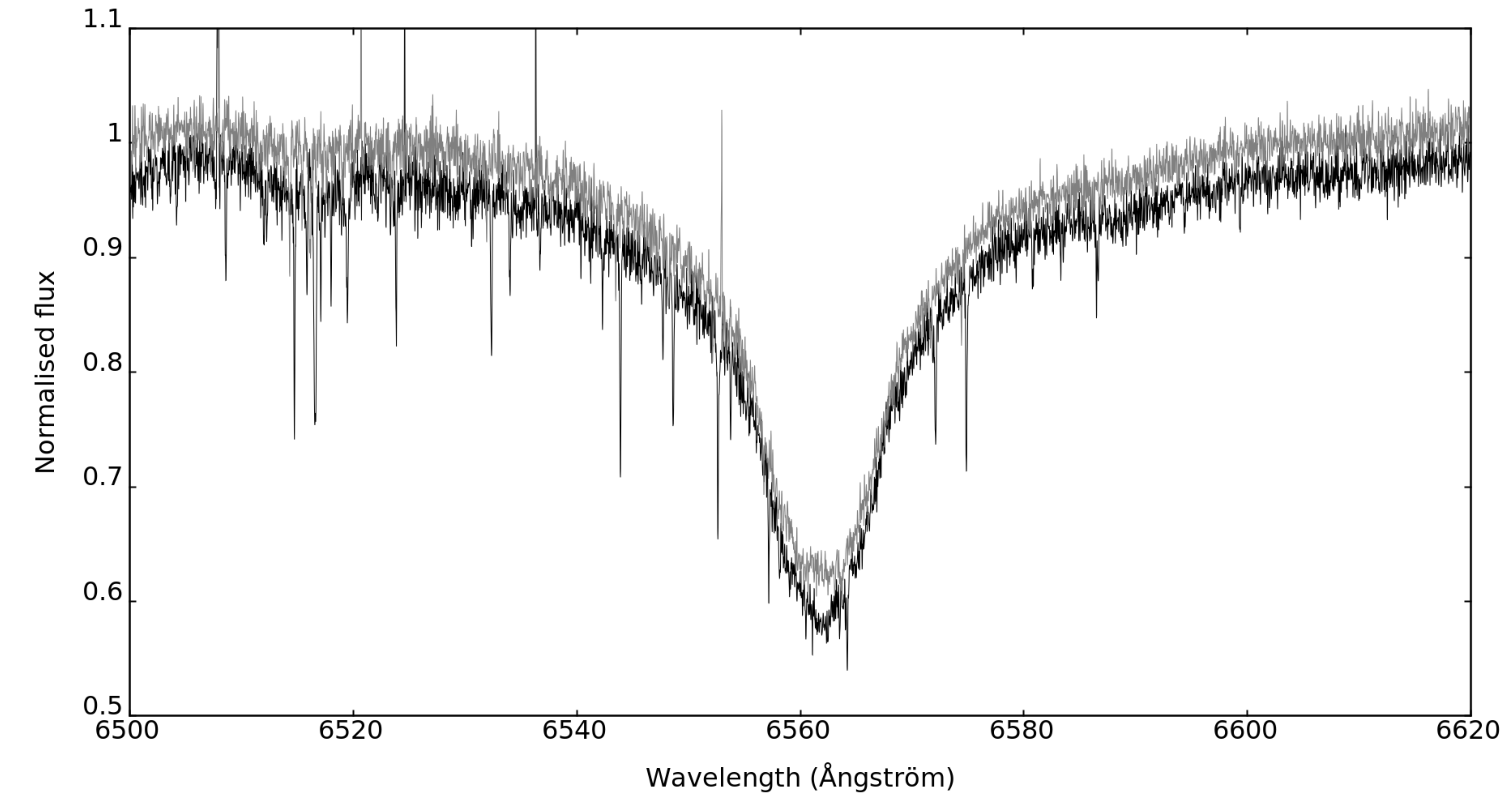}
\end{center}
\caption{Comparison of two H$\alpha$-profiles: KIC~6756481 (plotted in black) versus KIC~6670742 (in grey).} 
\label{fig:KIC6756481_Half}
\end{figure}

The CCFs present an extremely broad profile but with an additional sharp-peaked feature (cf. Appendix~A). 
Unlike the case of the very fast rotator KIC~5724440, we verified that the narrow central feature of 
KIC~6756481 remains constant on the velocity scale, suggesting the possible presence of a circumstellar 
shell. In such a case, the broad lines would correspond to the stellar photosphere and the narrow central
ones to a circumstellar shell \citep[e.g.][]{Mantegazza1996A&A...312..855M}. 
In the case of KIC~7119530, this feature is near central and appears to be stronger in the red part of the spectrum. 
The radial velocities of the broad absorption profiles are constant (cf. Appendix~B). In Fig.~\ref{fig:KIC6756481_Half}, 
we compare the H$\alpha$ profile of KIC~6756481 with that of KIC~6670742 of the same \teff and \vsini to look 
for a sharp absorption core \citep[e.g.][]{Henry2003AJ....126.3058H}. We see that both profiles are almost identical 
except in the small core region where KIC~6756481's profile is sharper and a bit deeper. In the case of KIC~7119530, 
we found a perfect match with the H$\alpha$ profile of KIC~4989900 (of the same \teff and \vsini).   
We classified both stars as ``CMP'' (which stands for composite spectrum). Their atmospheric parameters are 
close though not exactly identical, see e.g. the large projected rotational velocity of KIC~6756481. Due to its 
fast rotation, the uncertainties on the radial velocities are larger than usual, which makes it very difficult to 
detect any trace of variability. Similar cases among the late A/early F-type stars known in the literature are 
seven objects which were reported as possible shell stars by \citet{Fekel2003AJ....125.2196F}. An alternative 
explanation is that these objects consist of (at least) two components of similar type and without any detectable 
RV variation on the short term scale. Such physical cause was eventually confirmed \citep{Henry2003AJ....126.3058H}, 
as well as more recently by \citet{Fekel2015AJ....149...83F} for the triple system HD~207561.\\  

\subsection{KIC 7756853 }
\label{sub:KIC775}
The CCF shows a highly variable, sometimes asymmetric, profile due to the presence of a companion (cf. Fig.~\ref{fig:CCF1}). 
The variations of the RVs with time confirm the detection as a new double-lined spectroscopic binary (SB2). 
Both components present a \vsini difference of about 20~\kms (Table~\ref{tab:param1}). We applied the 2-D version 
of \textsc{Girfit} in different regions of each spectrum to find the solution which best minimizes the residuals. 
We explored the parameter space by varying the number as well as the initial values of the input parameters. To 
restrict the number of free parameters, we used our previous estimations of \logg and projected rotational 
velocity. The latter appeared to be reliable and very consistent, in such a way that we kept their values fixed in 
most fits. The \logg values of the models were fixed to the value 4 (MS phase). This also provided estimates 
of the uncertainties on the adopted input parameters. \\ 

While a good match was found with a model consisting of an A1-type primary with \vsinii$_1$ = 50~\kms and an A5-type secondary 
with \vsinii$_2$ = 30~\kms during the initial analysis, the actual process converged towards a pair of solutions signalling 
the existence of (at least) two minima. Imposing \logg = 4 and \vsinii, a first stable solution was found in the region 
[415 - 450]\,nm with \teffi$_1$ = 9600 $\pm$ 100~K and \teffi$_2$ = 8260 $\pm$ 60~K and a light factor l$_1$ equal to 0.61 
$\pm$ 0.03, but another stable solution was also found with a light factor l$_1$ equal to 0.54 after reversing the 
assignment of \vsini to the components. This smaller light factor indicates compensation for the fact that the \vsini 
value corresponding to the component with the lower \teff is larger.\\ 

\begin{figure}
\begin{center}
\includegraphics[width=8.9cm]{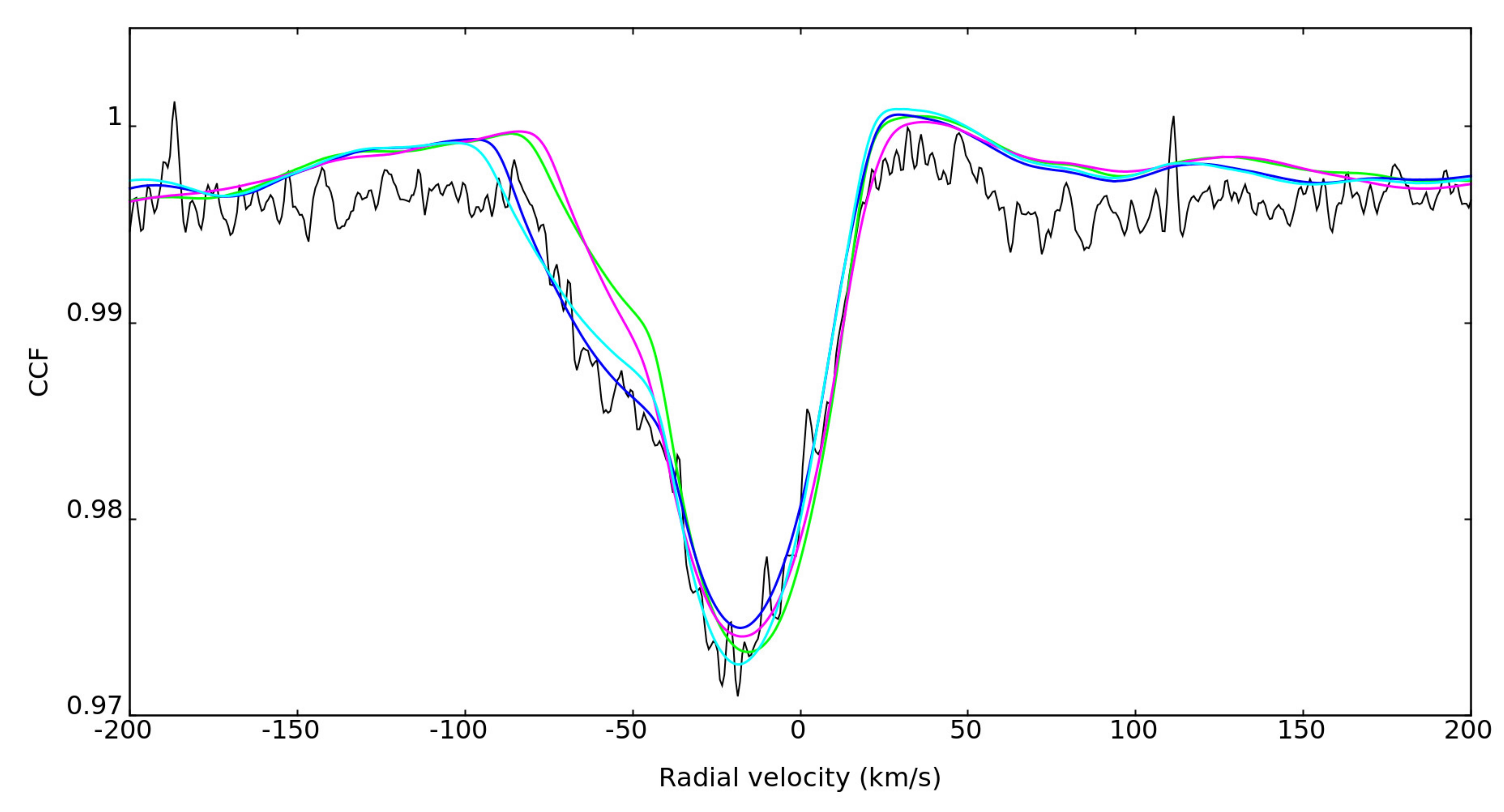}
\end{center}
\caption{Observed versus modelled CCF in the case of KIC~7756853. The spectral range used is [415 - 450]\,nm.
Model A is shown in green/lilac while model B is shown in (light/dark) blue. } 
\label{fig:KIC77_ccfcomb}
\end{figure}

In the region [500 - 520]\,nm, where the residuals were smallest, a first stable solution with a larger temperature differences
was found with \teffi$_1$ = 9880 $\pm$ 100~K and \teffi$_2$ = 7480 $\pm$ 170~K and a light factor l$_1$ equal to 0.73 $\pm$ 0.04 
(model A). However, exchanging the values of \vsinii, we found a second stable solution with \teffi$_1$ = 9980 $\pm$ 120~K and 
\teffi$_2$ = 7300 $\pm$ 140~K and a light factor l$_1$ equal to 0.66 $\pm$ 0.04 fitting all our spectra equally well (model B). 
Including \vsini as extra free parameters, similar values of \teff were retrieved. Model B thus consists of an A1-type primary 
with \vsinii$_1$ = 27 $\pm$ 5~\kms and an A5/F0-type secondary with \vsinii$_2$ = 57 $\pm$ 5 \kmsi. The distinction between the different 
solutions is hard to make as the computed composite spectra look very similar except for a few blended absorption lines.
We compared both possible models with a single observed spectrum in the range [415 - 450]\,nm and observed that 
the noise level in the spectrum does not allow to distinguish between them. To allow for better discrimination, 
we simulated the composite cross-correlation function for each model using a mask of spectral type A4 
(Fig.~\ref{fig:KIC77_ccfcomb}). This comparison shows that model B represents the observed composite CCF more 
adequately than model A. This is also confirmed by the residual values. In addition, we compared two spectra with 
the solutions of model B in the spectral range [386 - 404]\,nm. In 
this range, the solution with the higher temperature (\teffi$_2$ = 8400~K) reproduces the \ion{Ca}{ii}~line ($\lambda$ = 393\,nm) very well. 
This binary is clearly a difficult study case due to the large blends between the components. We tentatively adopted the solution 
named 'model B' found in the region [500 - 520]\,nm. \\

\subsection{KIC 7770282 }

\begin{figure}
\begin{center}
\includegraphics[width=8.9cm]{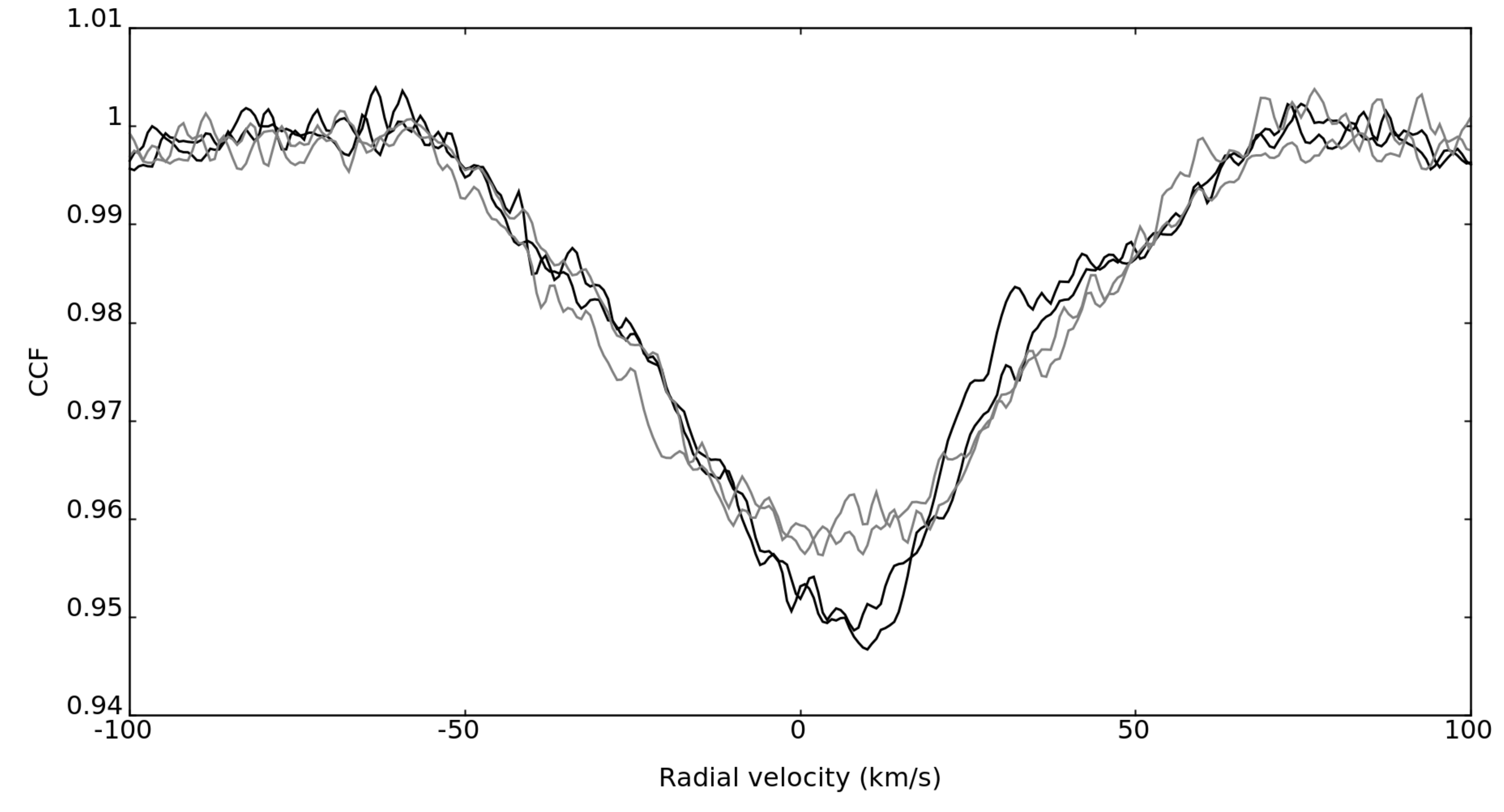}
\end{center}
\caption{CCF profiles of four different spectra of KIC~7770282. They illustrate the largest observed distortions.} 
\label{fig:K7770282_4ccf}
\end{figure}

The CCF profiles show clear asymmetric variations which were first attributed to the presence of a companion. 
Upon closer inspection, however, we found that the CCF shape is maximally distorted in the core and the red 
wing only, while the blue wing remains apparently unaffected (cf. Fig.~\ref{fig:K7770282_4ccf}). The RV data 
indicate a scatter higher than normal and thus possible variability of amplitude smaller than a few \kmsi,  
but we attribute this RV scatter to the line profile variations (cf. Appendix~B). We believe that the distortions 
seen in the CCFs are caused by features which are located at the stellar surface. They could be the signature 
of pulsations (''bumps``) or rotational modulation (''spots``). It should also be remarked that the dominant 
period found in the {\it Kepler} photometry is almost equal to one~day. We conclude that this object is (most)
probably not a spectroscopic binary, but should rather be considered as ``P''. It deserves to be re-examined 
in the light of several high(er) quality spectra.\\ 

\subsection{KIC 8975515 }

This object is another new double-lined spectroscopic system (SB2) with an orbital period of at least 1000~days. 
The RV curve presents a clear, well-defined modulation of the slowly rotating component with an amplitude 
smaller than 10~\kmsi. An orbital solution for this long-period binary system is presented in Sect.~\ref{sect:orb}.\\ 

Unlike the case of KIC~5219533, both components have dissimilar projected rotational velocities (cf. Fig~\ref{fig:KIC89}
and CCF profiles in Appendix~A). 
Therefore, we let the component effective temperatures, the \vsinii's and the luminosity ratio be free parameters but 
fixed the \logg values to 4. As for KIC~5219533, a wide variety of initial values and spectral regions was 
explored in the search for a consistent solution. In this case also, we observed that the resulting values depend 
on the chosen spectral region. In the interval [415 - 450]\,nm, the mean effective temperatures of \teffi$_1$ = 
6800 $\pm$ 20~K and \teffi$_2$ = 8250 $\pm$ 4~K coupled to \vsinii's of respectively 161 $\pm$ 1 and 32 $\pm$ 1\,\kms 
with a light ratio l$_1$ equal to 0.49 $\pm$ 0.02 were obtained. In the interval [500 - 520]\,nm, the mean effective 
temperatures of \teffi$_1$ = 7440 $\pm$ 20~K and \teffi$_2$ = 7380 $\pm$ 21~K coupled to \vsinii's of respectively 164 
$\pm$ 0.5 and 31 $\pm$ 1\,\kms with a light ratio l$_1$ equal to 0.65 $\pm$ 0.03 were obtained. In the interval 
[640 - 670]\,nm, the mean effective temperatures of \teffi$_1$ = 6960 $\pm$ 150~K and \teffi$_2$ = 7850 $\pm$ 150~K 
coupled to \vsini's of respectively 154 $\pm$ 2 and 32 $\pm$ 1\,\kms with a light ratio l$_1$ equal to 0.51 $\pm$ 
0.04 were obtained. Since the residual sum of squares (RSS) of all our spectra is much smaller in the [500 - 520]\,nm 
interval, we decided to adopt the atmospheric parameters derived from this region. This indicates that the pair might 
consist of two nearly identical stars of type A8. The projected rotational velocities \vsinii$_1$ and \vsinii$_2$ are 
safely determined to be 162 $\pm$ 2 and 32 $\pm$ 1 \kmsi. In the ranges [415 - 450]\,nm and [640 - 670]\,nm, however, 
a more pronounced temperature difference (of $\sim$ 1000~K) between the components was found, as in some other cases. 
Follow-up as well as high(er) S/N spectra will be required to obtain more accurate component properties of this system.\\   

\subsection{KIC 9700679 }
The few {\sc Hermes} spectra that were acquired indicate an obvious shift in the radial velocity (cf. 
Appendix~B). This object is another single-lined spectroscopic system (SB1). Since it has a KIC temperature 
of about 5070~K, it is (much) too cool to be considered as a potential A/F-type hybrid star.\\

\subsection{KIC 9775454 }
The CCF (computed with a mask of type K0 instead of F0) presents clear distortions in shape, accompanied by the presence 
of a narrow feature which moves but remains close to the central position. Furthermore, we detected a change of small 
amplitude in RV (cf. Appendix~A). This is most certainly another long-term spectroscopic binary system. We classified 
this object as ``SB1'', though some lines due to the secondary component obviously affect the observed spectrum (the companion 
is difficult to detect, until we obtain radial velocities for the secondary, we will not use the ``SB2'' classification). 
Based on the appearance of the RV curve (cf. Appendix~B), we suggest a simple estimate of the order of 1700~days 
for the period. We plan to continue the RV monitoring of this system and to search for an orbital solution of type SB2. 
Since the time-delay (TD) analysis supports the existence of a similar periodicity (cf. Sect.~\ref{sect:timeDelay}), we 
furthermore intend to combine both data types into a joint analysis in the future. This object is one of the coolest 
objects of the sample. \\

\subsection{KIC 9790479 }
The CCF presents a probable stable profile accompanied by long-term variations in RV. We classified it 
as a single-lined spectroscopic binary (SB1). A possible orbital period is of the order of 230~days 
(see Fig.~\ref{fig:KIC97_orb} in Sect.~\ref{sect:orb}). This system needs follow-up observations 
during the next two years.\\ 

\subsection{KIC 10537907 }
The CCF profiles clearly show the signature of pulsations (cf. Appendix~A), also accompanied by a small shift in RV 
on a time scale of $\sim$ 1500 days (cf. Appendix~B). We classified it as a probable single-lined spectroscopic 
binary with pulsations (``P+SB1''). The system also needs additional follow-up observations.\\ 

\subsection{KIC 10664975 }
Some CCF profiles of higher quality display features in the form of ''moving bumps`` probably caused by pulsations 
(cf. Appendix~A), while the RV data are stable (cf. Appendix~B). We thus classified it as a pulsator (''P``).\\ 

\subsection{KIC 11180361 }
The CCF shows an extremely broad and noisy profile which also appears to be stable in RV (cf. Appendix~B). This target 
(KOI-971) is a new Kepler eclipsing binary system \citep{Slawson2011AJ....142..160S} 
with an orbital period of 0.2665~days (instead of the \textit{Kepler} value of 0.5330~days). It has been classified 
as stable (``S'') based on our adopted criteria. This means that the secondary component does not visibly affect 
the observed spectrum.\\

\subsection{KIC 11445913 }
\begin{figure}
\begin{center}
\includegraphics[width=8.9cm]{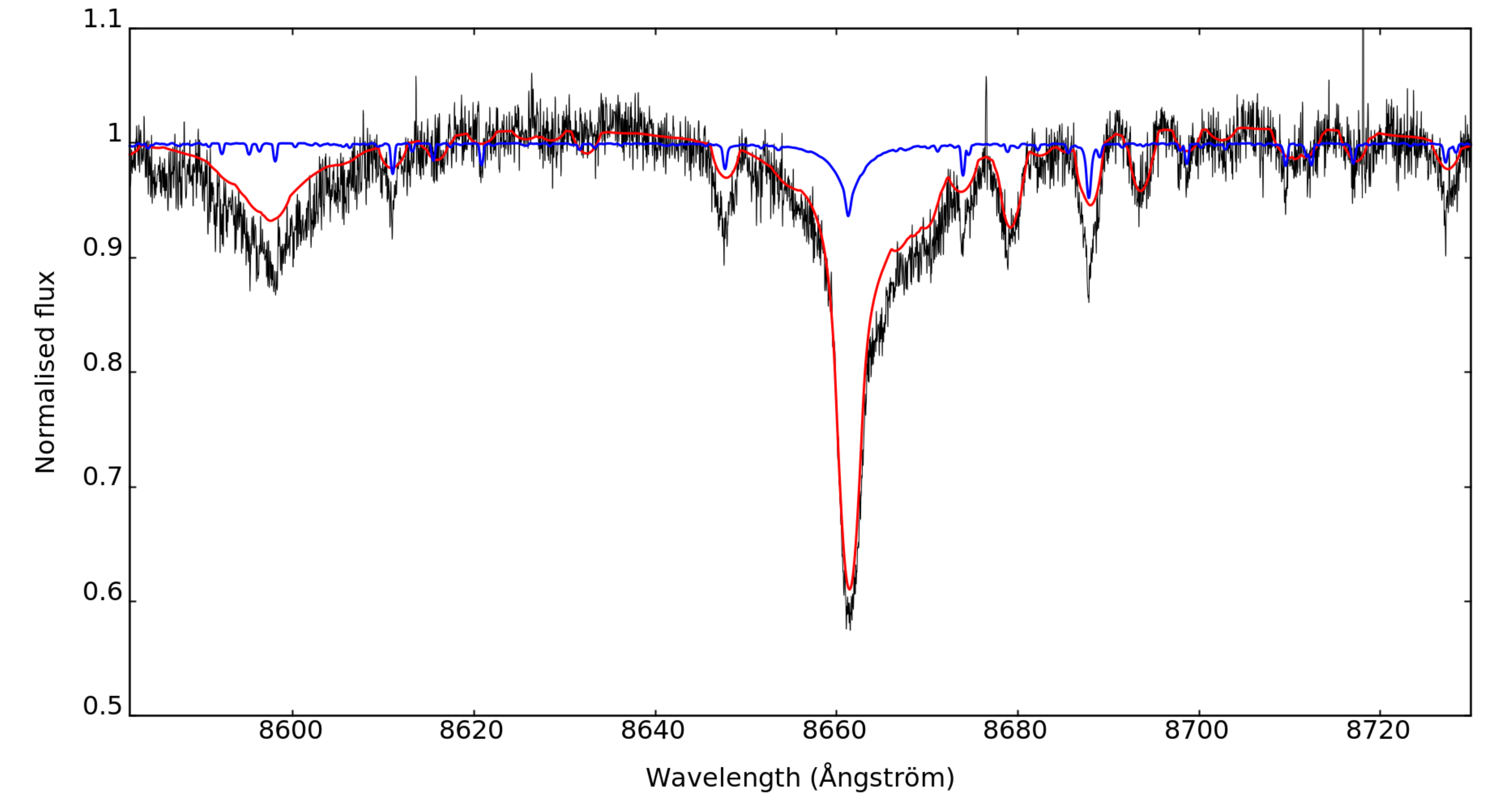}
\end{center}
\caption{Observed versus modelled spectrum for KIC~11445913 showing the two distinct contributions in red and blue. 
The plotted range includes the Paschen series as well as the \ion{Ca} triplet. } 
\label{fig:KIC11445913_sp1}
\end{figure}

The CCFs show a superposition of two components (cf. Appendix~A). This is a new double-lined spectroscopic 
system (SB2) consisting of an early F-type star with a K-type companion, both with a low \vsini 
(Fig.~\ref{fig:KIC11445913_sp1}). The primary component is probably of type Am. The variations in RV indicate a 
long-term change, mostly due to one older measurement and another measurement apparently in anti-phase with 
respect to the remainder of the data (cf. Appendix~B). The orbital period is not yet known.\\ 

A variety of initial values and spectral regions was explored in the search for a consistent match 
between the observed spectra and the models. We let the component effective temperatures, the \vsini's 
and the luminosity ratio be free parameters while fixing the \logg values to 4. Here again, the best fits 
were obtained in the [500 - 520]\,nm interval. Thus, we decided to adopt the atmospheric parameters derived 
from this region. The mean effective temperatures of \teffi$_1$ = 7180 $\pm$ 20~K and \teffi$_2$ = 5750 
$\pm$ 21~K coupled to \vsinii's of respectively 55 $\pm$ 1 and 8 $\pm$ 4\,\kms with a light ratio l$_1$ 
equal to 0.95 $\pm$ 0.01 were obtained. This solution was found to be consistent with all our spectra. In the 
[630 - 680]\,nm interval, the temperature difference between the components appeared to be about 500~K larger, 
though the quality (in terms of RSS) is worse. This system definitely deserves more follow-up observations in 
the next years.\\ 

\subsection{KIC~11572666 }

\begin{figure}
\begin{center}
\includegraphics[width=8.5cm]{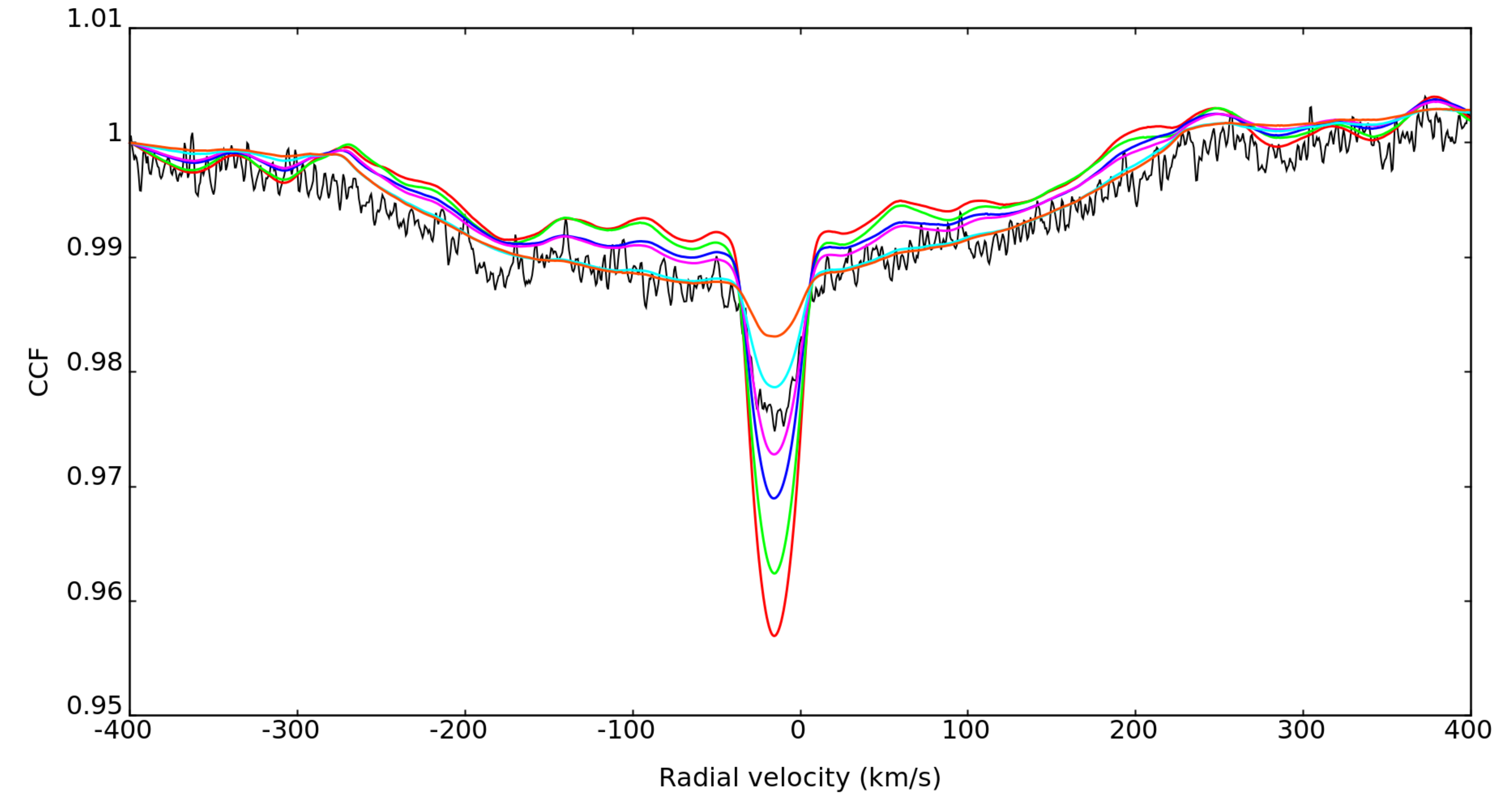}
\end{center}
\caption{Observed versus modelled CCF in the case of KIC~11572666. The combinations used for the models are F0-G2, F0-F3, A7-G2, 
A7-F3, A5-G2 and A5-F3. The closest match is found for the model with two early F-type components of very different \vsini (shown in 
light blue). The next best match is for the A7-G2 model (shown in pink). }
\label{fig:KIC115_ccfcomb}
\end{figure}

KIC~11572666 is another double-lined spectroscopic binary whose lines are strongly blended, with one extremely broad 
and one narrow component (SB2). It is very similar to the systems analysed by \citet{Fekel2003AJ....125.2196F}. We used 
the extended version of {\sc Girfit} to derive the component's spectral characteristics. For each observed spectrum, 
we chose several models and used different initial values based on our previous estimations of the spectral types and 
the projected rotational velocities. Here, we also let the projected rotational velocities as well as the \logg values 
be free parameters. For consistency, we repeated the fitting process with both \loggi's fixed to 4 (MS phase). 
In this case, a unique model appeared to be compatible with all our spectra. In the spectral range [415 - 450]\,nm, we 
found the mean parameters \teffi$_1$ = 7950 $\pm$ 40 K, (\loggi$_1$ = 3.9 $\pm$ 0.2 or 4 (fixed)), \vsinii$_1$ = 267 $\pm$ 5\,\kms 
in combination with \teffi$_2$ = 6000 $\pm$ 150 K, \loggi$_2$ = 4 (fixed) and \vsinii$_2$ = 22 $\pm$ 3\,\kmsi. In the spectral 
interval [500 - 520]\,nm, we derived \teffi$_1$ = 7900 $\pm$ 80 K, \vsini$_1$ = 253 $\pm$ 2\,\kms in combination with 
\teffi$_2$ = 6150 $\pm$ 200 K and \vsinii$_2$ = 20 $\pm$ 2 \kmsi. Both solutions agree with each other, the model consists of 
a rapidly rotating F0-type primary with \vsinii$_1$ = 250~\kms and a F/G-type secondary with \vsinii$_2$ = 20~\kmsi. We also 
simulated the composite cross-correlation function using a mask of spectral type A4 (Fig.~\ref{fig:KIC115_ccfcomb}). 
This plot shows that the adopted model represents the observed composite CCF extremely well. The RVs were previously determined 
using the model of an A5-type primary with \vsinii$_1$ = 250~\kms and an F3-type secondary with \vsinii$_2$ = 20~\kmsi. 
The RVs of the secondary component are very well-defined, whereas those of the primary component are only poorly determined 
due to its fast rotation (cf. Fig.~\ref{fig:KIC115_orb1}). This system contains a primary component of type Am and requires 
further RV monitoring.\\        

\section{Multiplicity rate and orbital solutions}
\label{sect:orb}

We repeatedly observed 50 targets as well as KIC~3429637 (previously classified as a $\delta$ Scuti star) with \textsc{Hermes}
and \textsc{Ace}, and we found direct evidence for spectroscopic duplicity in ten cases. This concerns the single-lined (SB1) 
systems KIC~9700679,~9775454 and~9790479, the double-lined (SB2) systems KIC~7756853,~8975515,~11445913 and~11572666 and the 
triple-lined (SB3) systems KIC~4480321,~5219533 and~6381306. The number of well-detected new spectroscopic systems 
represents 20\% of the total sample. Furthermore, we also identified two extremely fast rotators where a narrow and almost 
central feature can be noticed superposed onto a very broad stellar profile which we classified as ``C(o)MP(osite)'' (i.e. KIC~6756481 
and~7110530). In both cases, the central feature might be linked to the existence of a shell-like contribution or to an unknown 
stellar companion \citep{2015AJ....149...83F}.\\  

With respect to the sample of A/F-type candidate hybrid stars, we will not consider KIC~9700679 which is too
cool for an A/F-type star in the following discussion. Thus, we detected nine spectroscopic systems among 49 targets. 
In addition, we should consider the targets with a long-term and low-amplitude variability of their radial velocities 
classified as ``VAR'', as these may turn out to be (mostly single-lined) long-period systems. The number of such detections 
is three, which gives a total number of 12 (out of 49), corresponding to a spectroscopic multiplicity fraction of 24\%. 
If we add to this the known eclipsing binary KIC~11180361, we derive a global multiplicity fraction of 27\%. This number 
indicates that {\it at least} 1/4 of our sample of candidate hybrid stars belongs to a binary or a multiple system of stars.\\

For four systems (with periods smaller than 100~days), our RV measurements have sufficient phase coverage to allow
a reliable determination of the orbital period and a search for an orbital solution. In six more cases (of which 
three refer to different systems), we propose a preliminary or plausible orbital solution only. In all the remaining cases, 
the RVs plotted as a function of time are found in Appendix~B.\\ 

\subsection{KIC~4480321 }
\begin{figure}
\begin{center}
\includegraphics[width=8.9cm]{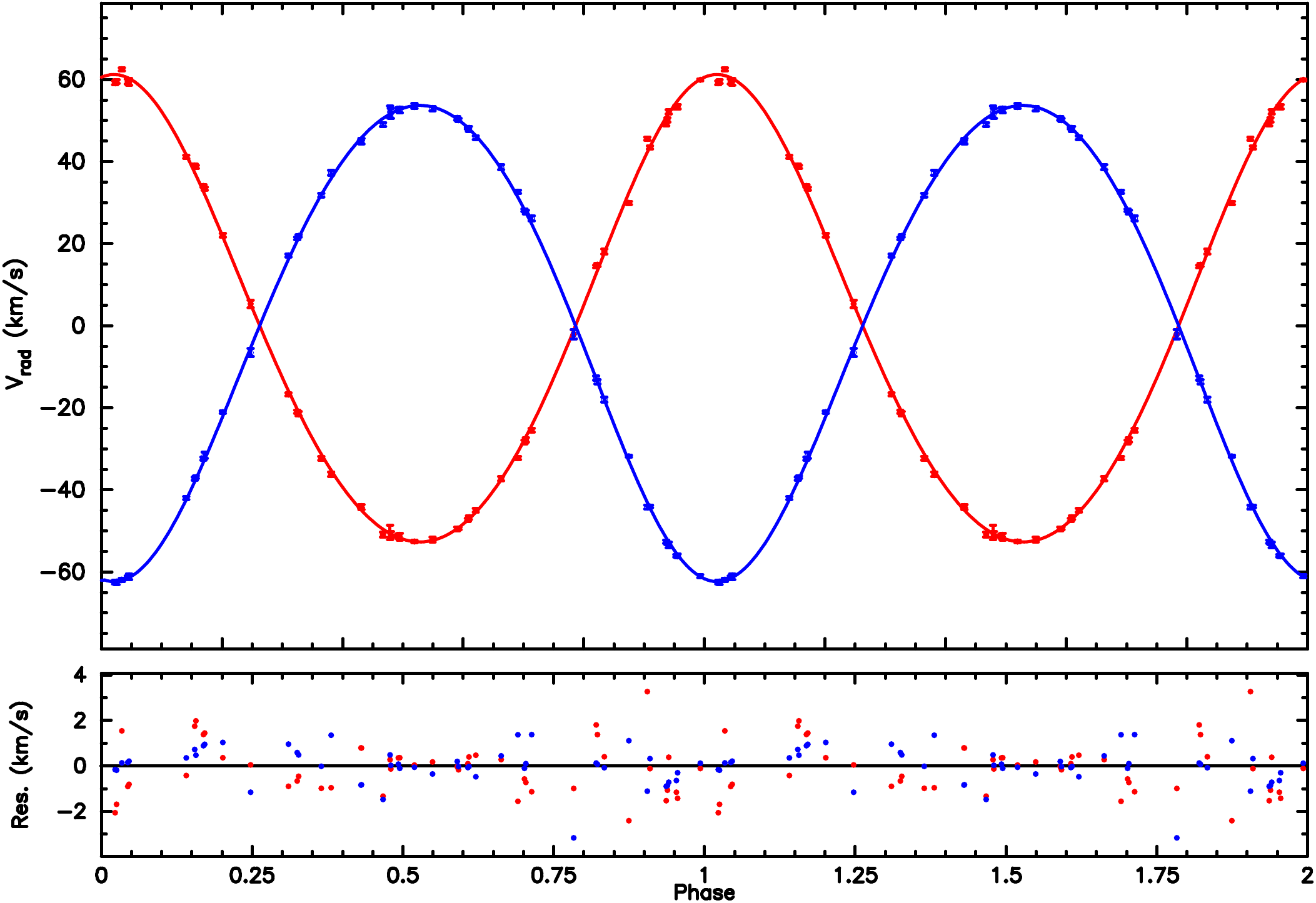}
\end{center}
\caption{Radial velocities for components A and B of the new triple-lined system KIC~4480321 plotted together with the orbital 
solution of the inner binary (AB) system and adjusted systemic velocity offsets. The residuals are shown in the bottom part.} 
\label{fig:KIC44_sb2}
\end{figure}

We computed a best-fitting orbital period of 9.2~days using \textsc{Period04} \citep{Lenz2014ascl.soft07009L}. 
This result was subsequently refined using an updated version of the Fortran code {\it vcurve\_SB} for 
double-lined systems (priv. commun. IvS, Leuven). The presence of the third component influences the systemic velocity
of the inner binary, thus we allowed for systematic offsets of its value with time. Both component RV curves 
are illustrated in Fig.~\ref{fig:KIC44_sb2}. The residuals are small and homogeneously distributed around null. 
The final orbital parameters are listed in the upper panel of Table~\ref{tab:vop1}. 

Concerning the wide system (AB-C), we found that orbital periods of the order of the time span (e.g. a period 
of 1500~days) or longer provided convincing solutions. Based on the currently available RVs, the best possible outer 
orbital solution (in terms of rms) has a period of about 2280~days. The RV curves of the centre of mass of the close 
pair AB, together with that of component C, illustrating the solution are displayed in Fig.~\ref{fig:KIC44_sb3} (Appendix~B). 
The parameters of this tentative orbital solution are listed in the bottom panel of Table~\ref{tab:vop1}.
We can derive a limitation on both inclinations if we consider that each component of the SB3 system should have a mass 
in the range [1 - 3.5] M$_\odot$ (following the conclusion in Sect.~\ref{spec_KIC4480321}). We thus would obtain the 
following conditions on $i$ and $i_{out}$: 36 < $i$ < 63$\degr$ and 42 < $i_{out}$ < 57$\degr$. 
In Sect.~\ref{sect:timeDelay}, we will show how the outer orbital solution can be confirmed and improved. However, to 
further constrain the parameters of the wide orbit, we will continue the long-term RV monitoring of this interesting 
system.\\ 

\begin{table}
\center
\setcounter{table}{3}
\begin{minipage}{8.9cm}
\centering \caption[]{\label{tab:vop1} Values and standard deviations of the constrained 
parameters of the orbital solutions for KIC~4480321. Note that from hereon, for all similar tables, 
$T_{\rm{0}}$ is expressed in Hel. JD - 2,400,000.}
\begin{tabular}{lr@{}l@{}lr@{}l@{}l}
\hline
\hline
\multicolumn{7}{c}{Solution A-B}\\
\hline
Orbital parameter &\multicolumn{3}{c}{Value} & \multicolumn{3}{c}{Std. dev.}\\
\hline
$P$ (days)&9&.&16592 &0&.&00006\\
$T_{\rm{0}}$ (Hel. JD)& 56523&.&25 &0&.&03\\
$e$  &0&.&0757 &0&.&0020\\
$\omega$ (\degr)&351&.&3 &1&.&4\\
$V_0$ (\kmsi)& (var.)& & & & &\\
$K_{\rm{A}}$ (\kmsi) & 56&.&94 & 0&.&15\\
$K_{\rm{B}}$ (\kmsi) & 58&.&00 & 0&.&15\\
$a_{\rm{A}} \mathrm{sin}\,i$ (AU) &0&.&04784 &0&.&00013\\
$a_{\rm{B}} \mathrm{sin}\,i$ (AU) &0&.&04873 &0&.&00013\\
$M_{\rm{A}} \mathrm{sin}\,i^3$ (M$_\odot$) &0&.&722 &0&.&004\\
$M_{\rm{B}} \mathrm{sin}\,i^3$ (M$_\odot$) &0&.&708 &0&.&004\\
$rms_{\rm{A}}$ (\kmsi) & 0&.&809 & & &\\
$rms_{\rm{B}}$ (\kmsi) & 0&.&473 & & &\\
\hline
\hline
\multicolumn{7}{c}{Preliminary solution AB-C}\\
\hline
Orbital parameter &\multicolumn{3}{c}{Value} & \multicolumn{3}{c}{Std. dev.}\\
\hline
$P$ (days)&2280&.& &29&.&\\
$T_{\rm{0}}$ (Hel. JD)& 54544&.& &46&.&\\
$e$  &0&.&09 &0&.&02\\
$\omega$ (\degr)&39&.& &5&.&\\
$V_0$ (\kmsi)&-19&.&28 & 0&.&12\\
$K_{\rm{AB}}$ (\kmsi) & 10&.&26 & 0&.&09\\
$K_{\rm{C}}$ (\kmsi) & 11&.&1 & 0&.&3\\
$a_{\rm{AB}} \mathrm{sin}\,i_{out}$ (AU) &2&.&14 &0&.&03\\
$a_{\rm{C}} \mathrm{sin}\,i_{out}$ (AU) &2&.&31 &0&.&06\\
$M_{\rm{AB}} \mathrm{sin}\,i_{out}^3$ (M$_\odot$) &1&.&17 &0&.&07\\
$M_{\rm{C}} \mathrm{sin}\,i_{out}^3$ (M$_\odot$) &1&.&09 &0&.&05\\
$rms_{\rm{AB}}$ (\kmsi) & 0&.&324 & & &\\
$rms_{\rm{C}}$ (\kmsi) & 1&.&658 & & &\\
\hline
\end{tabular}
\end{minipage}
\end{table}

\subsection{KIC~5219533 }

\begin{figure}
\begin{center}
\includegraphics[width=8.9cm]{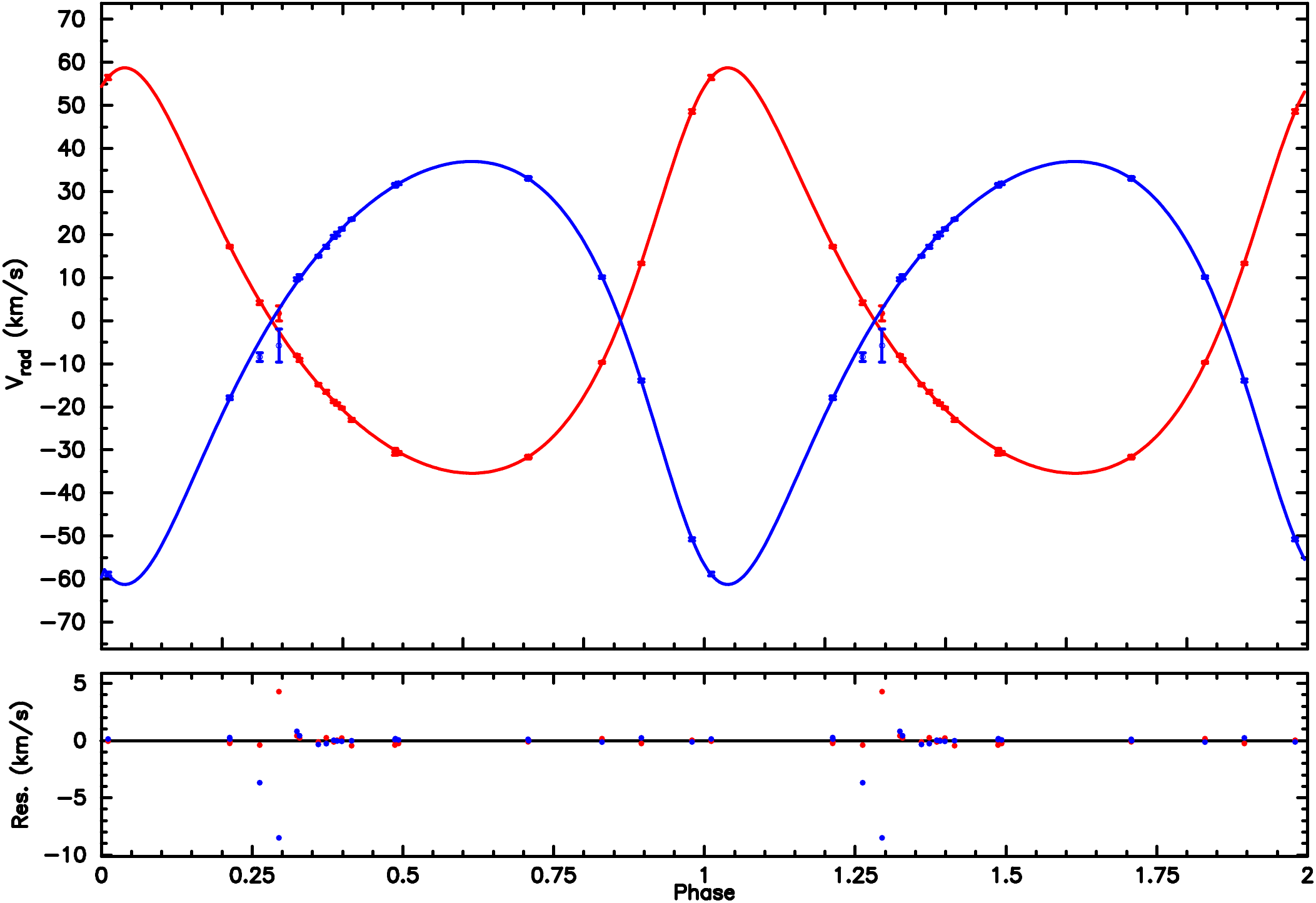}
\end{center}
\caption{Radial velocities for components A and B of the new triple-lined system KIC~5219533 plotted with a 
plausible orbital solution for the inner binary (AB) system. The residuals are shown in the bottom part.} 
\label{fig:KIC52_sb2}
\end{figure}

From the RV plot, we estimated an orbital period of the order of 32~days and derived a corresponding 
plausible orbital solution for the inner binary (AB) of this new SB3 system. Our solution confirms the twin 
character of the inner pair. The orbital parameters are listed in the upper panel of Table~\ref{tab:vop2}. 
Due to the presence of the (diluted) third component which influences the systemic velocity of the inner binary, 
we allowed for variability of this parameter. Both component RV curves are illustrated in Fig.~\ref{fig:KIC52_sb2}. 
We find a limitation on the inclination $i$ if we consider that each component of the inner binary should have a mass 
in the range [1.5 - 3.5] M$_\odot$ (following the conclusion in Sect.~\ref{spec_KIC5219533}). We then obtain 
the condition 46.5 < $i$ < 72$\degr$.
A preliminary solution for the systemic radial velocity of the close pair indicates that a period of the order 
of 1600~days accomodates the current RV data well (see Fig.~\ref{fig:KIC52_sb1}, Appendix~B). The parameters of 
this tentative orbital solution are listed in the bottom panel of Table~\ref{tab:vop2}. One data point is an obvious outlier 
caused by the presence of a blend. Additional RVs are planned to determine a more accurate orbital solution for the AB pair, 
as well as to better constrain the wide orbit of this system. In Sect.~\ref{sect:timeDelay}, we will present new evidence 
for the outer orbital solution.\\

\begin{table}
\center
\begin{minipage}{8.9cm}
\centering \caption[]{\label{tab:vop2} Values and standard deviations of the constrained parameters of the orbital 
solutions for KIC~5219533.}
\begin{tabular}{lr@{}l@{}lr@{}l@{}l}
\hline
\hline
\multicolumn{7}{c}{Solution A-B}\\
\hline
Orbital parameter &\multicolumn{3}{c}{Value} & \multicolumn{3}{c}{Std. dev.}\\
\hline
$P$ (days)&31&.&9181 &0&.&0006\\
$T_{\rm{0}}$ (Hel. JD)& 57467&.&41 &0&.&04\\
$e$  &0&.&273 &0&.&002\\
$\omega$ (\degr)&335&.&1 &0&.&4\\
$V_0$ (\kmsi)& (var.)& & & & &\\
$K_{\rm{A}}$ (\kmsi) & 47&.&1 & 0&.&4\\
$K_{\rm{B}}$ (\kmsi) & 49&.&1 & 0&.&5\\
$a_{\rm{A}} \mathrm{sin}\,i$ (AU) &0&.&133 &0&.&001\\
$a_{\rm{B}} \mathrm{sin}\,i$ (AU) &0&.&139 &0&.&001\\
$M_{\rm{A}} \mathrm{sin}\,i^3$ (M$_\odot$) &1&.&34 &0&.&03\\
$M_{\rm{B}} \mathrm{sin}\,i^3$ (M$_\odot$) &1&.&28 &0&.&03\\
$rms_{\rm{A}}$ (\kmsi) & 0&.&203 & & &\\
$rms_{\rm{B}}$ (\kmsi) & 0&.&315 & & &\\
\hline
\hline
\multicolumn{7}{c}{Tentative solution AB-C}\\
\hline
Orbital parameter &\multicolumn{3}{c}{Value} & \multicolumn{3}{c}{Std. dev.}\\
\hline
$P$ (days)&1595&.& &5&.&\\
$T_{\rm{0}}$ (Hel. JD)& 58815&.& &9&.&\\
$e$  &0&.&57 &0&.&04\\
$\omega$ (\degr)& 30&.& &4&.&\\
$V_0$ (\kmsi)& 10&.&6 & 0&.&3\\
$K_{\rm{AB}}$ (\kmsi) & 12&.& & 1&.&\\
$a_{\rm{AB}} \mathrm{sin}\,i$ (AU) &1&.&5 &0&.&2\\
$f(m_{\rm{C}})$ (M$_\odot$) &0&.&19 &0&.&08\\
$rms_{\rm{AB}}$ (\kmsi) & 0&.&228 & & &\\
\hline
\end{tabular}
\end{minipage}
\end{table}

\subsection{KIC~6381306 }
\begin{figure}
\begin{center}
\includegraphics[width=8.9cm]{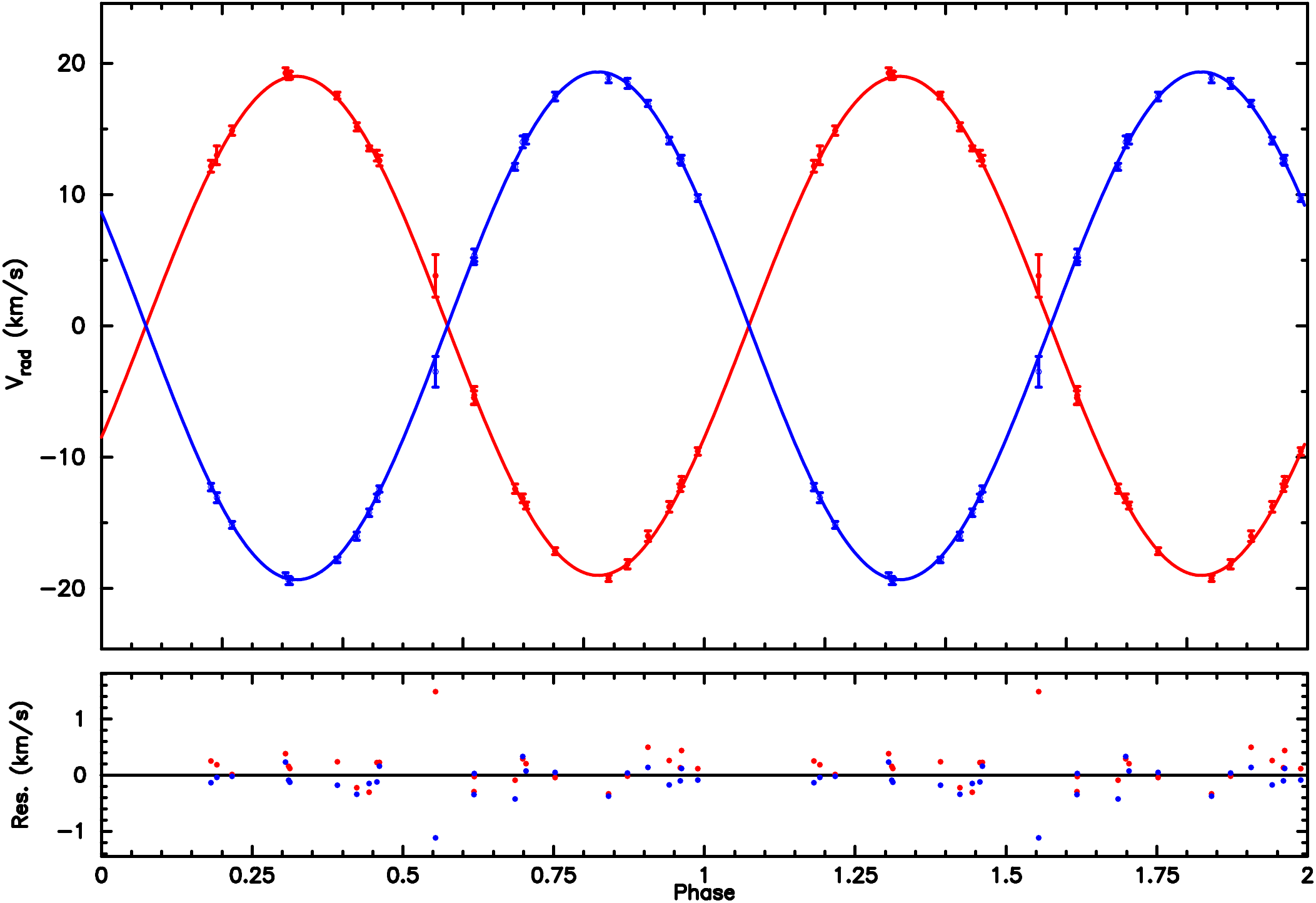}
\end{center}
\caption{Radial velocities for components A and B of the new triple-lined system KIC~6381306 plotted together with the orbital 
solution of the inner (AB) binary system and adjusted systemic velocity offsets. The residuals are shown in the bottom part.} 
\label{fig:KIC63_sb2}
\end{figure}

We computed the best-fitting orbital period using {\sc Period04} and obtained the value of 3.9~days. This value was
subsequently refined using the code {\it vcurve\_SB} for double-lined systems. 
The presence of the third component influences the systemic velocity of the inner binary, thus we allowed for systematic 
offsets of its value with time. Both component RV curves are represented in Fig.~\ref{fig:KIC63_sb2}. The residuals are 
very small and show no systematic trend. 
The orbital parameters of the close (AB) and the wide (AB-C) system are listed in Table~\ref{tab:vop3}. 
We remark that, since the eccentricity of the AB pair is consistent with null, the orbit is circular, therefore 
neither $\omega$ nor $T_{\rm{0}}$ are very meaningful. 
The phased RV curves of the centre of mass of the close pair AB, together with that of of component C, 
based on an orbital period of 212~days, are displayed in Fig.~\ref{fig:KIC63_sb3} (Appendix~B). We can 
furthermore derive strict limitations on both inclinations from the condition that each component of this
system should have a mass in the range [1 - 3.5] M$_\odot$ (following the conclusion in Sect.~\ref{spec_KIC6381306}). 
We thus obtain the following conditions: 8.5 < $i$ < 13\degr and 8.4 < $i_{outer}$ < 10\degr. Note that there is 
a high probability of co-planarity.\\

\begin{table}
\center
\begin{minipage}{8.9cm}
\centering \caption[]{\label{tab:vop3} Values and standard deviations of the constrained parameters of the 
orbital solutions for KIC~6381306.}
\begin{tabular}{lr@{}l@{}lr@{}l@{}l}
\hline
\hline
\multicolumn{7}{c}{Solution A-B}\\
\hline
Orbital parameter &\multicolumn{3}{c}{Value} & \multicolumn{3}{c}{Std. dev.}\\
\hline
$P$ (days)&3&.&91140 &0&.&00001\\
$T_{\rm{0}}$ (Hel. JD)& 56975&.&052 &0&.&002\\
$e$  &0&.&001 &0&.&002\\
$\omega$ (\degr)&&&--- &&&---\\
$V_0$ (\kmsi)&(var.)& & & & &\\
$K_{\rm{A}}$ (\kmsi) & 19&.&00 & 0&.&07\\
$K_{\rm{B}}$ (\kmsi) & 19&.&34 & 0&.&07\\
$a_{\rm{A}} \mathrm{sin}\,i$ (AU) &0&.&00683 &0&.&00003\\
$a_{\rm{B}} \mathrm{sin}\,i$ (AU) &0&.&00695 &0&.&00002\\
$M_{\rm{A}} \mathrm{sin}\,i^3$ (M$_\odot$) &0&.&01151 &0&.&00009\\
$M_{\rm{B}} \mathrm{sin}\,i^3$ (M$_\odot$) &0&.&01131 &0&.&00009\\
$rms_{\rm{A}}$ (\kmsi) & 0&.&174 & & &\\
$rms_{\rm{B}}$ (\kmsi) & 0&.&140 & & &\\
\hline
\hline
\multicolumn{7}{c}{Preliminary solution AB-C}\\
\hline
Orbital parameter &\multicolumn{3}{c}{Value} & \multicolumn{3}{c}{Std. dev.}\\
\hline
$P$ (days)&212&.&0 &0&.&3\\
$T_{\rm{0}}$ (Hel. JD)& 57168&.& &3&.&\\
$e$  &0&.&116 &0&.&020\\
$\omega$ (\degr)&204&.& &5&.&\\
$V_0$ (\kmsi)&-18&.&32 & 0&.&04\\
$K_{\rm{AB}}$ (\kmsi) & 5&.&02 & 0&.&06\\
$K_{\rm{C}}$ (\kmsi) & 5&.&03 & 0&.&18\\
$a_{\rm{AB}} \mathrm{sin}\,i_{out}$ (AU) &0&.&0972 &0&.&0012\\
$a_{\rm{C}} \mathrm{sin}\,i_{out}$ (AU) &0&.&097 &0&.&004\\
$M_{\rm{AB}} \mathrm{sin}\,i_{out}^3$ (M$_\odot$) &0&.&0109 &0&.&0008\\
$M_{\rm{C}} \mathrm{sin}\,i_{out}^3$ (M$_\odot$) &0&.&0109 &0&.&0005\\
$rms_{\rm{AB}}$ (\kmsi) & 0&.&123 & & &\\
$rms_{\rm{C}}$ (\kmsi) & 0&.&691 & & &\\
\hline
\end{tabular}
\end{minipage}
\end{table}

\subsection{KIC~7756853 }

\begin{figure}
\begin{center}
\includegraphics[width=8.9cm]{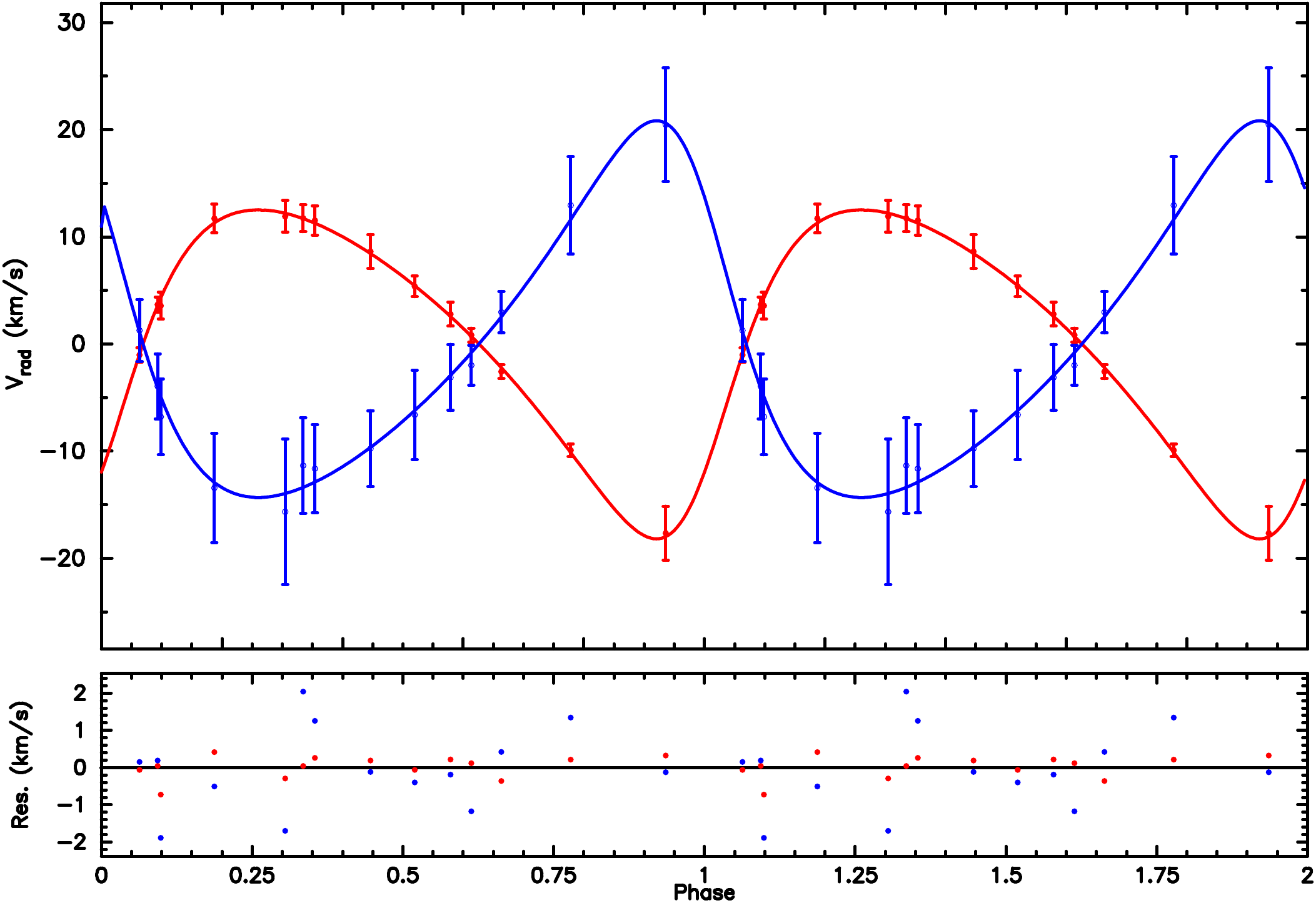}
\end{center}
\caption{Radial velocities for components A and B of the new double-lined system KIC~7756853 plotted with the proposed 
orbital solution of the binary. The residuals are shown in the bottom part.} 
\label{fig:KIC775}
\end{figure}

KIC~7756853 is an SB2 whose spectral lines are mostly blended. We recomputed its radial velocities { using synthetic
spectra} with the parameters of model B (cf. Sect.~\ref{sub:KIC775}) and compared them to the original data set. 
An orbital solution was derived for both cases. We estimated an orbital period of the order of $\sim$ 100~days based 
on the RV plot of 14 spectra. A slightly better agreement in terms of root mean squared residuals was found using model B: 
the mean residuals stay below 1\,\kms and are systematically smaller than with model A. We therefore consider that the best 
choice consists of an A1-type primary with \vsinii$_{1}$ = 30~\kms and an A5-type secondary with \vsinii$_{2}$ = 60~\kms 
(model B). The adopted orbital solution is illustrated by Fig.~\ref{fig:KIC775}. Table~\ref{tab:vop4} lists the corresponding 
orbital parameters with their uncertainties.\\  

\begin{table}
\center
\begin{minipage}{8.9cm}
\centering \caption[]{\label{tab:vop4} Values and standard deviations of the constrained parameters of the orbital 
solution for KIC~7756853.}
\begin{tabular}{lr@{}l@{}lr@{}l@{}l}
\hline
\hline
\multicolumn{7}{c}{Solution A-B}\\
\hline
Orbital parameter &\multicolumn{3}{c}{Value} & \multicolumn{3}{c}{Std. dev.}\\
\hline
$P$ (days)&99&.&32 &0&.&01\\
$T_{\rm{0}}$ (Hel. JD)& 57305&.&8 &0&.&5\\
$e$  &0&.&311 &0&.&010\\
$\omega$ (\degr)& 53&.&5 &1&.&9\\
$V_0$ (\kmsi)&-21&.&51 & 0&.&13\\
$K_{\rm{A}}$ (\kmsi) & 15&.&4 & 0&.&2\\
$K_{\rm{B}}$ (\kmsi) & 17&.&6 & 0&.&3\\
$a_{\rm{A}} \mathrm{sin}\,i$ (AU) &0&.&133 &0&.&002\\
$a_{\rm{B}} \mathrm{sin}\,i$ (AU) &0&.&152 &0&.&003\\
$M_{\rm{A}} \mathrm{sin}\,i^3$ (M$_\odot$) &0&.&168 &0&.&007\\
$M_{\rm{B}} \mathrm{sin}\,i^3$ (M$_\odot$) &0&.&147 &0&.&006\\
$rms_{\rm{A}}$ (\kmsi) & 0&.&198 & & &\\
$rms_{\rm{B}}$ (\kmsi) & 0&.&696 & & &\\
\hline
\end{tabular}
\end{minipage}
\end{table}

\subsection{KIC~8975515 }

\begin{figure}
\begin{center}
\includegraphics[width=8.9cm]{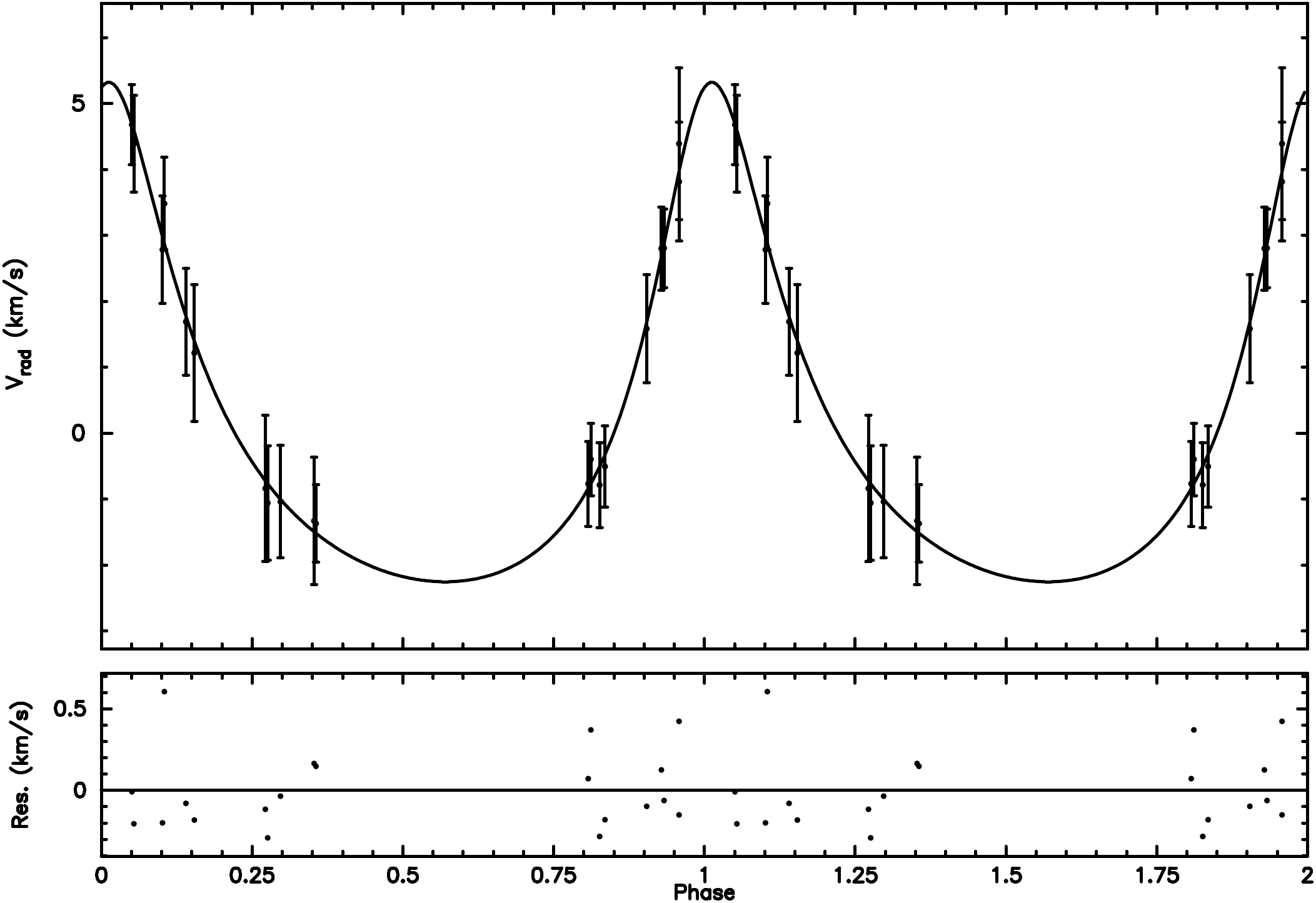}
\end{center}
\caption{Radial velocities for the companion of KIC~8975515 plotted with the preliminary orbital solution of the binary 
(type SB1). The residuals are shown in the bottom part.} 
\label{fig:KIC89_orb}
\end{figure}

The RV curve of the slowly rotating (named secondary) component presents a clearly defined modulation with an amplitude well 
below 10~\kmsi. From this plot, we estimated an orbital period of the order of 1000~days (possibly longer) for this new 
SB2 system with dissimilar components. Due to the extreme scatter on the RVs of the fast rotating (named primary) component, 
we applied the code {\it vcurve\_SB} for single-lined systems to the data of the secondary (only), and found possible
orbital solutions with a period close to either 800~or 1600~days (twice as large). Since the time-delay 
analysis supports the existence of a 1600-days long variation (cf. Sect.~\ref{sect:timeDelay}), we chose the second possibility. 
A tentative set of orbital parameters is listed in Table~\ref{tab:vop5}. The preliminary orbital solution is illustrated 
in Fig.~\ref{fig:KIC89_orb}. It is obvious that this solution accomodates very well the currently available radial 
velocities of component B since the residuals are really small. In Sect.~\ref{sect:timeDelay}, we will present new evidence
to confirm this solution. We will also extend the RV monitoring for another season, in order to further improve the orbital 
parameters, whilst offering the benefit of a solution of type SB2.\\  

\begin{table}
\center
\begin{minipage}{8.9cm}
\centering \caption[]{\label{tab:vop5} Values and standard deviations of the constrained parameters of the orbital 
solution for KIC~8975515.}
\begin{tabular}{lr@{}l@{}lr@{}l@{}l}
\hline
\hline
\multicolumn{7}{c}{Preliminary solution A-B}\\
\hline
Orbital parameter &\multicolumn{3}{c}{Value} & \multicolumn{3}{c}{Std. dev.}\\
\hline
$P$ (days)&1582&.& &7&.&\\
$T_{\rm{0}}$ (Hel. JD)& 57080&.& &7&.&\\
$e$  &0&.&41 &0&.&02\\
$\omega$ (\degr)& 348&.& &2&.&\\
$V_0$ (\kmsi)&-20&.&47 & 0&.&07\\
$K_{\rm{2}}$ (\kmsi) & 3&.&79 & 0&.&09\\
$a_{\rm{2}} \mathrm{sin}\,i$ (AU) &0&.&502 &0&.&013\\
$f(m_{\rm{1}})$ (M$_\odot$) &0&.&0067 &0&.&0005\\
$rms_{\rm{2}}$ (\kmsi) & 0&.&159 & & &\\
\hline
\end{tabular}
\end{minipage}
\end{table}

\subsection{KIC~9790479 }
\begin{figure}
\begin{center}
\includegraphics[width=7.5cm]{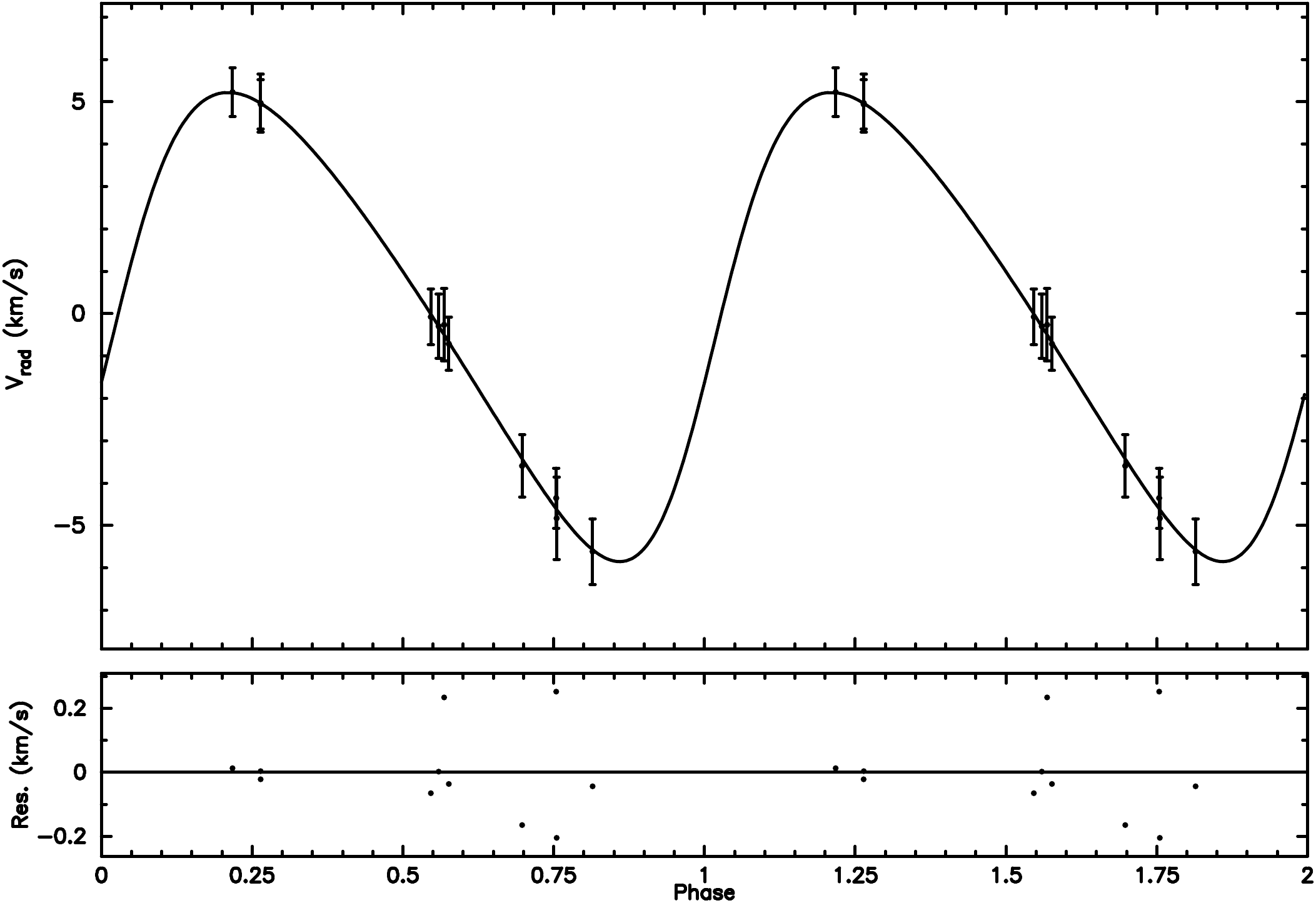} 
\end{center}
\caption{Radial velocities for the single-lined system KIC~9790479 showing a preliminary orbital solution.
 The residuals are plotted in the bottom part.} 
\label{fig:KIC97_orb}
\end{figure}

The radial velocity plot of this single-lined system and a preliminary orbital solution based on an
approximated period of 231~days are illustrated by Fig.~\ref{fig:KIC97_orb}. A tentative set of 
orbital parameters is listed in Table~\ref{tab:vop6}.\\ 

\subsection{KIC~11572666 }
\begin{figure}
\begin{center}
\includegraphics[width=7.5cm]{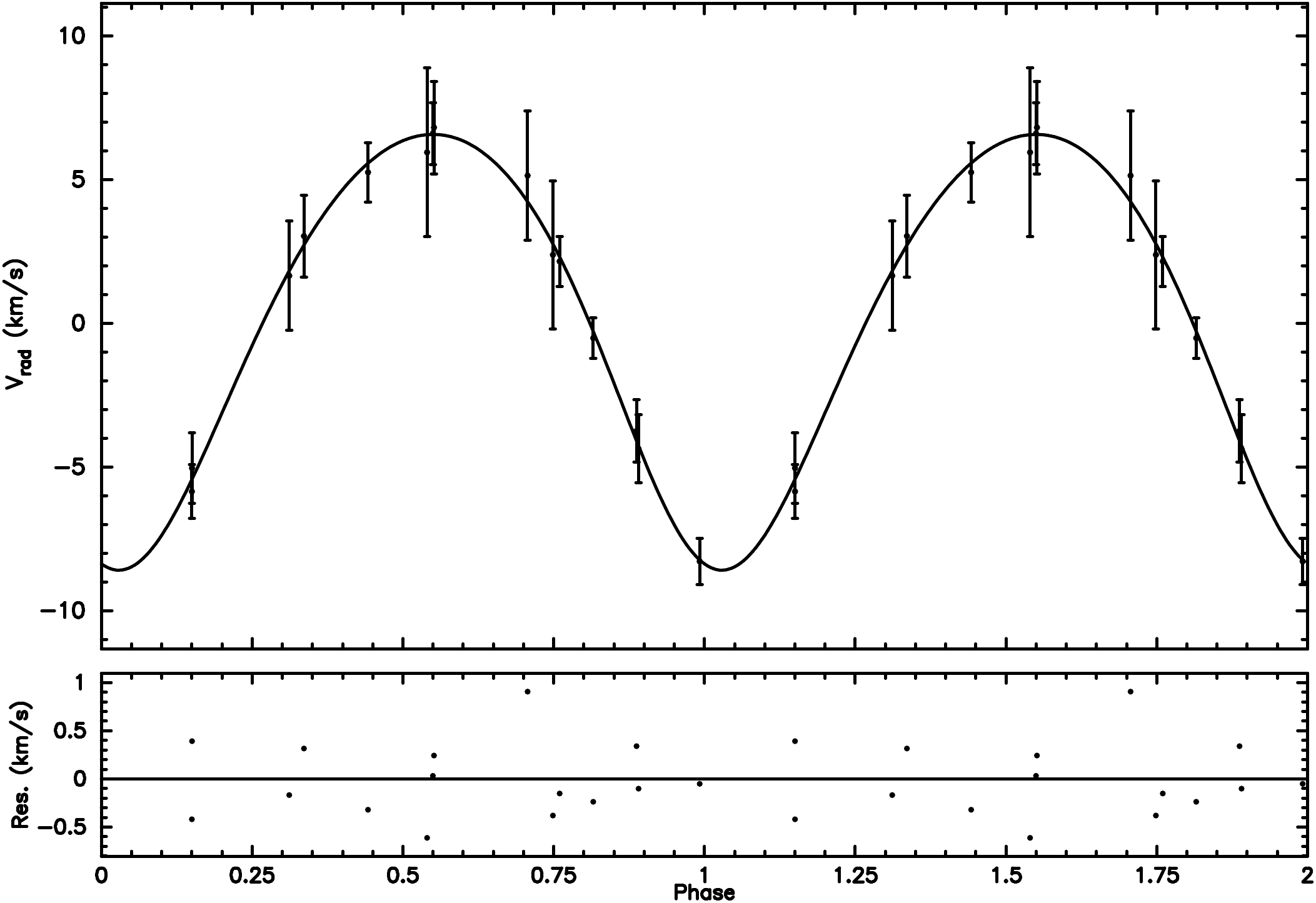}
\end{center}
\caption{Radial velocities for the companion of KIC~11572666 showing a preliminary orbital solution (type SB1).
 The residuals are plotted in the bottom part.} 
\label{fig:KIC115_orb1}
\end{figure}

The clearly defined  plot for the cooler secondary component of this new SB2 system enabled us to derive 
an estimated orbital period of 611~days, as well as a tentative set of orbital parameters which is listed in Table~\ref{tab:vop7}. 
This preliminary orbital solution is illustrated in Fig.~\ref{fig:KIC115_orb1}. 
Due to the huge scatter of the RVs of the primary component, a modelling of type SB2 based on the (weighted) velocities of 
both components did not allow to improve upon this preliminary solution.\\ 

\begin{table}
\center
\begin{minipage}{8.9cm}
\centering \caption[]{\label{tab:vop6} Values and standard deviations of the constrained parameters of the 
orbital solution proposed for KIC~9790479.}
\begin{tabular}{lr@{}l@{}lr@{}l@{}l}
\hline
\hline
\multicolumn{7}{c}{Plausible solution A-B}\\
\hline
Orbital parameter &\multicolumn{3}{c}{Value} & \multicolumn{3}{c}{Std. dev.}\\
\hline
$P$ (days)&230&.&9 &0&.&5\\
$T_{\rm{0}}$ (Hel. JD)& 56649&.& &9&.&\\
$e$  &0&.&24 &0&.&03\\
$\omega$ (\degr)&76&.& &7&.&\\
$V_0$ (\kmsi)& 5&.&2 & 0&.&3\\
$K_{\rm{1}}$ (\kmsi) &  5&.&5 & 0&.&2\\
$a_{\rm{1}} \mathrm{sin}\,i$ (AU) &0&.&114 &0&.&004\\
$f(m_{\rm{2}})$ (M$_\odot$) &0&.&0037 &0&.&0004\\
$rms_{\rm{1}}$ (\kmsi) & 0&.&084 & & &\\
\hline
\end{tabular}
\end{minipage}
\end{table}

\begin{table}
\center
\begin{minipage}{8.9cm}
\centering \caption[]{\label{tab:vop7} Values and standard deviations of the constrained parameters of the 
orbital solution proposed for KIC~11572666.}
\begin{tabular}{lr@{}l@{}lr@{}l@{}l}
\hline
\hline
\multicolumn{7}{c}{Plausible solution A-B}\\
\hline
Orbital parameter &\multicolumn{3}{c}{Value} & \multicolumn{3}{c}{Std. dev.}\\
\hline
$P$ (days)&611&.& &4&.&\\
$T_{\rm{0}}$ (Hel. JD)& 57307&.& &16&.&\\
$e$  &0&.&14 &0&.&02\\
$\omega$ (\degr)&166&.& &9&.&\\
$V_0$ (\kmsi)& -20&.&2 & 0&.&1\\
$K_{\rm{2}}$ (\kmsi) &  7&.&6 & 0&.&2\\
$a_{\rm{2}} \mathrm{sin}\,i$ (AU) &0&.&422 &0&.&010\\
$f(m_{\rm{1}})$ (M$_\odot$) &0&.&0268 &0&.&0019\\
$rms_{\rm{2}}$ (\kmsi) & 0&.&216 & & &\\
\hline
\end{tabular}
\end{minipage}
\end{table}

\section{Time-delay analysis}
\label{sect:timeDelay}

The orbital motion of a system with a pulsating component introduces a periodic shift of the pulsation frequencies (or 
their phases) which is caused by variations of the light-travel-time along the orbit. This is known as the 'light-travel-time effect' 
(hereafter LiTE). In the case of a multi-mode pulsator, we expect that all the pulsation frequencies show the same cyclical 
variability pattern. This approach was successfully applied by \citet{Murphy2014MNRAS.441.2515M}
to a sample of $\delta$ Sct stars observed by the \textit{Kepler} satellite. We applied the same method to the \textit{Kepler} 
light curves of all the targets in our sample.

First, we determined all the frequencies with a significance level \emph{S/N} $>4$ \citep[see e.g.][]{Breger1993A&A...271..482B}
on the basis of each full data set. Next, we selected the 20 frequencies with the highest \emph{S/N}. Then, each data set was 
divided into segments of random length between 9 and 11\,days in order to prevent unwanted aliasing. Subsequently, the data 
of each segment was fitted using a non-linear least-squares fitting routine \textsc{LCfit} \citep{Sodor2012KOTN...15....1S} 
with fixed frequencies and starting epoch. In this way, we derived the time-dependent amplitudes and phases for all 20 frequencies. 
The time delay $\Delta t_{ij}$ for a given frequency $\nu_{j}$ and time segment \emph{i} is then easily computed using
\begin{equation}
	\Delta t_{ij}=\frac{\Delta \phi_{ij}}{2\pi \nu_{j}}
\end{equation} 
where $\Delta \phi_{ij}$ is the difference between $\phi_{ij}$ and the mean phase calculated from all the segments.

The analysis of the time delays (TDs) is complicated by several factors. Firstly, the phases of the low frequencies (those which might 
correspond to $\gamma$ Dor frequencies) are often highly scattered, since the corresponding periods are only slightly shorter than 
the length of the segments and since the frequency spectra often show close peaks that cannot be resolved in the subsets. This is 
the main reason why we detected correlated phase variations almost exclusively among the higher $\delta$ Sct frequencies. Secondly, 
some frequencies show significant changes in amplitude and/or phase (usually both) and are thus unstable \citep{Bowman2016MNRAS.460.1970B}. 
Such changes will often conceal the overall phase variations caused by LiTE. The analysis can furthermore be affected by the 
frequency content and distribution. For example, KIC~5965837 shows two most dominant modes with amplitudes of a few mmag in the 
$\gamma$ Dor regime, while the rest of the frequencies has amplitudes about 30$\times$ lower. 
In order to reduce the problems with the close frequencies, we also constructed and analysed subsets with a length of 20, 50 and 100~days. 
Though generally less scattered, these results still remained very close to those of the 10-days long subsets. We adopted the LiTE 
interpretation as the cause for the detected phase variations when at least three independent frequencies showed the same time-delay 
pattern.\\ 

\begin{figure*}[htbp]
	\begin{center}
	\includegraphics[width=\hsize]{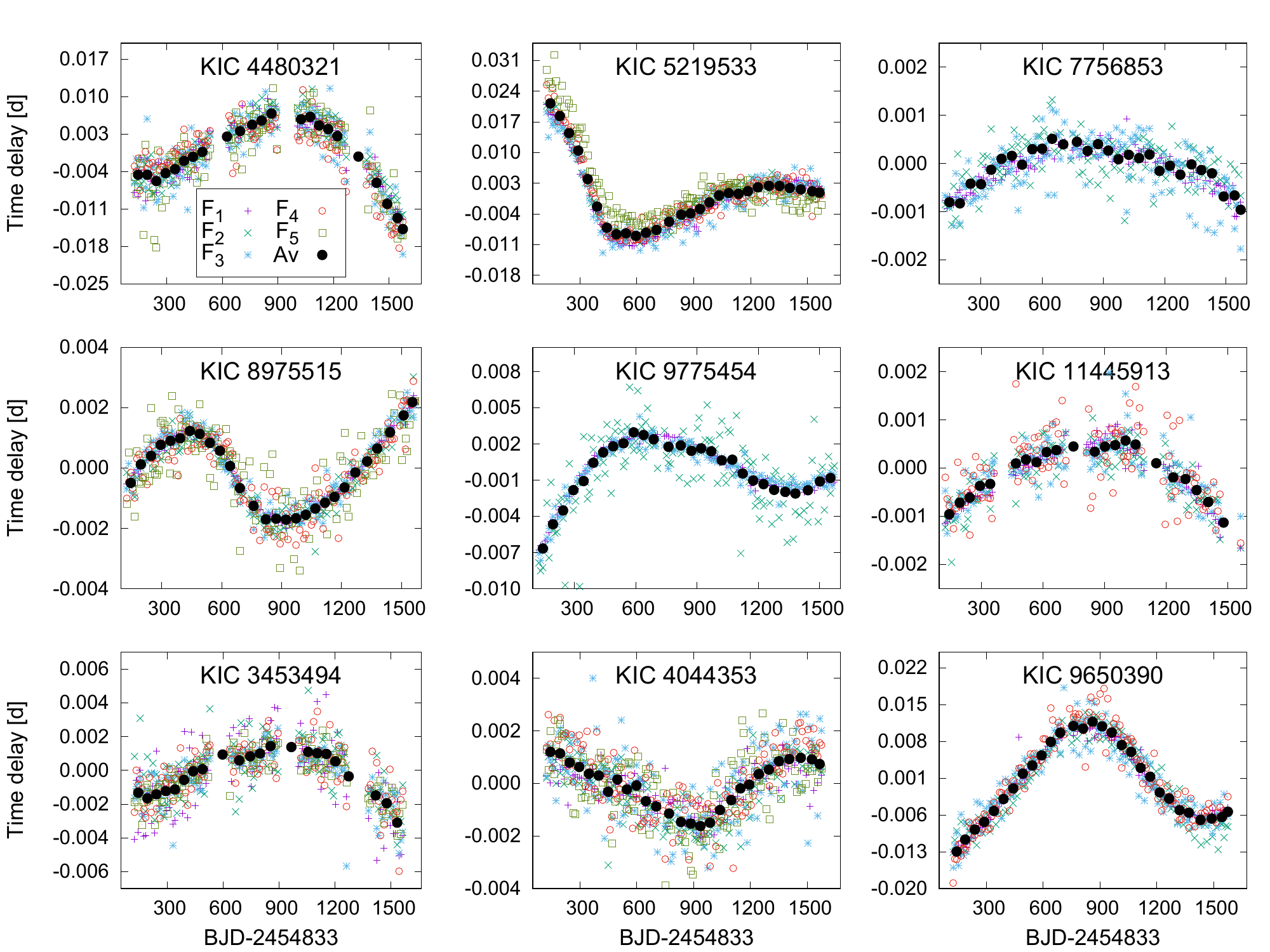}
	\caption{Time delays suggesting orbital motion in nine \textit{Kepler} candidate hybrid stars. The frequencies are not explicitly 
	listed - their values are listed in Table~\ref{Tab:Frequencies}. The symbols and colours in all panels
	are the same as in the top-left panel. The black circles show the weighted (50\,days) averages.}\label{Fig:Mosaic}	
	\end{center}
\end{figure*}

We discovered correlated variations of comparable amplitude in the time delays of nine objects. In Fig.~\ref{Fig:Mosaic},
we illustrate the detection of this effect (LiTE). For the sake of clarity, we show only a few frequencies having the lowest scatter. 
To highlight the general patterns, we plotted the mean value of five points based on a weighted average computed for 10-days long 
segments\footnote[3]{As weight, we took the value of the frequency $\cdot$ amplitude.}. Whenever the low ($\gamma$-Dor like) frequencies 
showed some apparent trend, these always followed the trend of the higher ($\delta$-Sct like) frequencies suggesting that both frequency 
regimes arise in the same star (e.g. f$_2$ in KIC~9775454). This is also confirmed by the detection of combination peaks of low and 
high frequencies in the periodograms. 

Fig.~\ref{Fig:Mosaic} furthermore shows that the time scales of the detected variations are comparable to the length of the full data sets 
(i.e. of the order of 4\,yrs or longer). In six cases (cf. the upper two rows), RV variability due to orbital motion was also 
detected (cf. Sect.~\ref{sect:orb}). In three cases (cf. the bottom row), the time delays provide evidence for (undetermined) orbital 
motions with a long period. However, these targets were classified by us as spectroscopically stable (class 'S' or 'S?'). 
It is relevant to note that two of these show extremely broad profiles indicative of very fast rotation which could hamper 
the detection of spectroscopic multiplicity (KIC\,3453494 and~9650390). 

In four cases, the LiTE amplitude is large ($\geq 0.01$\,days, i.e. KIC~4480321, 5219533, 9775454, and~9650390). 
In five other cases (three are SB2 systems, e.g. KIC~8975515), the LiTE amplitude is of order of a few thousandths 
of a day at most. The shortest orbital periods, including 3.9\,(KIC~6381306), 9.2\,(KIC~4480321), 32\,(KIC~5219533), 
99\,(KIC~7756853) and 212\,days (KIC~6381306), were not detected. 
To increase the chances of detecting the tiny time delays produced by the shortest-period orbits (with an expected 
total LiTE amplitude of the order of a few $10^{-4}$\,days, such as in KIC~6381306 and KIC~4480321), we divided each data set
into bins of width 0.10 in orbital phase. We thus obtained ten data sets containing several thousands of points with a time 
span almost equal to that of the full set for each target. However, even with this modification, the phase variations 
remained undetected in these two particular cases. This is not due to the method itself, since the smallest LiTE amplitude 
thus far found in \textit{Kepler} data equals 8$\cdot10^{-5}$\,days \citep{Murphy2016ApJ...827L..17M}, while the shortest LiTE 
period yet detected in such light curves is 9.15\,days \citep{Murphy2016MNRAS.461.4215M}. 
We conclude that the non-detections may be the consequence of the true LiTE amplitude in combination with the pulsation 
characteristics of a particular star or system. An alternative and physical explanation is that the detected pulsations 
might not arise in the close binary itself but in the outer, third component.\\ 

\begin{figure}[htbp]
	\begin{center}
	\includegraphics[width=7.5cm]{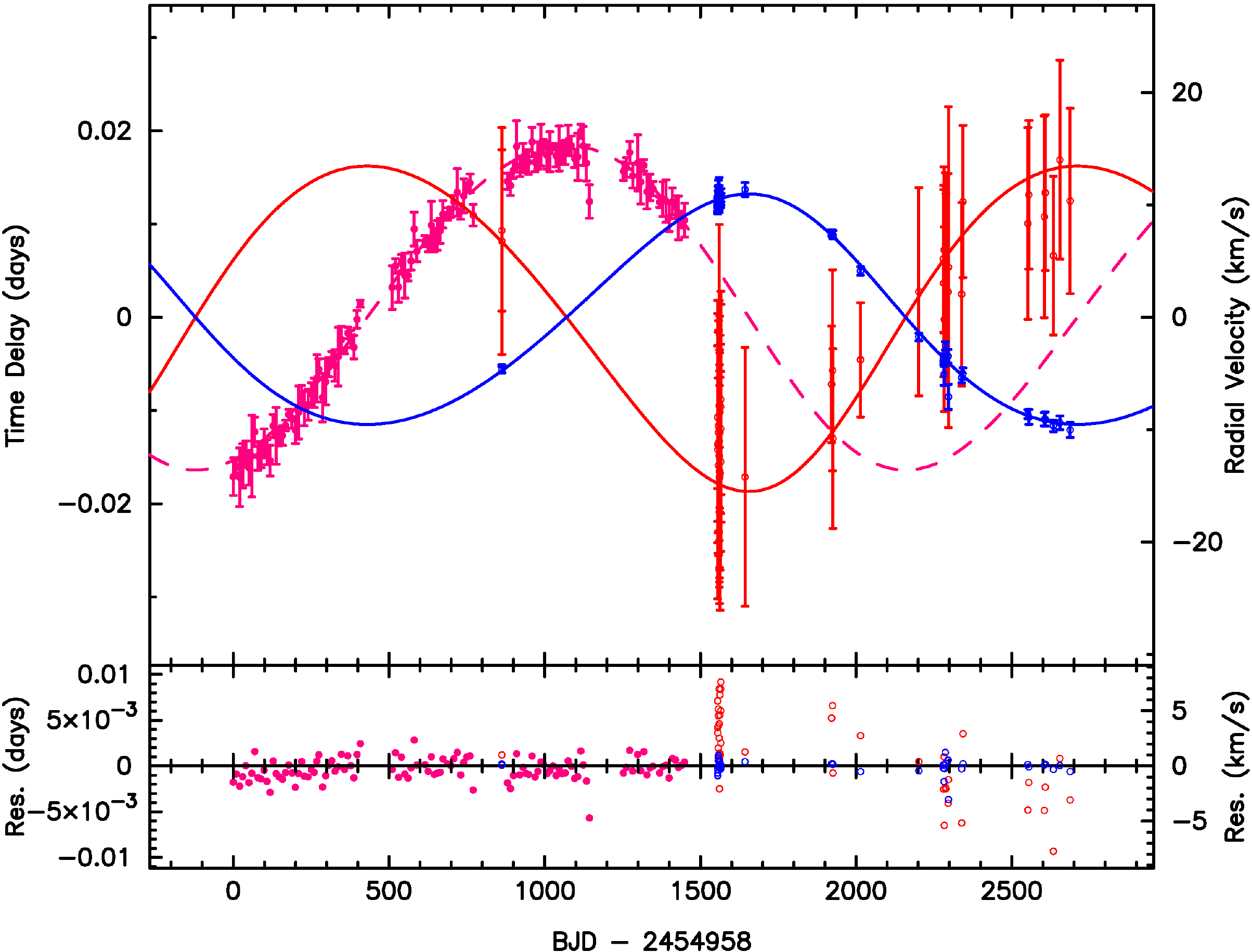}
	\caption{LiTE versus RV orbital solution for KIC~4480321. Note that the TDs (pink symbols) were corrected 
	by removing a linear trend from the original data. The blue and red solid lines refer to the RV solution (see Fig.~\ref{fig:KIC44_sb3}). 
	The dashed line shows the computed model.\label{KIC4480321_RV2+TD} }	
	\end{center}
\end{figure}

\begin{figure}[htbp]
	\begin{center}
	\includegraphics[width=7.5cm]{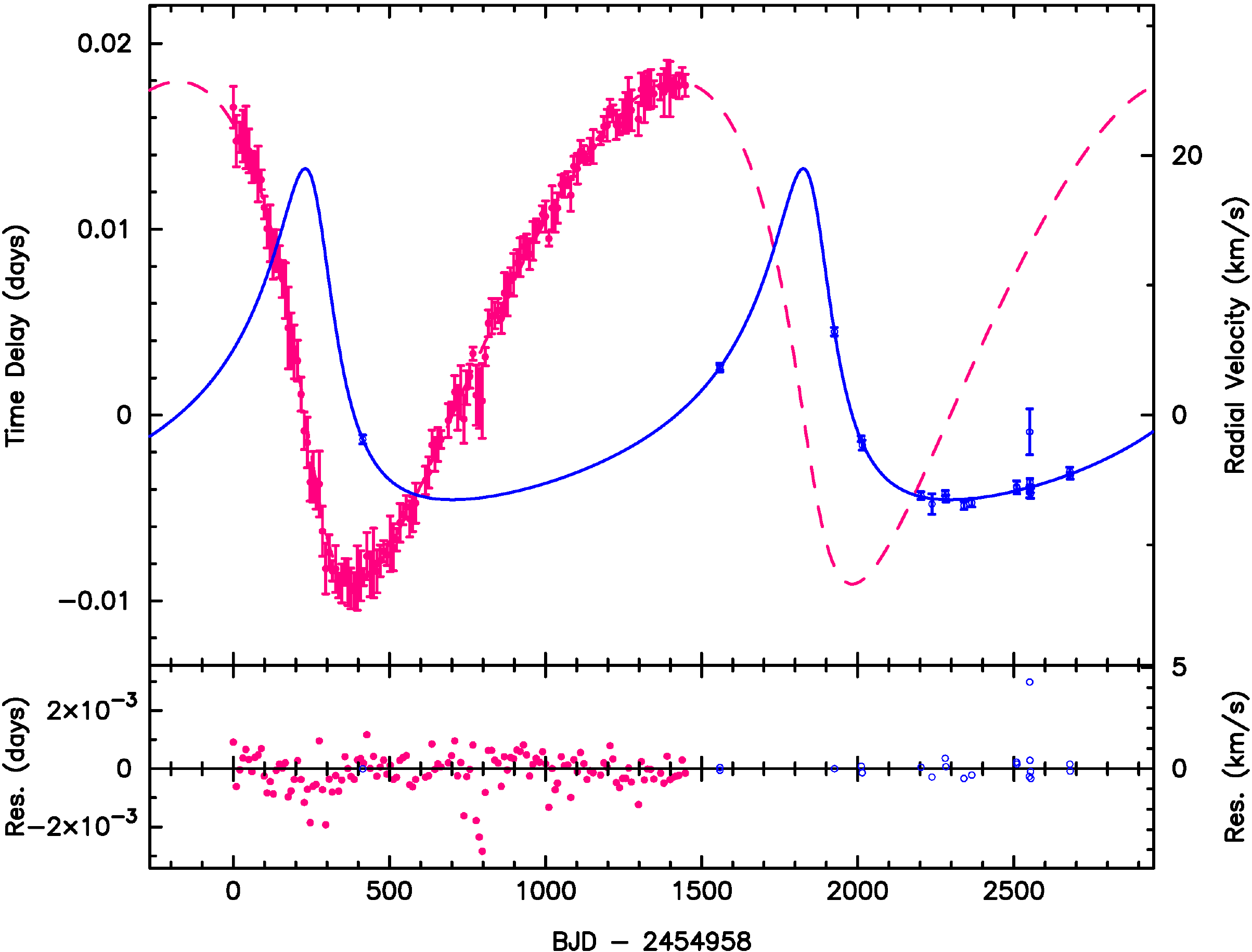}
	\caption{LiTE versus RV orbital motion for KIC~5219533. Note that the TDs (pink symbols) were corrected 
	by removing a linear trend from the original data. The solid line refers to the RV solution (see Fig.~\ref{fig:KIC52_sb1}).
	The dashed line shows the computed model.\label{KIC5219533_RV+TD} }	
	\end{center}
\end{figure}

\begin{figure}[htbp]
	\begin{center}
	\includegraphics[width=7.5cm]{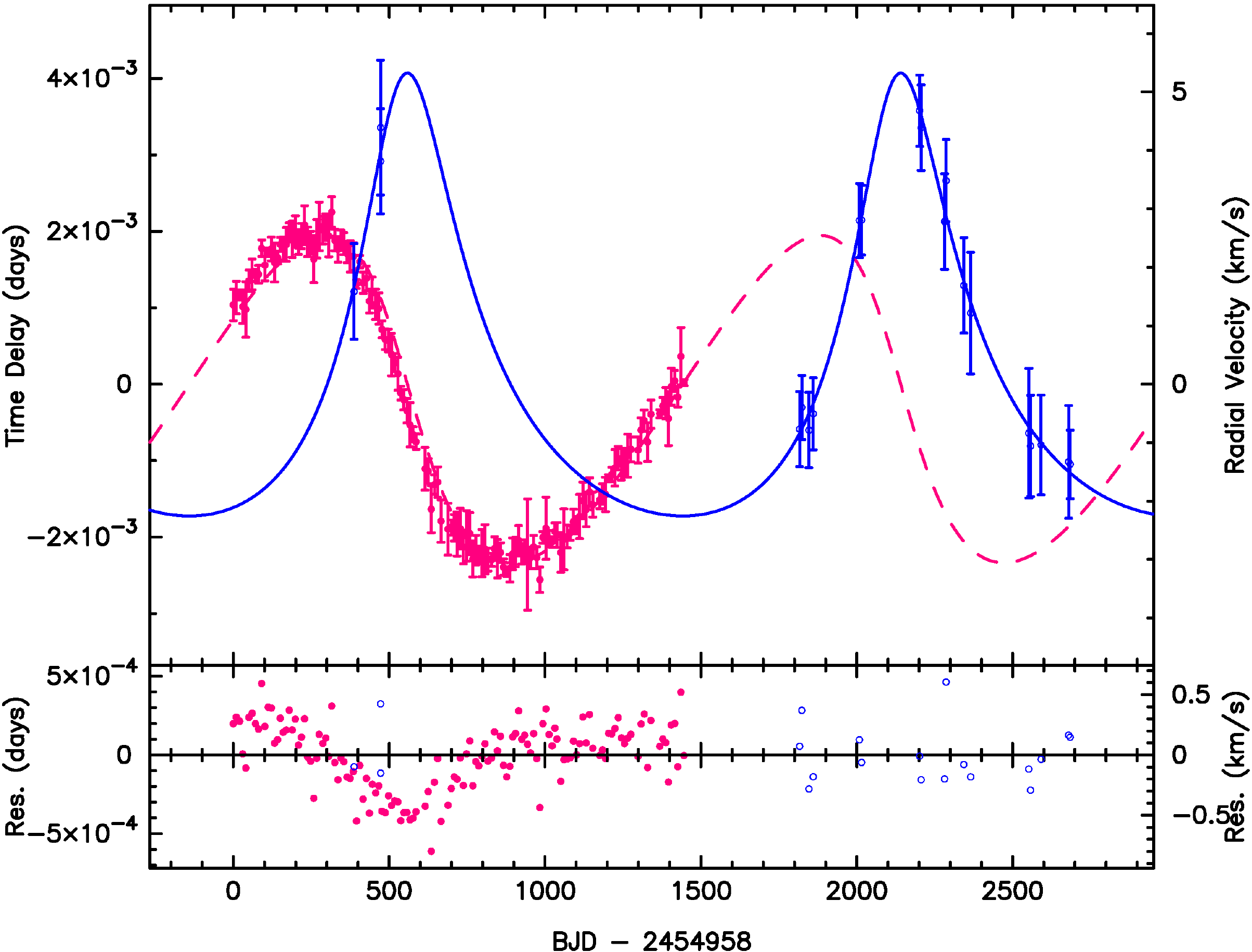}
	\caption{LiTE versus RV orbital motion for KIC~8975515. Note that the TDs (pink symbols) were corrected 
	by removing a linear trend from the original data. The solid line refers to the RV solution (see Fig.~\ref{fig:KIC89_orb}). 
	The dashed line shows the computed model.\label{KIC8975515_RV+TD} }	
	\end{center}
\end{figure}

In order to show that the detected LiTE has the same cause as the long-term RV variations, we performed a simple modelling 
based on both data types for the following systems: KIC~4480321, 5219533, and~8975515. We have different orbital solutions 
for each: KIC~4480321 is a triple-lined system with two SB2 solutions, KIC~5219533 is a triple-lined system with one SB2 
and one SB1 solution (for the outer system) while KIC~8975515 is a double-lined system with one SB1 solution (for the 
slower rotating component).
KIC~4480321 AB-C shows the longest detected orbital period (P$_{orb}\sim$ 2300 d). The inclusion of the 
time delays allows us to confirm and improve the previous orbital solution. To this purpose, we modelled the TDs adopting 
the orbital parameters of the RV solution except for the parameter ($a_{C}$. sin $i_{out}$) which was fitted to match the 
TDs only (cf. Table~\ref{tab:comb}). 
Fig.~\ref{KIC4480321_RV2+TD} shows the excellent agreement between the two data types. The TD residuals in the sense
(observed minus modelled) are small and homogeneously distributed around zero. The advantage is a more accurate 
determination of the mass ratio q$_{out}$ $\equiv \frac{a_{AB}}{a_{C}}$ = 0.708 $\pm$ 0.004. 
KIC~5219533 AB-C has P$_{orb}\sim$ 1600~d. We modelled the TD variations adopting all orbital parameters of the RV solution
but fitted an additional parameter ($a_{C}$. sin $i_{out}$) (Table~\ref{tab:comb}). 
Fig.~\ref{KIC5219533_RV+TD} shows the excellent agreement between the two data types. The TD residuals are small and 
homogeneously distributed around zero. The advantage is the determination of the new ratio q$_{out}$ $\equiv  
\frac{a_{AB}}{a_{C}}$ = 0.56 $\pm$ 0.07.  
KIC~8975515 AB also has P$_{orb}\sim$ 1600~d. We modelled the TD variations adopting all orbital parameters of the RV solution
but fitted an additional parameter ($a_{A}$. sin $i$) (Table~\ref{tab:comb}). 
Fig.~\ref{KIC8975515_RV+TD} shows the good agreement between the two data types. The TD residuals are small though somewhat 
inhomogeneously distributed around zero, as they show a small discrepancy in eccentricity. The advantage here also is the 
determination of the new ratio q $\equiv \frac{a_{A}}{a_{B}}$ = 0.81 $\pm$ 0.01. In all three cases, 
we confirm that the observed TDs are consistent with the previously derived RV orbital solutions.\\

\begin{table}
\centering
\caption{\label{tab:comb} Results of a fit of the LiTE adopting the previously derived RV orbital solutions. 
}
\begin{tabular}{llcc}
\hline\hline
\multicolumn{1}{c}{\rm KIC Nr} & \multicolumn{1}{c}{Fitted parameter} & \multicolumn{1}{c}{Value} & \multicolumn{1}{c}{Uncertainty}\\
\hline
\small 4480321 & $a_C$ sin $i_{out}$ & 3.023 & 0.058\\
\small 5219533 & $a_C$ sin $i_{out}$ & 2.682 & 0.026\\
\small 8975515 & $a_A$ sin $i$ & 0.4055 & 0.0036\\
\hline
\end{tabular}
\end{table}

Such examples illustrate that the TD method can identify long-term changes which nicely complement the results of the 
spectroscopic analyses. Although the argument for long-term variations caused by orbital motion is quite strong, we should not 
exclude the possibility that the parabola-shaped variations of the { TDs} as e.g. observed in KIC\,7756853 
might also be caused by some { intrinsic} evolutionary effect. Indeed, such a behaviour was reported in the $\delta$ Sct star 4~CVn. 
Although 4~CVn is also a short-period spectroscopic binary system \citep{Schmid2014A&A...570A..33S}, the secular variations detected 
in (only) some pulsation modes follow different patterns related to a change of the rotational splitting \citep{Breger2016A&A...592A..97B}. 
In the case of the above discussed \textit{Kepler} stars, however, the reported patterns of the secular variations are common 
to all investigated frequencies.\\

\setlength\tabcolsep{2pt}
\begin{table}
\centering
\caption{Frequencies (expressed in d$^{-1}$) for which the time delay is plotted in Fig.~\ref{Fig:Mosaic}.}		
\begin{tabular}{rrrrrr}
\hline \hline 
\mcol{1}{c}{\rm KIC Nr} & \mcol{1}{c}{$F_{1}$} & \mcol{1}{c}{$F_{2}$} & \mcol{1}{c}{$F_{3}$} & \mcol{1}{c}{$F_{4}$} & \mcol{1}{c}{$F_{5}$}\\
\hline
\small 3453494  & \small 7.4989  & \small 9.9099  & \small 19.6236 & \small 16.4034 & \small 24.8482 \\
\small 4044353  & \small 17.8971 & \small 17.1562 & \small 21.7553 & \small 19.2751 & \small 27.5050 \\
\small 4480321  & \small 23.0027 & \small 14.9304 & \small 13.9569 & \small 22.9501 & \small 13.0734 \\
\small 5219533  & \small 10.2853 & \small 17.1820 & \small 13.2421 & \small 15.1066 & \small 19.4854 \\
\small 7756853  & \small 21.1416 & \small 20.7177 & \small 18.9791 & \small 22.3150 & \\
\small 8975515  & \small 13.9724 & \small 16.0763 & \small 17.7345 & \small 15.5147 & \\
\small 9650390  & \small 17.3726 & \small 16.3018 & \small 12.8837 & \small 14.3490 & \\
\small 9775454  & \small 14.9387 & \small 4.6108  & \small 14.7482 & & \\
\small 11445913 & \small 31.5579 & \small 25.3771 &\small  22.1331 & & \\\hline
\end{tabular}\label{Tab:Frequencies}
\end{table}

\section{An observational H-R diagram}
\label{sect:HRD}

\begin{figure}
\begin{center}
\includegraphics[width=8.9cm,viewport = 8 27 350 255,clip=]{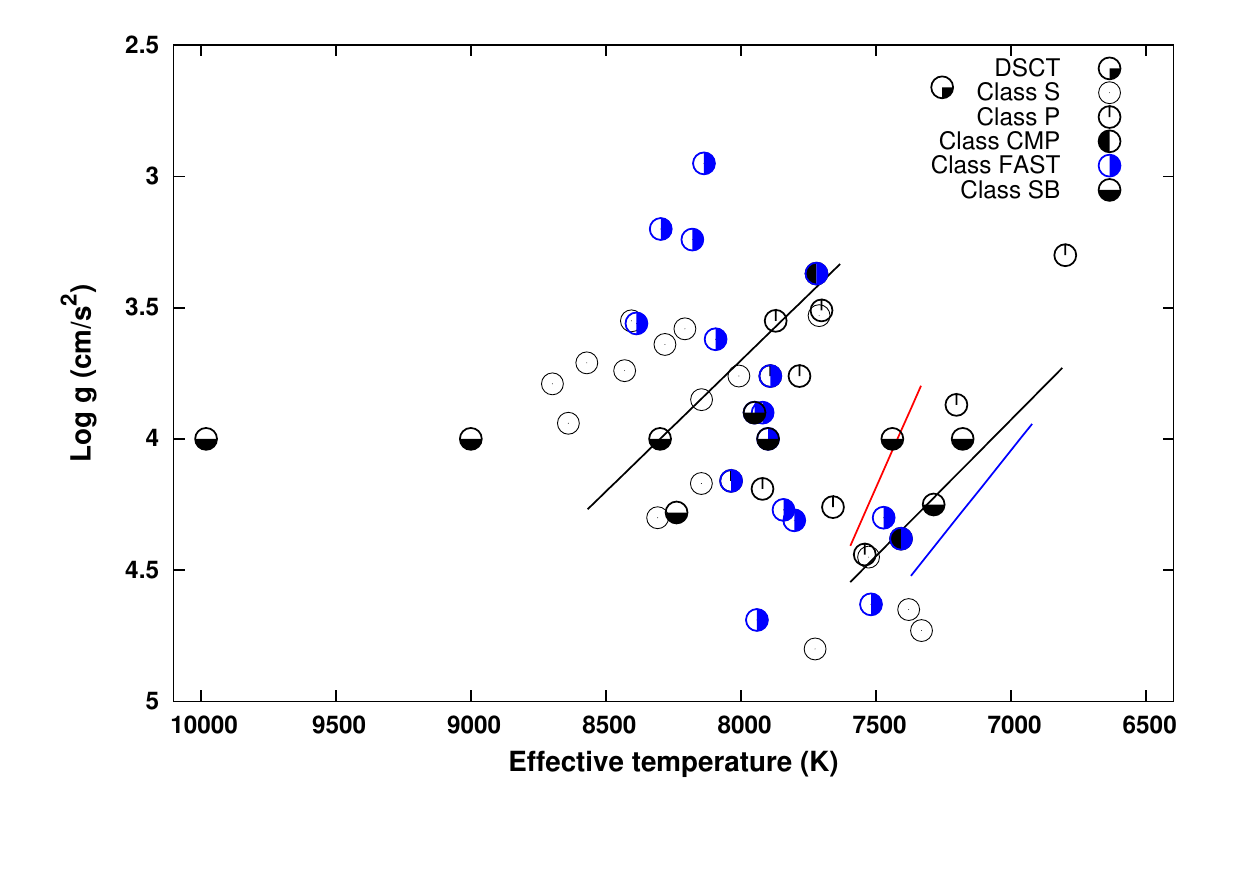}
\end{center}
\caption{A (\teffi, \loggi)-representation for the 50 candidate hybrid stars and one evolved $\delta$ Scuti star. We show 
the location of the objects belonging to the classes S, P(+VAR), CMP and SB, as well as the rapidly rotating ones (FAST). The 
empirical borders of the classical $\delta$ Scuti (in black) and $\gamma$ Dor (in colour) instability strips are represented 
by solid lines. }
\label{fig:HR1}
\end{figure}

In Fig.~\ref{fig:HR1}, we plotted the distribution of the updated atmospheric parameters for our sample of A/F-type candidate hybrid stars. 
In this observational (\teffi, \loggi)-diagram, we represented the five categories of our classification using different symbols. Based on the 
new \vsini measurements, we furthermore indicated the objects which are subject to (very) fast rotation (i.e. with \vsini > 200~\kmsi). Of 
course, both CMP-targets are fast rotators. The borders of the classical $\delta$ Scuti and the $\gamma$ Doradus instability strips reproduced 
by \citet{Uytterhoeven2011A&A...534A.125U} are also shown. It is remarkable that all the P-targets are located within these limits. The 
SB objects located (far) outside the instability strips are KIC~6381306 (\teff = 9000~K) and KIC~7756853 (\teff = 9980~K). \\

\section{Discussion and conclusions}
\label{sect:con}

\begin{figure}[ht]
\begin{center}
\includegraphics[width=5.5cm]{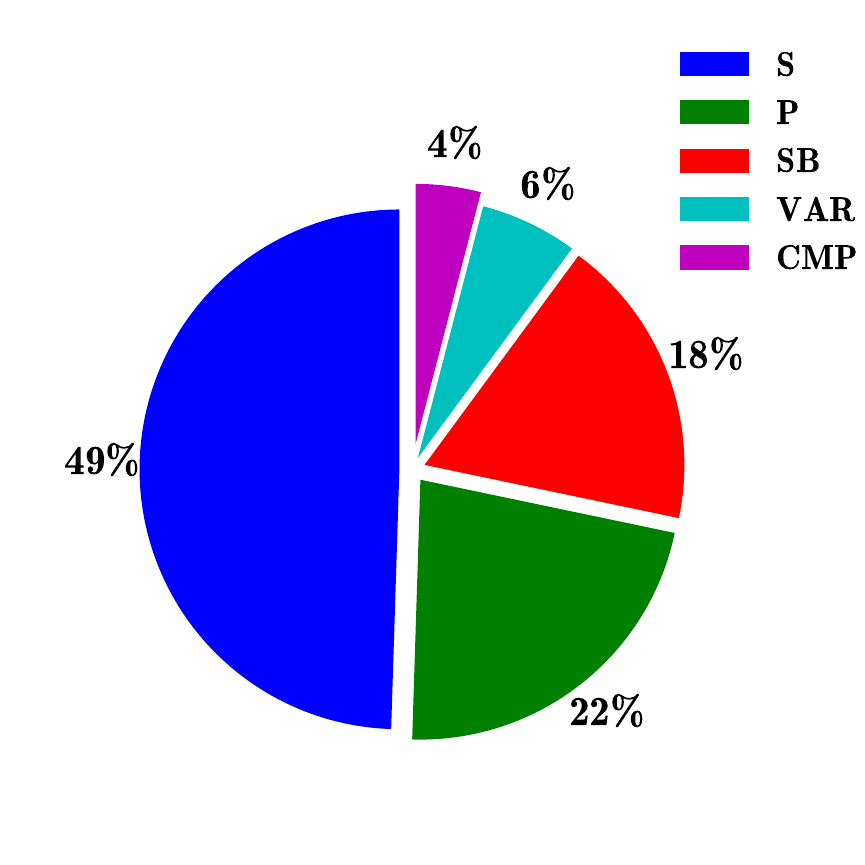}
\end{center}
\caption{Classification results of the A/F-type candidate hybrid stars. }
\label{fig:pie}
\end{figure}

We classified a sample of 50 regularly observed bright A/F-type candidate hybrid stars and one $\delta$ Scuti star of the \textit{Kepler} 
mission into different classes according to the shape (in some cases, the model) of their cross-correlation profiles and the evolution of 
their radial velocities with time. Classes were defined as S (for stable), SB (for spectroscopic binary or triple system), VAR (for long-term 
radial-velocity variable star), P (for pulsating and/or rotating), and CMP (for composite spectrum). According to these spectroscopic criteria, 
we find the following distribution (without considering the cool object KIC~9700679 which is also SB1): 49\% of S-stars, 22\% of P-stars, 
18\% of SB-objects, 6\% of VAR-stars and 4\% of CMP-objects (Fig.~\ref{fig:pie}). In addition, the VAR-stars are (most) probably long-period SB1 systems. 
Including the known eclipsing binary KIC~11180361 (KOI-971), we find a global multiplicity fraction of 27\%. This indicates that {\it at 
least 25\%} of our sample of potential hybrid stars belongs to a binary or a multiple system.\\
 
We (re)determined the fundamental atmospheric properties such as \teffi, \logg and \vsini based on the high-resolution \textsc{Hermes} spectra, 
which are important physical parameters, for all of our targets including - where feasible - for the components of the newly detected 
systems. A comparison with the published data showed that our values agree in most cases, though some evident discrepancies were also 
reported. Evidently, in the case of the multiple systems where the ambiguity is higher, several possible models were found. In these cases, 
we usually adopted the solution from the [500 - 520]\,nm interval. Consequently, we presented an updated observational H-R diagram.\\

\begin{figure*}[htbp]
	\begin{center}
	\includegraphics[width=\hsize]{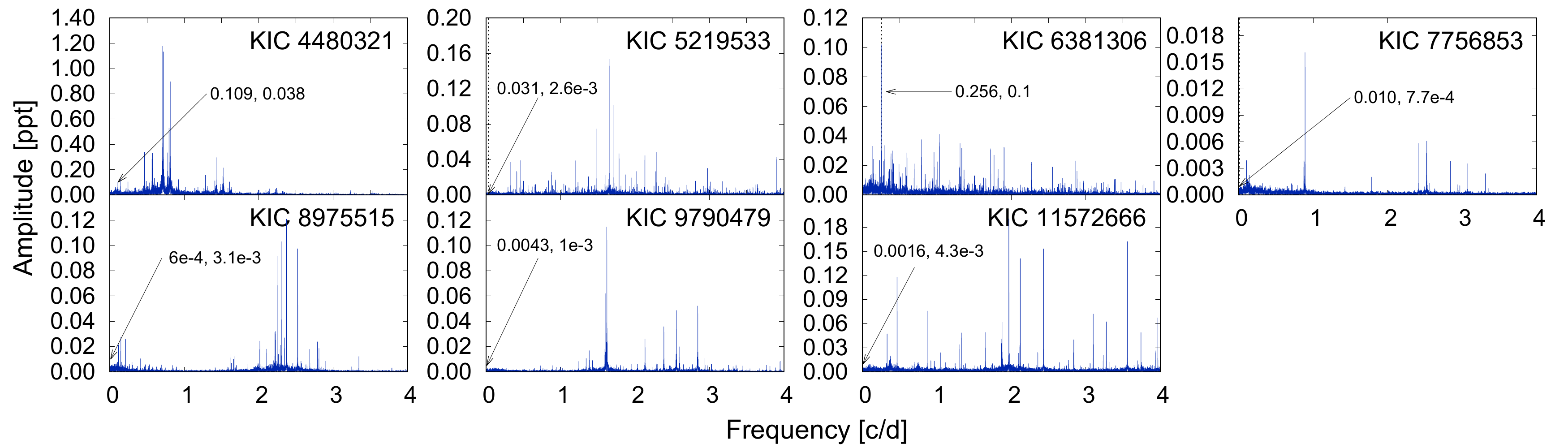}
	\caption{Periodograms of the \textit{Kepler} light curves for the seven SB systems discussed in Sect.~\ref{sect:orb}. For each object, 
	the position corresponding to the newly revealed orbital frequency is plotted as a dashed line indicated by an arrow, and the frequency 
	value as well as its amplitude {(expressed in relative intensity units)} are mentioned. }
	\label{Fig:Mosaic2}	
	\end{center}
\end{figure*}

At this stage, we can start to address our longer term goal by taking a closer look at the frequency content of the newly reported systems
in the region of interest, i.e. in the $\gamma$ Dor regime. 
In Fig.~\ref{Fig:Mosaic2}, we present the periodograms in the low-frequency range, i.e. [0-4] d$^{-1}$, based on the \textit{Kepler} data sets 
for the seven spectroscopic { (binary and multiple)} systems discussed in Sect.~\ref{sect:orb}. The aim is to search for the presence of the orbital 
frequencies in the original periodograms. We see that, in most cases, the amplitudes in the vicinity of the orbital frequency are either much lower 
than the amplitudes of the obvious peaks or fully embedded in the noise. There are two exceptions, however: KIC~4480321 and~6381306. In the case of 
KIC~4480321 (SB3), which shows the highest power in the explored range, there is a well-detected peak (with an amplitude of 0.04 parts per thousand 
(ppt)) which is close though not identical to the orbital frequency. In the case of KIC~6381306 (SB3), the dominant peak (with an amplitude of 0.1 ppt) 
agrees perfectly with the spectroscopically derived orbital frequency. In all other cases, we can find no clear sign of the presence of the orbital 
frequency in the periodograms. The figure furthermore demonstrates that all these systems are rather strongly multi-periodic in the low-frequency regime.\\ 

\begin{figure}[ht]
	\begin{center}
	\includegraphics[width=8.9cm]{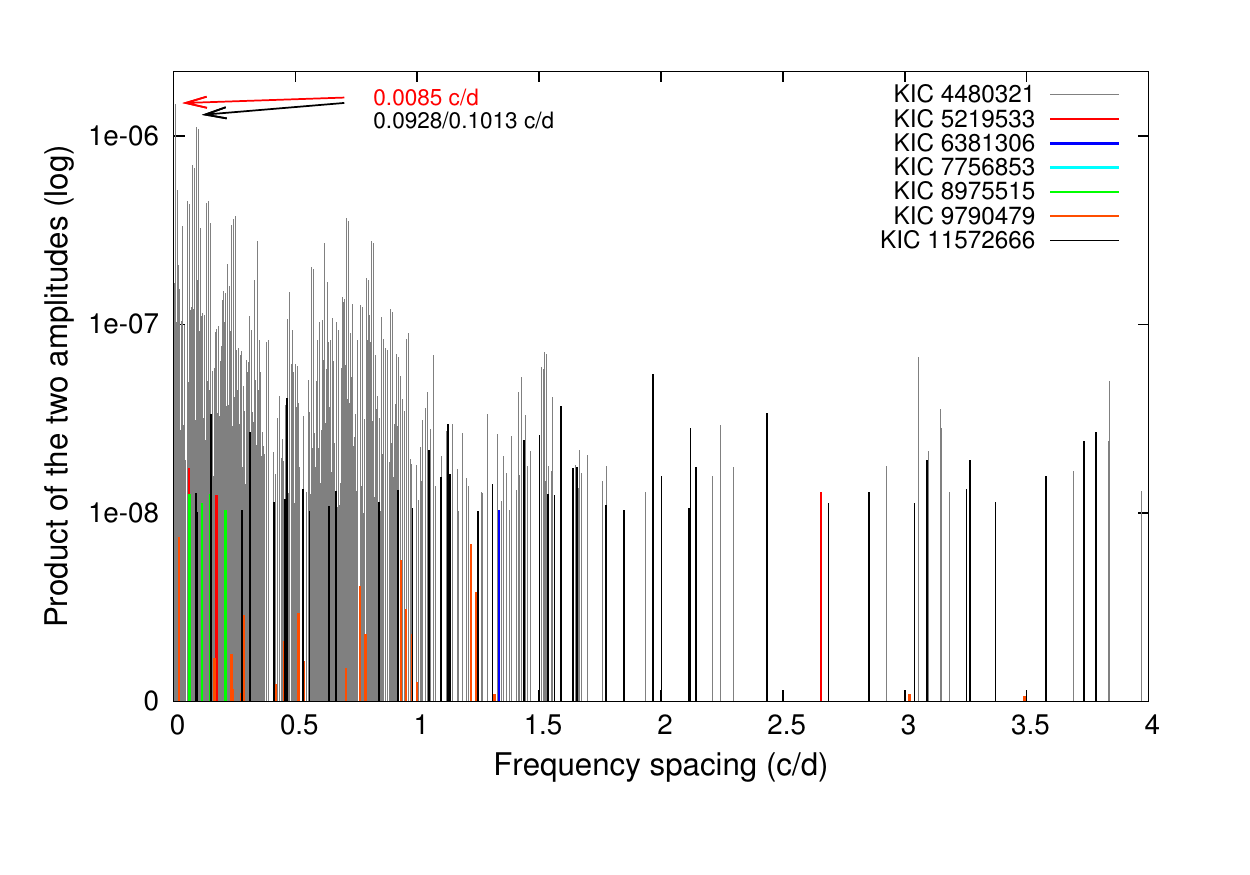}
	\caption{Frequency spacings for the SB systems discussed in Sect.~\ref{sect:orb}. The weights (plotted on the Y-axis) were computed as the product 
	of the amplitudes of both frequencies. For each object, a different colour is used. KIC~4480321 presents by far the richest spacing pattern. Its dominant 
	spacings are located in the vicinity of the orbital frequency. The arrows indicate two of the most frequent spacings as well as their difference.}
	\label{Fig:Fspacing-all}	
	\end{center}
\end{figure}

We also searched for regular frequency spacings in the full-scale periodograms of these objects which might be linked to the orbital 
frequency. In an eccentric binary system, the tidal potential can indeed be represented by a Fourier series with many forcing 
frequencies. When one of the forcing frequencies happens to be close to an eigenfrequency, the oscillation mode is resonantly excited 
(tidally induced) \citep{2003MNRAS.346..968W}. A well-known example of detected tidal excitation is the {\it Kepler} main-sequence
binary KOI-54, where the strongest oscillations occur at 90 and 91 times the orbital frequency \citep{Fuller2012MNRAS.420.3126F}. 
Table~\ref{Tab:Fspacings} lists the relevant spacings for each target. In Fig.~\ref{Fig:Fspacing-all}, we plotted the frequency 
spacings in the range [0-4]~d$^{-1}$ on a logarithmic scale for the SB systems discussed in Sect.~\ref{sect:orb}. We see that KIC~4480321 
has the most dominant and most densely distributed spacing pattern. In comparison, the spacing density of the other systems is very poor. 
In the case of KIC~4480321, the most frequent spacing (S$_1$ = 0.0085 d $^{-1}$) corresponds to the exact difference of the next two most frequent 
ones, i.e. 0.0928 and 0.1013 d$^{-1}$, which are both very close but not equal to the orbital frequency (0.1091 d$^{-1}$). We conclude that this 
signals the influence of the orbital frequency of the close (AB) pair on the other frequencies of KIC~4480321. In the case of KIC~6381306, we 
highlight the presence of a single significant spacing in this range (S$_1$ = 1.3355 d$^{-1}$). \\

\setlength\tabcolsep{2pt}
\begin{table}
\centering
\caption{Most relevant frequency spacings (expressed in d$^{-1}$) sorted according to decreasing importance for the SB systems of Sect.~\ref{sect:orb}.}		
\begin{tabular}{rrrrrr}
\hline\hline 
\mcol{1}{c}{\rm KIC Nr} & \mcol{1}{c}{\rm $S_{1}$} & \mcol{1}{c}{$S_{2}$} & \mcol{1}{c}{$S_{3}$} & \mcol{1}{c}{$S_{4}$} & \mcol{1}{c}{$S_{5}$}\\
\hline
\small 4480321  & \small  0.0085 & \small  0.0928 & \small  0.1013 & \small  0.0787 & \small  0.0872 \\
\small 5219533  & \small  8.6315 & \small  8.5682 & \small  0.0632 & \small  5.9755 & \small  8.8068 \\
\small 6381306  & \small  1.3355 & \small  5.5596 & \small  5.0208 & \small  4.7808 & \small  4.4781 \\
\small 7756853  & \small  0.4240 & \small  2.1626 & \small 20.2503 & \small 15.4084 & \small  -- \\
\small 8975515  & \small 11.5981 & \small 11.6622 & \small 11.4488 & \small  0.0641 & \small  0.1493 \\
\small 9790479  & \small  0.0217 & \small  1.2210 & \small  0.9324 & \small 10.3974 & \small  0.7654 \\
\small 11572666 & \small 16.3120 & \small  1.9671 & \small 15.8459 & \small 14.7212 & \small 10.9377 \\\hline
\end{tabular}\label{Tab:Fspacings}
\end{table}

This study represents an unavoidable step in the investigation of the new A/F-type hybrid phenomenon. We find a significant 
rate of both short- and long-period binary/multiple systems among our targets. The multiplicity comes with different flavours. 
If we add the positive new detections from the photometric time-delay analysis, we may include the targets KIC~3453494,~4044353 
and~9650390 as (most probably) belonging to the class of single-lined, long-period binary systems. The multi-year
monitoring of the RVs enabled the detection of several long-period systems with orbital periods of the order of 4-6 years, e.g. 
KIC~4480321,~5219533,~8975515 and~9775454. It is our intention to continue the monitoring for these interesting systems. 
Moreover, since the photometric time-delay analysis - which is based on the 4-years time span of the \textit{Kepler} 
mission - shows the existence of periodicities of the same order, we intend to combine both data types into a simultaneous modelling 
in the near future. Indeed, we already showed that the orbital solutions currently based on the RVs only are confirmed by a 
simple modelling of the LiTE in three cases. 
Overall, we may safely conclude that {\it about one third} of our sample of candidate hybrid stars belongs to a binary or a 
multiple system. This number is very close to the fraction of binary and multiple systems suggested from a spectroscopic-photometric 
analysis for a sample of 32 candidate SX~Phe stars in the \textit{Kepler} field \citep[][]{Nemec2017MNRAS.466.1290N}.\\

This result implies that the physical cause of the low frequencies which are detected in the \textit{Kepler} periodograms could be related 
to the orbital motions, and that these so-called hybrid stars may not be genuine hybrid pulsators. They may have forced oscillations 
coupled to the orbital frequency (its harmonics), or show ellipsoidal modulations in their light curves. In two of the seven 
SB systems, we found straightforward evidence for the presence of the orbital frequency in the periodograms (KIC~4480321 and~6381306). 
In KIC~6381306, the orbital frequency is the most dominant peak. It is worth noting that such interaction was detected in the two 
short-period systems with an orbital period smaller than 10~d. In the remaining cases (those with orbital periods beyond 30~d), 
there is no influence at first sight, the more dominant peaks are situated in the range [0.5-3]~d$^{-1}$, as is usual for the g-mode 
pulsators of $\gamma$ Dor type. The multiperiodicity and apparent complexity of the frequency content in this region still need an 
explanation in most of the (long-period) systems.\\ 

We are not yet able to differentiate between the true hybrid pulsators and those stars where spots or fast rotation introduce inhomogeneities 
on the stellar surface. On the one hand, an extended analysis of the \textit{Kepler} data including a precise identification of the g-mode period 
spacing patterns may help out, provided there are enough visible g-modes to reliably detect such patterns \citep[e.g.][]{VanReeth2015ApJS..218...27V}. 
On the other hand, detailed spectropolarimetric investigations \citep[e.g.][]{Neiner2015MNRAS.454L..86N} can provide true physical 
insight regarding the origin of the low frequencies in some cases. In conclusion, the resulting orbital periods and elements of the newly 
identified systems, projected rotational velocities and homogeneous atmospheric properties provide relevant and essential information 
for a more reliable and deeper physical understanding of the \textit{Kepler} periodograms.\\ 

\section*{Acknowledgements}

This research is based on high-resolution spectra obtained with the \textsc{Hermes} \'echelle spectrograph (2013-2016) installed at the 
Mercator telescope, operated by the IvS, KULeuven, funded by the Flemish Community, and located at the Observatorio del Roque de los Muchachos, 
La Palma, Spain, of the Instituto de Astrofisica de Canarias, and with the \textsc{Ace} \'echelle spectrograph (2014-2015)
attached to the 1-m RCC telescope of the Konkoly Observatory at Piszk\'es-tet\H o, Hungary. The observations performed at Piszk\'es-tet\H o 
were supported by the Lend\"ulet grant LP2012-31 of the Hungarian Academy of Sciences.
We furthermore made usage of the very high-quality data collected by the \textit{Kepler} satellite. 
Funding for the \textit{Kepler} mission is provided by the NASA Science Mission directorate.\\

The authors gratefully acknowledge the financial support from the Royal Observatory of Belgium to the \textsc{Hermes} 
Consortium, as well as the help of the \textsc{Hermes} observers J. Boulangier, W. Homan, R. Karjalainen, R. Lombaert, 
R. Manick, C. Paladini, A. Tkachenko, M. Van de Sande and H. Van Winckel. We thank the referee for most useful 
comments and suggestions for improvements. M.S. acknowledges the support of the postdoctoral fellowship programme of the 
Hungarian Academy of Sciences and the Konkoly Observatory as a host institution. \'A.S., M.S. and Zs.B. acknowledge the 
support of the Hungarian NKFIH Grants K-113117, K-115709 and K-119517. \'A.S. received a J\'anos Bolyai Research Scholarship 
(Hungarian Academy of Sciences). Zs.B. acknowledges the support of the Hungarian NKFIH Grant PD-123910. {H.L. acknowledges 
the support of the DFG grant LE 1102/3-1.} 
This work was performed using the POLLUX database (\textit{http://pollux.graal.univ-montp2.fr}) operated by LUPM, Université 
Montpellier - CNRS, France with the support of the PNPS and INSU, as well as the SAO/NASA Astrophysics Data System (ADS) and 
the SIMBAD database operated at the CDS, Strasbourg, France (\textit{http://simbad.u-strasbg.fr/Simbad}).\\ 

\bibliographystyle{aa}
\bibliography{Lampens_v8}

\setcounter{table}{0}
\longtab{
\begin{landscape}
\begin{longtable}{lllccrccccccccrr}
\caption{Classification, model parameters and atmospheric stellar parameters obtained from the \textsc{Hermes} spectra. \label{tab:param1}}\\
\hline\hline
\mcol{1}{c}{\rm KIC} & \mcol{1}{l}{\rm Period} & \mcol{1}{l}{\rm Sp T} & \mcol{1}{c}{$T_{\mathrm{eff}}$} & \mcol{1}{c}{$\mathrm{log}\,g$} & \mcol{1}{r}{Kp} & {\textsc H} & Comments & Category & Param. Mod. & \mcol{1}{c}{Mean $T_{\mathrm{eff}}$} & \mcol{1}{c}{Mean $\mathrm{log}\,g$} & \mcol{1}{c}{GIR $T_{\mathrm{eff}}$} & \mcol{1}{c}{GIR $\mathrm{log}\,g$} & \mcol{1}{c}{GIR $v\,\mathrm{sin}\,i$}\\
\mcol{1}{c}{Nr} & \mcol{1}{l}{KIC} & \mcol{1}{l}{KIC} & \mcol{1}{c}{KIC} & \mcol{1}{c}{KIC} & \mcol{1}{r}{mag} & &&&\mcol{1}{c}{Sp,$v\,\mathrm{sin}\,i$} &&& &&&\\ 
\hline
\endfirsthead
\caption{continued.}\\
\hline\hline
\mcol{1}{c}{\rm KIC} & \mcol{1}{l}{\rm Period} & \mcol{1}{l}{\rm Sp T} & \mcol{1}{c}{$T_{\mathrm{eff}}$} & \mcol{1}{c}{$\mathrm{log}\,g$} & \mcol{1}{r}{Kp} & {\textsc H} & Comments & Category & Param. Mod. & \mcol{1}{c}{Mean $T_{\mathrm{eff}}$} & \mcol{1}{c}{Mean $\mathrm{log}\,g$} & \mcol{1}{c}{GIR $T_{\mathrm{eff}}$} & \mcol{1}{c}{GIR $\mathrm{log}\,g$} & \mcol{1}{c}{GIR $v\,\mathrm{sin}\,i$}\\
\mcol{1}{c}{Nr} & \mcol{1}{l}{KIC} & \mcol{1}{l}{KIC} & \mcol{1}{c}{KIC} & \mcol{1}{c}{KIC} & \mcol{1}{r}{mag} & &&&\mcol{1}{c}{Sp,$v\,\mathrm{sin}\,i$} &&& &&&\\ 
\hline
\endhead
\setlength{\tabcolsep}{1.5mm}
\small 3097912 &\small  0.68213 &\small A5    &\small 7880 &\small 3.56 &\small 9.4  &\small 5 &\small  visual\tablefootmark{a} &\small S       &\small A5, 120 &\small 8000 &\small 4.1-4.3 &\small 8147 &\small 4.17 &\small 118 \struutup\\
\small 3429637 &\small  0.09673 &\small A9m   &\small 7200 &\small 4 &\small 7.7&\small 21 &\small visual\tablefootmark{d} Am &\small P+VAR &\small F0, 50 &\small 7350 &\small <=4.0 &\small 7256 &\small 2.66 &\small 49\\
\small 3437940 &\small  0.09673 &\small F0    &\small 7430 &\small 3.86 &\small 8.5  &\small 11 &\small                  &\small P       &\small A7, 110 &\small 7700 &\small <=3.9   &\small 7702 &\small 3.51 &\small 111\\
\small 3453494 &\small  0.13335 &\small A5    &\small 7810 &\small 3.84 &\small 9.6  &\small 5 &\small                   &\small S?      &\small A5, 220 &\small 7870 &\small >=4.1   &\small 7941 &\small 4.69 &\small 230\\
\small 3851151 &\small  0.03775 &\small A2    &\small 8190 &\small 4.1  &\small 9.8  &\small 5 &\small                   &\small S?      &\small A2, 110 &\small 8600 &\small 3.8     &\small 8571 &\small 3.71 &\small 124\\
\small 4044353 &\small  0.41841 &\small A2    &\small 8300 &\small 3.56 &\small 9.8  &\small 6 &\small                   &\small S       &\small A5, 110 &\small 8300 &\small 4.1-4.5 &\small 8309 &\small 4.3 &\small 107\\
\small 4281581 &\small  1.13379 &\small A2    &\small 8140 &\small 3.84 &\small 9.4  &\small 5 &\small                   &\small S       &\small A5, 110 &\small 8300 &\small 3.7-3.9 &\small 8406 &\small 3.55 &\small 104\\
\small 4480321 &\small  1.40845 &\small A3    &\small 7147 &\small 3.87 &\small 10.3 &\small 49 &\small                  &\small SB3     &\small 2F0,10/A5,160  & \hspace{1mm}\it{\small See text}  &\small  &\small  &\small  &\small \\
\small 4671225 &\small  0.11261 &\small A7    &\small 8230 &\small 3.27 &\small 10   &\small 6 &\small                   &\small S       &\small A5, 150 &\small 8000 &\small <=4.1   &\small 8008 &\small 3.76 &\small 152\\
\small 4989900 &\small  0.45683 &\small A2    &\small 7900 &\small 3.51 &\small 6.9  &\small 9 &\small                   &\small S/P?    &\small A5, 200 &\small 8100 &\small <=3.7   &\small 8180 &\small 3.24 &\small 207\\
\small 5219533 &\small  0.09723 &\small A8m   &\small 7410 &\small 3.94 &\small 9.2  &\small 21 &\small visual\tablefootmark{a} Am      &\small SB3   &\small 2A5,10/A7,{\it 115\tablefootmark{a}} & \hspace{1mm}\it{\small See text} &\small  &\small  &\small  &\small \\
\small 5437206 &\small  0.07686 &\small A2    &\small 7710 &\small 3.67 &\small 8.4  &\small 12 &\small                   &\small P       &\small A5, 110 &\small 7930 &\small <=3.7   &\small 7872 &\small 3.55 &\small 115\\
\small 5473171 &\small  0.13203 &\small A2    &\small 7450 &\small 3.63 &\small 9    &\small 5 &\small                   &\small P       &\small A5, 160 &\small 7670 &\small <=4.3   &\small 7784 &\small 3.76 &\small 164\\
\small 5724440 &\small  0.05195 &\small A5    &\small 7290 &\small 3.57 &\small 7.9  &\small 9 &\small                   &\small P?      &\small A5, 240 &\small 7730 &\small >= 3.9  &\small 7749 &\small 4.64 &\small 259\\
\small 5965837 &\small  0.29481 &\small F2    &\small 6520 &\small 4.15 &\small 9.2  &\small 10 &\small                   &\small P       &\small F3, 15  &\small --   &\small  --     &\small 6800 &\small 3.3\tablefootmark{c}  &\small 15\\
\small 6032730 &\small  0.06147 &\small A2    &\small 7110 &\small 3.79 &\small 8.7  &\small 5 &\small                   &\small S       &\small A7, 260 &\small 7530 &\small >=4.7   &\small 7519 &\small 4.63 &\small 257\\
\small 6289468 &\small  0.07325 &\small A2    &\small 8270 &\small 3.74 &\small 9.4  &\small 7 &\small                   &\small S       &\small A5, 160 &\small 8300 &\small 3.7-3.9 &\small 8208 &\small 3.58 &\small 164\\
\small 6381306 &\small  0.172   &\small A0    &\small 8060 &\small 3.63 &\small 8.7  &\small 25&\small                   &\small SB3     &\small 2A7,0/A5,90 & \hspace{1mm}\it{\small See text} &\small  &\small  &\small  &\small \\
\small 6432054 &\small  0.0946  &\small F0    &\small 7090 &\small 3.85 &\small 8.2  &\small 10 &\small                   &\small P       &\small A7, 180 &\small 7530 &\small 4.5     &\small 7542 &\small 4.44 &\small 184\\
\small 6587551 &\small  0.05740 &\small A0    &\small 8380 &\small 3.93 &\small 9.8  &\small 4 &\small                   &\small S       &\small A2, 150 &\small 8500 &\small 4.1-4.3 &\small 8639 &\small 3.94 &\small 151\\
\small 6670742 &\small  0.07819 &\small A5    &\small 7450 &\small 3.61 &\small 9.3  &\small 7 &\small                   &\small S/P?    &\small A7, 260 &\small 7470 &\small <=4.5   &\small 7472 &\small 4.3 &\small 261\\
\small 6756386 &\small  0.11156 &\small A2    &\small 7990 &\small 3.51 &\small 8.7  &\small 7 &\small                   &\small P+VAR   &\small A5, 200 &\small 7900 &\small 4.1     &\small 8037 &\small 4.16 &\small 206\\
\small 6756481 &\small  0.17343 &\small F0    &\small 7310 &\small 3.52 &\small 9.3  &\small 7 &\small                   &\small CMP     &\small A7, 220 &\small --   &\small         &\small 7408 &\small 4.38 &\small 245\\
\small 6951642 &\small  1.38696 &\small A5    &\small 7180 &\small 3.37 &\small 9.7  &\small 11 &\small                   &\small P+VAR   &\small F0, 120 &\small 7200 &\small 3.7-3.9 &\small 7203 &\small 3.87 &\small 121\\
\small 7119530 &\small  0.23849 &\small A3    &\small 7780 &\small 3.49 &\small 8.5  &\small 7 &\small                   &\small CMP     &\small A5, 220 &\small --   &\small         &\small 7721 &\small 3.37 &\small 249\\
\small 7668791 &\small  0.04844 &\small A2    &\small 8150 &\small 3.89 &\small 9.3  &\small 7 &\small                   &\small S       &\small A5, 50  &\small 8300 &\small 3.7-3.9 &\small 8146 &\small 3.85 &\small 49\\
\small 7748238 &\small  0.4363  &\small A5    &\small 7230 &\small 3.47 &\small 9.5  &\small 6 &\small                   &\small S       &\small A7, 120 &\small 7400 &\small 4.5     &\small 7380 &\small 4.65 &\small 123\\
\small 7756853 &\small  0.04730 &\small A0    &\small 8060 &\small 3.94 &\small 9    &\small 14 &\small       Am         &\small SB2     &\small A1,30/A5,50 & \hspace{1mm}\it{\small See text} &\small  &\small  &\small  &\small \\
\small 7770282 &\small  0.99602 &\small F0    &\small 7450 &\small 3.5  &\small 9.7  &\small 20 &\small                  &\small P?      &\small A5,  50 &\small 7800 &\small 4.3     &\small 7921 &\small 4.19 &\small 52\\
\small 7827131 &\small  0.09956 &\small A2    &\small 8290 &\small 3.49 &\small 8    &\small 9 &\small                  &\small S       &\small A2, 240 &\small 8100 &\small <=3.5   &\small 8137 &\small 2.95 &\small 277\\
\small 7959867 &\small  0.12965 &\small A2    &\small 8480 &\small 3.9  &\small 9.8  &\small 6 &\small                  &\small S       &\small A3, 150 &\small 8350 &\small 3.9-4.1 &\small 8431 &\small 3.74 &\small 151\\
\small 8738244 &\small  0.0684  &\small A3V   &\small 8170 &\small 4.15 &\small 8.2  &\small 7 &\small                  &\small S       &\small A5, 140 &\small 8300 &\small 3.8-3.9 &\small 8282 &\small 3.64 &\small 143\\
\small 8915335 &\small  0.11328 &\small A2    &\small 7770 &\small 3.48 &\small 9.6  &\small 6 &\small                  &\small S       &\small A5, 200 &\small 8100 &\small 3.7-3.9 &\small 8094 &\small 3.62 &\small 211\\
\small 8975515 &\small  0.07157 &\small A2    &\small 7180 &\small 3.9  &\small 9.5  &\small 20 &\small                 &\small SB2     &\small A7,150/A7,30 & \hspace{1mm}\it{\small See text} &\small  &\small  &\small  &\small \\
\small 9351622 &\small  0.16611 &\small F0    &\small 7450 &\small 3.53 &\small 9.1  &\small 9 &\small                  &\small S       &\small A7, 80  &\small 7600 &\small <=3.8   &\small 7711 &\small 3.53 &\small 77\\
\small 9413057 &\small  0.07382 &\small A2    &\small 8470 &\small 3.87 &\small 9.6  &\small 4 &\small                  &\small S       &\small A2, 180 &\small 8600 &\small 3.9-4.1 &\small 8698 &\small 3.79 &\small 190\\
\small 9509296 &\small  0.05657 &\small ---   &\small 7400 &\small 3.61 &\small 9.9  &\small 8 &\small                  &\small S       &\small A7, 110 &\small 7400 &\small 4.3-4.7 &\small 7332 &\small 4.73 &\small 117\\
\small 9650390 &\small  0.93371  &\small A0   &\small 8390 &\small 3.68 &\small 9.4  &\small 7 &\small                  &\small S       &\small A2, 240 &\small 8100 &\small <=3.7   &\small 8297 &\small 3.2 &\small 272\\
\small 9700679 &\small  3.81679  &\small G2III &\small 5070 &\small 4.45 &\small 9.9  &\small 4 &\small                 &\small SB1     &\small G2, 5      &\small --   &\small   --    &\small  &\small  &\small \\
\small 9764965 &\small  0.03679  &\small A5mp &\small 7460 &\small 4.09 &\small 8.9  &\small 6 &\small        Am        &\small S       &\small A7, 85  &\small 7670 &\small 4.1-4.7 &\small 7726 &\small 4.8 &\small 83\\
\small 9775454 &\small  0.24033  &\small F1IV &\small ---  &\small ---  &\small 8.2  &\small 13 &\small                  &\small SB1     &\small F0, 70 &\small 7200 &\small 3.9     &\small 7287 &\small 4.25 &\small 65\\
\small 9790479 &\small  0.61805  &\small A2   &\small 7840 &\small 4.04 &\small 9.9  &\small 11 &\small                  &\small SB1     &\small A5, 40  &\small 8300 &\small 3.9-4.1 &\small 8239 &\small 4.28 &\small 40\\
\small 9970568 &\small  0.29343  &\small A2   &\small 7790 &\small 3.64 &\small 9.6  &\small 5 &\small                  &\small S       &\small A5, 240 &\small 7750 &\small 3.9-4.5 &\small 7843 &\small 4.27 &\small 252\\
\small 10264728 &\small  0.29595 &\small A2   &\small 7790 &\small 3.85 &\small 9.9  &\small 4 &\small                  &\small S       &\small A5, 240 &\small 7730 &\small 4.5-4.7 &\small 7803 &\small 4.31 &\small 256\\
\small 10537907 &\small  0.08654 &\small F0   &\small 7500 &\small 3.45 &\small 9.9  &\small 6 &\small                  &\small P+VAR?  &\small A5, 110 &\small 7600 &\small 4.1-4.3 &\small 7660 &\small 4.26 &\small 118\\
\small 10664975 &\small  0.38835 &\small A2   &\small 7950 &\small 3.58 &\small 7.6  &\small 10 &\small                  &\small P       &\small A5, 240 &\small 7800 &\small <=4.3   &\small 7892 &\small 3.76 &\small 243\\
\small 11180361 &\small  0.2665\tablefootmark{b}  &\small A2/3V &\small 8330 &\small 3.55 &\small 7.7  &\small 7 &\small KEB\tablefootmark{b}\     &\small S/P? &\small A5, 200 &\small 8350 &\small 3.7-3.9 &\small 8387 &\small 3.56 &\small 196\\
\small 11193046 &\small  0.78431 &\small A2   &\small 8170 &\small 3.7  &\small 9.6  &\small 5 &\small                  &\small S       &\small A5, 200 &\small 7900 &\small >=3.9   &\small 7920 &\small 3.9 &\small 208\\
\small 11445913 &\small  0.03169 &\small A9m  &\small 6950 &\small 3.89 &\small 8.5  &\small 22 &\small         Am      &\small SB2     &\small F0,55/K0,10  &\hspace{1mm}\it{\small See text} &\small  &\small  &\small  &\small \\
\small 11572666 &\small  0.05471 &\small ---  &\small 7040 &\small 3.49 &\small 9.9  &\small 14 &\small                 &\small SB2     &\small A5,250/F3,20 &\hspace{1mm}\it{\small See text} &\small  &\small  &\small  &\small \\
\small 11602449 &\small  0.09453 &\small ---  &\small 7390 &\small 3.83 &\small 9.9  &\small 15 &\small                 &\small S       &\small A7, 130 &\small 7470 &\small >=3.9   &\small 7528 &\small 4.45 &\small 134 \struutdown\\
\end{longtable}
\tablefoot{
\tablefoottext{a}{\citet{Uytterhoeven2011A&A...534A.125U}}
\tablefoottext{b}{\citet{Slawson2011AJ....142..160S}}
\tablefoottext{c}{Range [510-520]~nm} }
\end{landscape} }

\appendix

\section{Cross-correlation functions of A/F-type candidate hybrid stars}
\label{sect:CCFs}

%
%

\begin{figure*}
\begin{center}
\includegraphics[width=16.01cm]{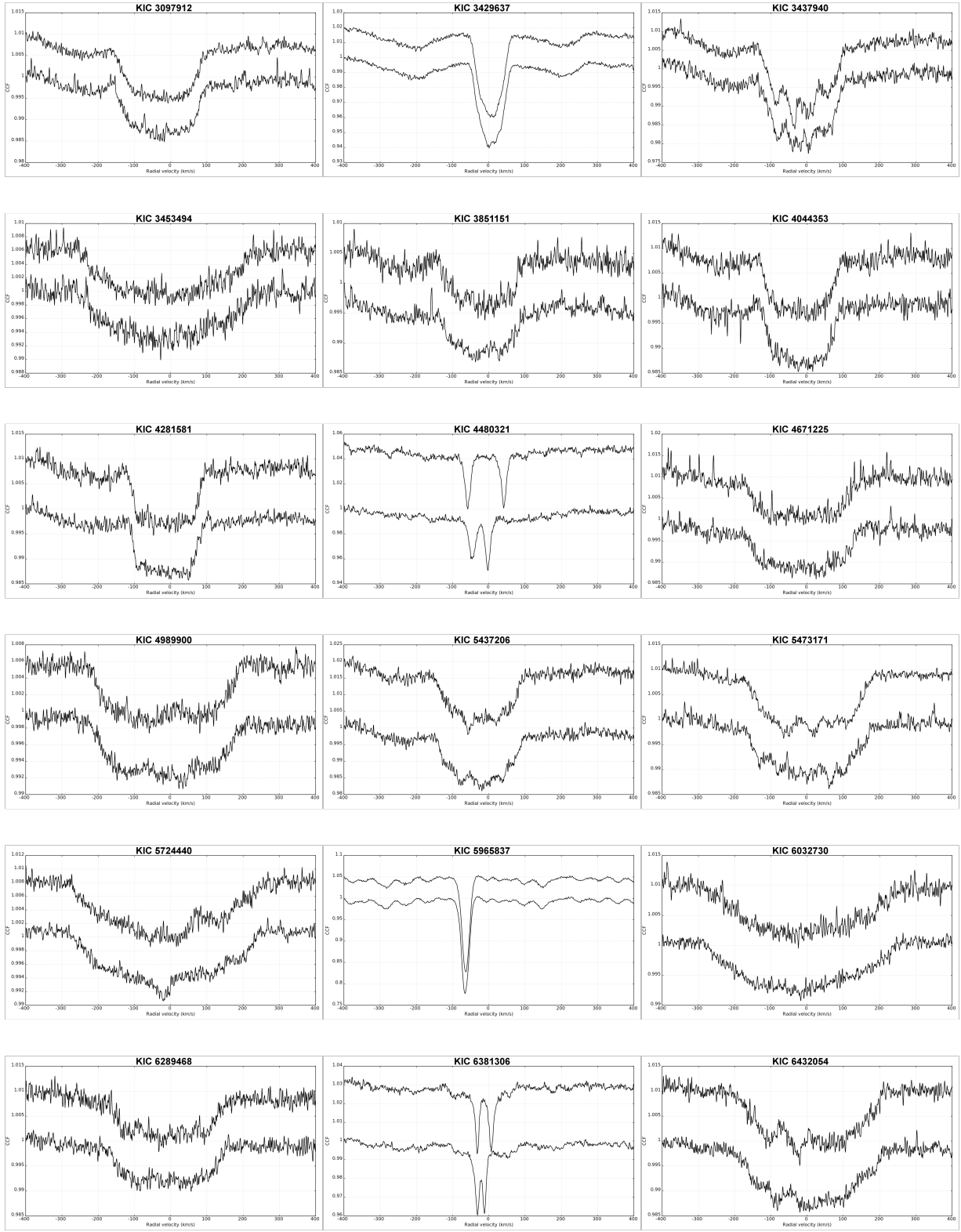}
\end{center}
\caption{Cross-correlation functions for 48 Kepler objects. The epochs were chosen so as to show as much contrast as possible between two CCFs.}
\label{ccf:3x6fig11}
\end{figure*}

\setcounter{figure}{0}
\begin{figure*}
\begin{center}
\includegraphics[width=16.01cm]{lampens_ccf_3x6_lbl_fig11.pdf}
\end{center}
\caption{Cross-correlation functions for 48 Kepler objects (cont'ed).}
\label{ccf:3x6fig12}
\end{figure*}

\setcounter{figure}{0}
\begin{figure*}
\begin{center}
\includegraphics[width=16.01cm]{lampens_ccf_3x6_lbl_fig12.pdf}
\end{center}
\caption{Cross-correlation functions for 48 Kepler objects (cont'ed).}
\label{ccf:3x6fig13}
\end{figure*}

\newpage
\section{Radial velocity curves}
\label{sect:RVs}

\begin{figure*}
\begin{center}
\includegraphics[width=14.21cm]{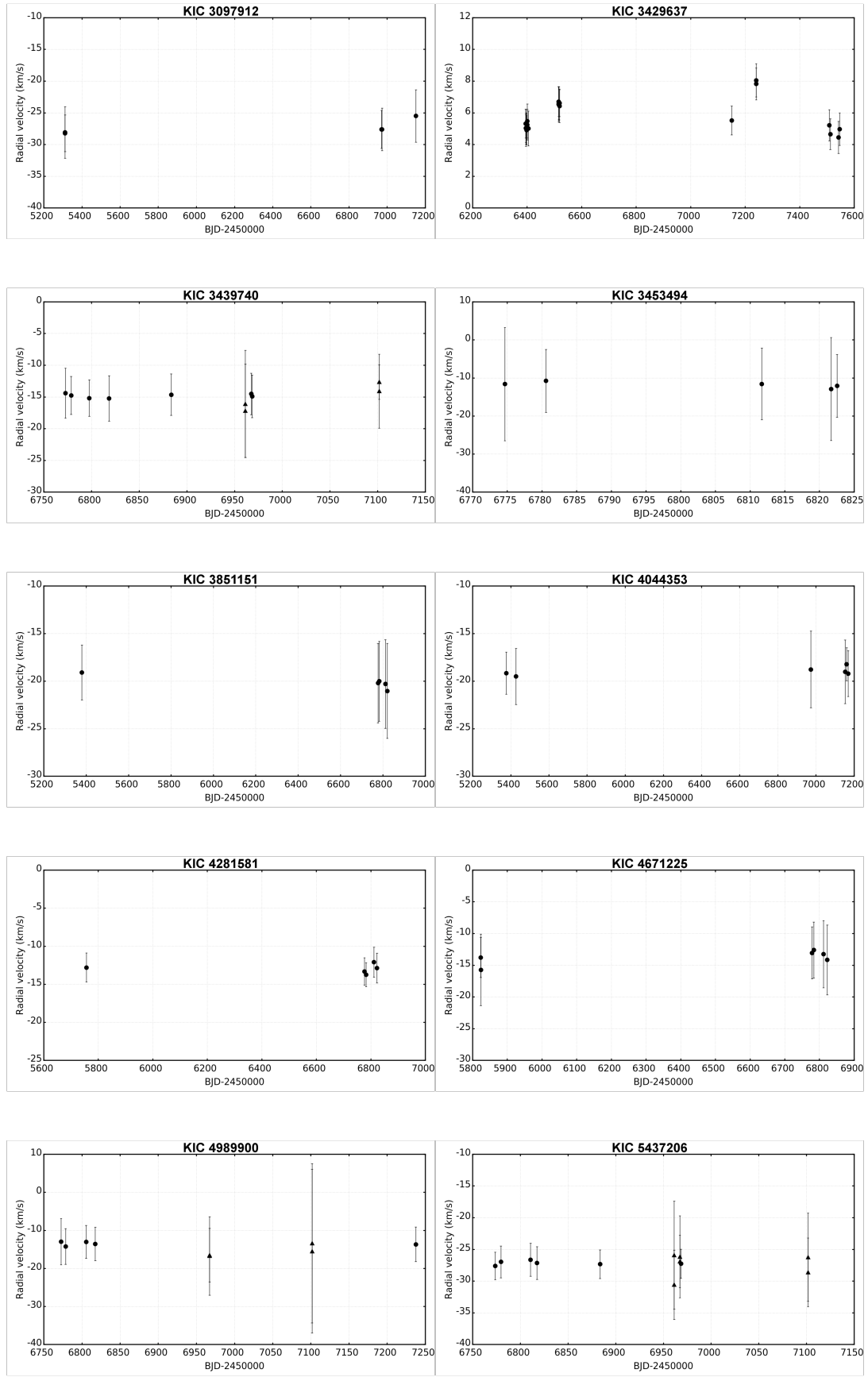}
\end{center}
\caption{Radial velocity plots for 44 Kepler objects. The symbols indicate data collected with the {\sc Hermes} ($\circ$, $\bullet$) or {\sc Ace} spectrographs ($\bigtriangleup$).}
\label{an:rvs_fig1}
\end{figure*}

\setcounter{figure}{0}
\begin{figure*}
\begin{center}
\includegraphics[width=14.21cm]{lampens_rvs_2x5_fig01_lbl.pdf}
\end{center}
\caption{Radial velocity plots for 44 Kepler objects (cont'ed). The same symbols as before are used.}
\label{an:rvs_fig2}
\end{figure*}

\setcounter{figure}{0}
\begin{figure*}
\begin{center}
\includegraphics[width=14.21cm]{lampens_rvs_2x5_fig02_lbl.pdf}
\end{center}
\caption{Radial velocity plots for 44 Kepler objects (cont'ed). The same symbols as before are used.}
\label{an:rvs_fig3}
\end{figure*}

\setcounter{figure}{0}
\begin{figure*}
\begin{center}
\includegraphics[width=14.21cm]{lampens_rvs_2x5_fig03_lbl.pdf}
\end{center}
\caption{Radial velocity plots for 44 Kepler objects (cont'ed). The same symbols as before are used.}
\label{an:rvs_fig4}
\end{figure*}

\setcounter{figure}{0}
\begin{figure*}
\begin{center}
\includegraphics[width=14.21cm]{lampens_rvs_2x5_fig04_lbl.pdf}
\end{center}
\caption{Radial velocity plots for 44 Kepler objects (cont'ed). The same symbols as before are used.}
\label{an:rvs_fig5}
\end{figure*}

\begin{figure}
\begin{center}
\includegraphics[width=7.4cm]{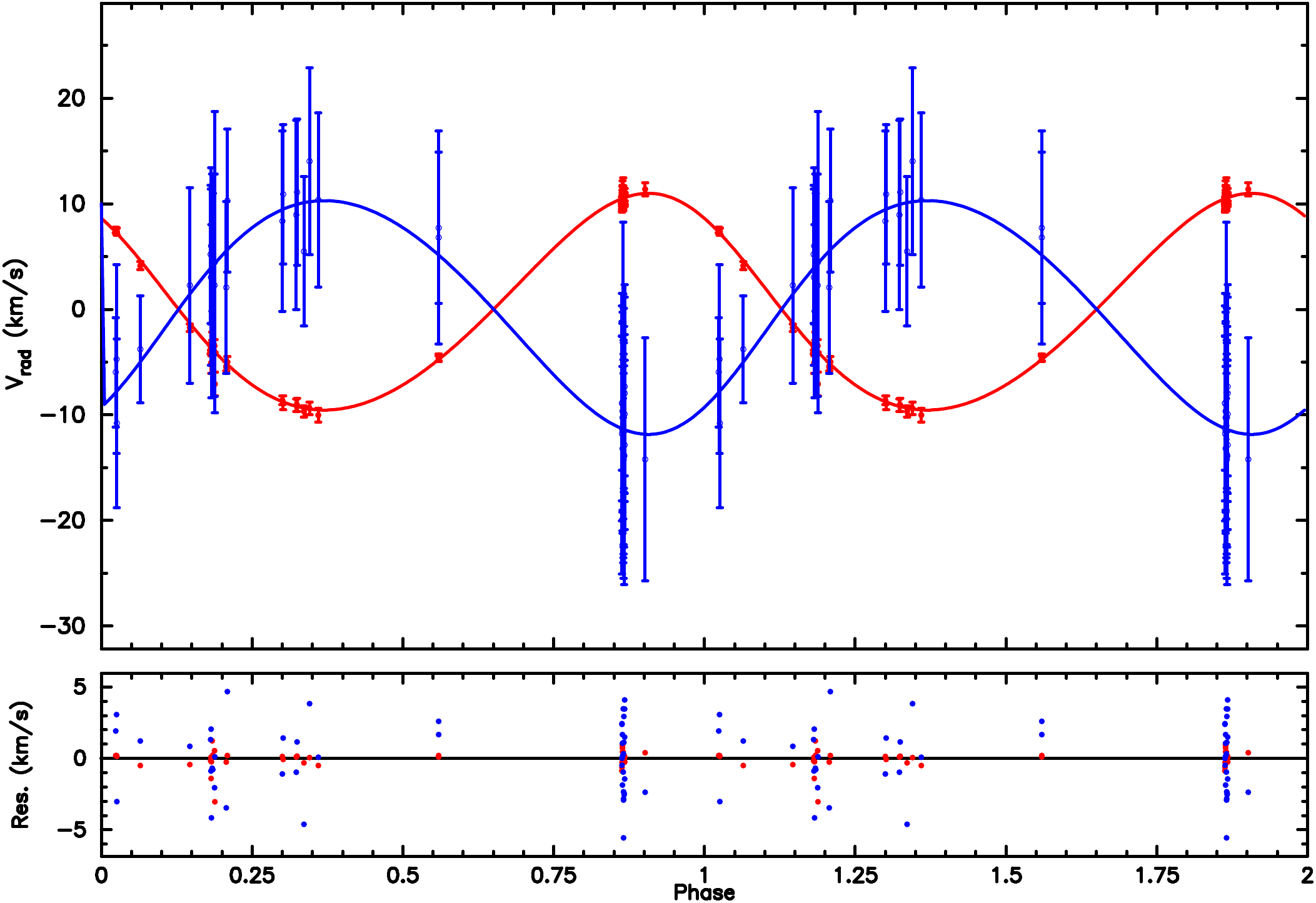}
\end{center}
\caption{Radial velocities for the mass centre of the inner binary (AB) and for component C of KIC~4480321, 
 plotted over a possible orbital solution with a period of 2280~days. The residuals are shown in the bottom part.} 
\label{fig:KIC44_sb3}
\end{figure}

\begin{figure}
\begin{center}
\includegraphics[width=7.4cm]{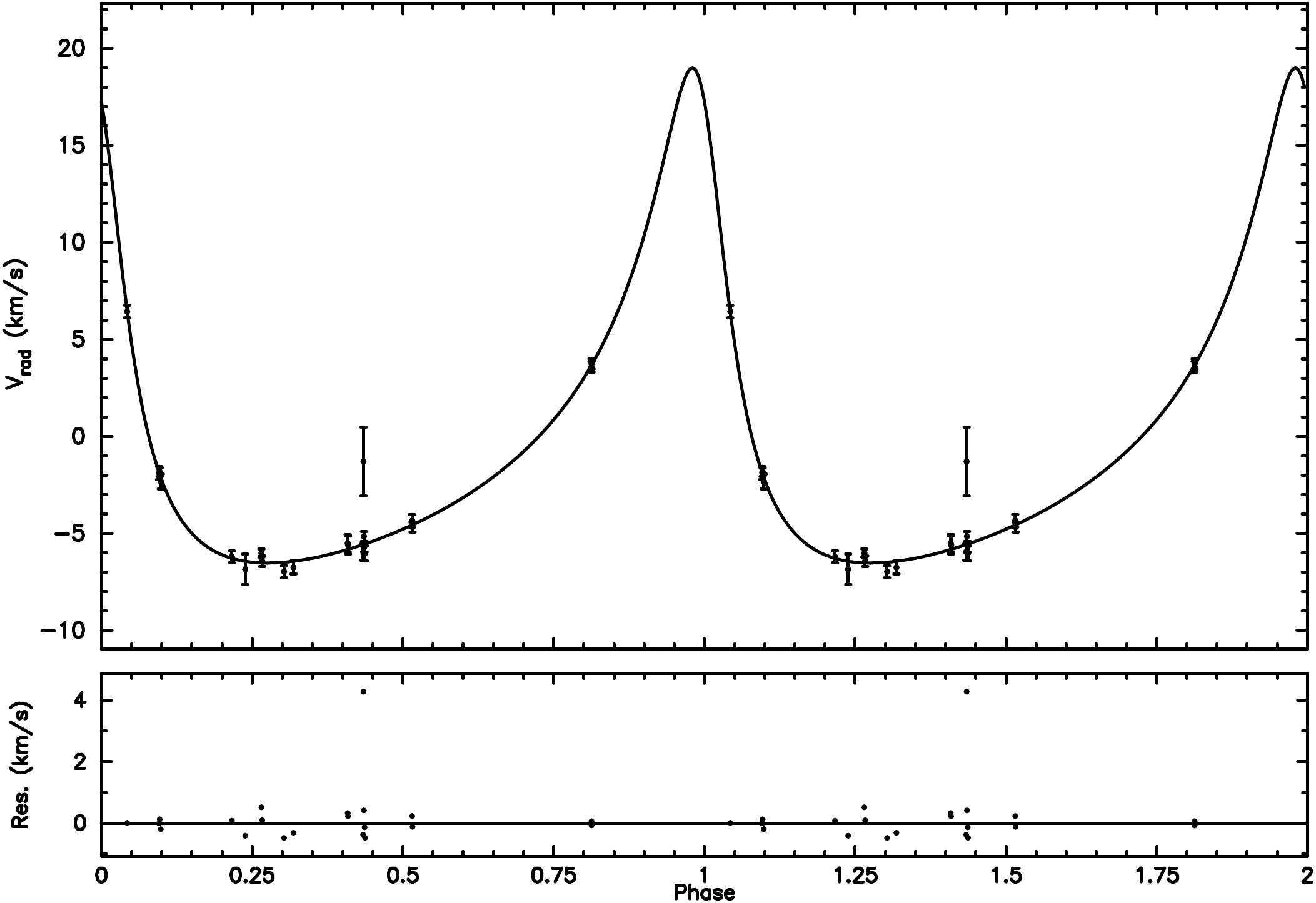}
\end{center}
\caption{Radial velocities for the mass centre of the inner binary (AB) in the triple system KIC~5219533 plotted 
 over a tentative orbital solution with a period of about 1600~days. The residuals are shown in the bottom part.} 
\label{fig:KIC52_sb1}
\end{figure}

\begin{figure}
\begin{center}
\includegraphics[width=7.4cm]{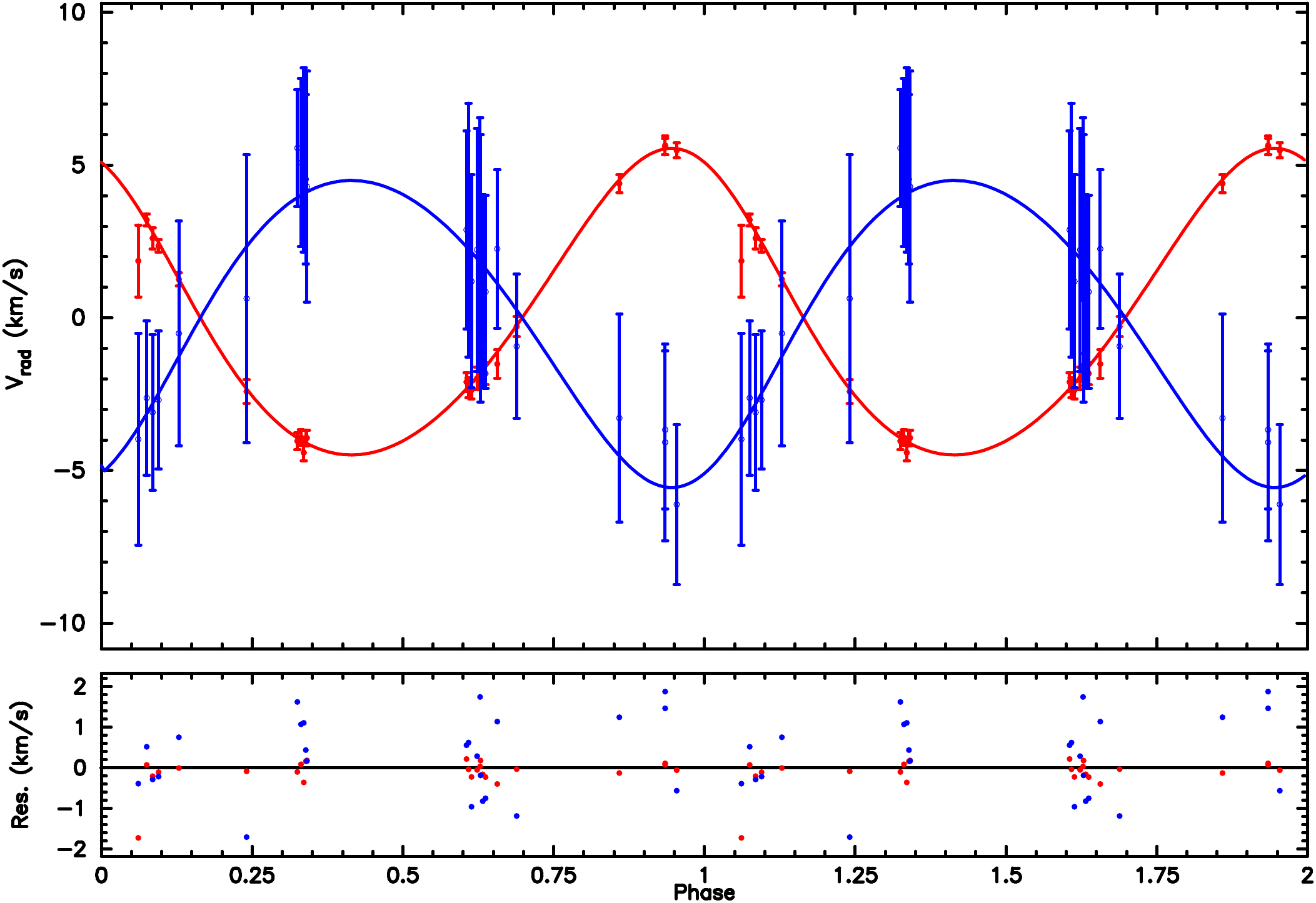}
\end{center}
\caption{Radial velocities for the mass centre of the inner binary (AB) and for component C of KIC~6381306, 
 plotted over the outer orbital solution with a period of 212~days. The residuals are shown in the bottom part.} 
\label{fig:KIC63_sb3}
\end{figure}

\section{List of radial velocity data }\label{sect:listRV}

\setcounter{table}{0}
\begin{table*}
\centering
\setlength{\tabcolsep}{1.5mm}
\caption{List of radial velocity data used in this work. I. Single objects and single-lined systems. 
\textit{H} stands for \textit{Hermes} while \textit{A} stands for \textit{Ace}.}
\begin{tabular}{ccrrc}
\hline
\hline
\mcol{1}{c}{\rm KIC} & \mcol{1}{c}{\rm HJD} & \mcol{1}{c}{\rm RV$_{1}$} & \mcol{1}{c}{$\sigma_{RV_{1}}$} & \mcol{1}{c}{Source}\\
\mcol{1}{c}{\rm Nr} & \mcol{1}{c}{ } & \mcol{1}{c}{\kms} & \mcol{1}{c}{\kms} & \mcol{1}{c}{[1:\textit{H}, 2:\textit{A}]}\\
\hline
\small 3097912  & \small 2455311.54168  & \small -28.232  & \small  2.894  & \small   1   \\
\small 3097912  & \small 2455311.56259  & \small -28.092  & \small  4.071  & \small   1   \\
\small 3097912  & \small 2456970.40110  & \small -27.605  & \small  2.986  & \small   1   \\
\small 3097912  & \small 2456973.30681  & \small -27.623  & \small  3.353  & \small   1   \\
\small 3097912  & \small 2457150.67141  & \small -25.497  & \small  4.127  & \small   1   \\
\small 3429637  & \small 2456394.66396  & \small   5.312  & \small  0.907  & \small   1   \\
\hline                                                                                                              
\end{tabular}
\label{tab:radvel1}
\end{table*}

\clearpage
\setcounter{table}{0}
\begin{table*}
\centering
\setlength{\tabcolsep}{1.5mm}
\caption{List of radial velocity data obtained in this work (cont'ed).}
\begin{tabular}{ccrrc}
\hline
\hline
\mcol{1}{c}{\rm KIC} & \mcol{1}{c}{\rm HJD} & \mcol{1}{c}{\rm RV$_{1}$} & \mcol{1}{c}{$\sigma_{RV_{1}}$} & \mcol{1}{c}{Source}\\
\mcol{1}{c}{\rm Nr} & \mcol{1}{c}{ } & \mcol{1}{c}{\kms} & \mcol{1}{c}{\kms} & \mcol{1}{c}{ [1:\textit{H}, 2:\textit{A}]}\\
\hline
\small 4671225  & \small 2455823.59410  & \small -15.743  & \small  5.634  & \small   1   \\
\small 4671225  & \small 2456778.64304  & \small -13.057  & \small  4.099  & \small   1   \\
\small 4671225  & \small 2456783.60829  & \small -12.604  & \small  4.422  & \small   1   \\
\small 4671225  & \small 2456811.70493  & \small -13.257  & \small  5.286  & \small   1   \\
\small 4671225  & \small 2456822.61118  & \small -14.156  & \small  5.508  & \small   1   \\
\small 4989900  & \small 2456772.64609  & \small -12.981  & \small  6.090  & \small   1   \\
\hline                                                                                                              
\end{tabular}
\label{tab:radvel2}
\end{table*}

\clearpage
\setcounter{table}{0}
\begin{table*}
\centering
\setlength{\tabcolsep}{1.5mm}
\caption{List of radial velocity data obtained in this work (cont'ed).}
\begin{tabular}{ccrrc}
\hline
\hline
\mcol{1}{c}{\rm KIC} & \mcol{1}{c}{\rm HJD} & \mcol{1}{c}{\rm RV$_{1}$} & \mcol{1}{c}{$\sigma_{RV_{1}}$} & \mcol{1}{c}{Source}\\
\mcol{1}{c}{\rm Nr} & \mcol{1}{c}{ } & \mcol{1}{c}{\kms} & \mcol{1}{c}{\kms} & \mcol{1}{c}{[1:\textit{H}, 2:\textit{A}]}\\
\hline
\small 6289468  & \small 2456820.50998  & \small  -6.006  & \small  7.884  & \small   1   \\
\small 6289468  & \small 2457160.62559  & \small  -6.491  & \small  8.570  & \small   1   \\
\small 6289468  & \small 2457238.40678  & \small  -7.420  & \small  7.749  & \small   1   \\
\small 6289468  & \small 2457241.57072  & \small  -6.086  & \small  6.765  & \small   1   \\
\small 6432054  & \small 2456772.68105  & \small  12.487  & \small  4.117  & \small   1   \\
\small 6432054  & \small 2456778.66024  & \small  11.371  & \small  3.466  & \small   1   \\
\hline                                                                                                              
\end{tabular}
\label{tab:radvel3}
\end{table*}

\clearpage
\setcounter{table}{0}
\begin{table*}
\centering
\setlength{\tabcolsep}{1.5mm}
\caption{List of radial velocity data obtained in this work (cont'ed).}
\begin{tabular}{ccrrc}
\hline
\hline
\mcol{1}{c}{\rm KIC} & \mcol{1}{c}{\rm HJD} & \mcol{1}{c}{\rm RV$_{1}$} & \mcol{1}{c}{$\sigma_{RV_{1}}$} & \mcol{1}{c}{Source}\\
\mcol{1}{c}{\rm Nr} & \mcol{1}{c}{ } & \mcol{1}{c}{\kms} & \mcol{1}{c}{\kms} & \mcol{1}{c}{[1:\textit{H}, 2:\textit{A}]}\\
\hline
\small 7668791  & \small 2456818.48741  & \small -28.210  & \small  0.629  & \small   1   \\
\small 7668791  & \small 2456883.61216  & \small -27.888  & \small  0.714  & \small   1   \\
\small 7668791  & \small 2456938.34404  & \small -28.299  & \small  0.721  & \small   1  \\
\small 7668791  & \small 2456971.39724  & \small -27.982  & \small  0.762  & \small   1  \\
\small 7748238  & \small 2456773.67912  & \small -20.810  & \small  4.052  & \small   1  \\
\small 7748238  & \small 2456778.68018  & \small -22.638  & \small  3.651  & \small   1  \\
\hline                                                                                                              
\end{tabular}
\label{tab:radvel4}
\end{table*}

\clearpage
\setcounter{table}{0}
\begin{table*}
\centering
\setlength{\tabcolsep}{1.5mm}
\caption{List of radial velocity data obtained in this work (cont'ed).}
\begin{tabular}{ccrrc}
\hline
\hline
\mcol{1}{c}{\rm KIC} & \mcol{1}{c}{\rm HJD} & \mcol{1}{c}{\rm RV$_{1}$} & \mcol{1}{c}{$\sigma_{RV_{1}}$} & \mcol{1}{c}{Source}\\
\mcol{1}{c}{\rm Nr} & \mcol{1}{c}{ } & \mcol{1}{c}{\kms} & \mcol{1}{c}{\kms} & \mcol{1}{c}{[1:\textit{H}, 2:\textit{A}]}\\
\hline
\small 9351622  & \small 2456773.74009  & \small  -7.317  & \small  1.415  & \small   1  \\
\small 9351622  & \small 2456778.70867  & \small  -7.010  & \small  1.620  & \small   1  \\
\small 9351622  & \small 2456810.72603  & \small  -7.512  & \small  1.566  & \small   1  \\
\small 9351622  & \small 2456822.59921  & \small  -7.728  & \small  1.519  & \small   1   \\
\small 9351622  & \small 2456886.46224  & \small  -7.195  & \small  1.860  & \small   1   \\
\small 9351622  & \small 2456971.42319  & \small  -7.795  & \small  2.076  & \small   1   \\
\hline                                                                                                              
\end{tabular}
\label{tab:radvel5}
\end{table*}

\clearpage
\setcounter{table}{0}
\begin{table*}
\centering
\setlength{\tabcolsep}{1.5mm}
\caption{List of radial velocity data obtained in this work (cont'ed).}
\begin{tabular}{ccrrc}
\hline
\hline
\mcol{1}{c}{\rm KIC} & \mcol{1}{c}{\rm HJD} & \mcol{1}{c}{\rm RV$_{1}$} & \mcol{1}{c}{$\sigma_{RV_{1}}$} & \mcol{1}{c}{Source}\\
\mcol{1}{c}{\rm Nr} & \mcol{1}{c}{ } & \mcol{1}{c}{\kms} & \mcol{1}{c}{\kms} & \mcol{1}{c}{[1:\textit{H}, 2:\textit{A}]}\\
\hline
\small 9790479  & \small 2457298.49499  & \small  -0.441  & \small  0.776  & \small   1   \\
\small 9790479  & \small 2457515.68719  & \small   0.345  & \small  0.970  & \small   1   \\
\small 9970568  & \small 2455756.69520  & \small   9.880  & \small  7.754  & \small   1   \\
\small 9970568  & \small 2456776.73133  & \small  11.896  & \small 11.861  & \small   1   \\
\small 9970568  & \small 2456781.71387  & \small   8.834  & \small  7.530  & \small   1   \\
\small 9970568  & \small 2456809.72125  & \small   9.053  & \small  7.642  & \small   1   \\
\hline                                                                                                              
\end{tabular}
\label{tab:radvel6}
\end{table*}

\setcounter{table}{1}
\begin{table*}
\centering
\setlength{\tabcolsep}{1.5mm}
\caption{List of radial velocity data used in this work. II. Double- and triple-lined systems. \textit{H} stands for 
\textit{Hermes} while \textit{A} stands for \textit{Ace}.}
\begin{tabular}{ccrrrrrrc}
\hline
\hline
\mcol{1}{c}{\rm KIC} & \mcol{1}{c}{\rm HJD} & \mcol{1}{c}{\rm RV$_{1}$} & \mcol{1}{c}{$\sigma_{RV_{1}}$} & \mcol{1}{c}{RV$_{2}$} & \mcol{1}{c}{$\sigma_{RV_{2}}$} & \mcol{1}{c}{RV$_{3}$} & \mcol{1}{c}{$\sigma_{RV_{3}}$} & \mcol{1}{c}{Source}\\
\mcol{1}{c}{\rm Nr} & \mcol{1}{c}{ } & \mcol{1}{c}{\kms} & \mcol{1}{c}{\kms} & \mcol{1}{c}{\kms} & \mcol{1}{c}{\kms} & \mcol{1}{c}{\kms} & \mcol{1}{c}{\kms} & \mcol{1}{c}{[1:\textit{H}, 2:\textit{A}]}\\
\hline
\small 4480321  & \small 2455820.44990  & \small -45.392  & \small  0.408  & \small -3.146  & \small  0.352  & \small-12.481  & \small 10.101  & \small1 \\
\small 4480321  & \small 2455820.47133  & \small -45.885  & \small  0.548  & \small -2.540  & \small  0.272  & \small-11.552  & \small  7.182  & \small1 \\
\small 4480321  & \small 2456513.50147  & \small  40.447  & \small  0.487  & \small-61.375  & \small  0.409  & \small-28.157  & \small  6.368  & \small1 \\
\small 4480321  & \small 2456513.51999  & \small  41.381  & \small  0.535  & \small-61.823  & \small  0.298  & \small-28.192  & \small  9.238  & \small1 \\
\small 4480321  & \small 2456513.65562  & \small  44.472  & \small  0.458  & \small-64.841  & \small  0.365  & \small-30.583  & \small  8.740  & \small1 \\
\small 4480321  & \small 2456513.67472  & \small  44.602  & \small  0.526  & \small-64.910  & \small  0.453  & \small-31.058  & \small 13.289  & \small1 \\
\hline                                                                                                             
\end{tabular}
\label{tab:radvel7}
\end{table*}

\clearpage
\setcounter{table}{1}
\begin{table*}
\centering
\setlength{\tabcolsep}{1.5mm}
\caption{List of radial velocity data obtained in this work (cont'ed).}
\begin{tabular}{ccrrrrrrc}
\hline
\hline
\mcol{1}{c}{\rm KIC} & \mcol{1}{c}{\rm HJD} & \mcol{1}{c}{\rm RV$_{1}$} & \mcol{1}{c}{$\sigma_{RV_{1}}$} & \mcol{1}{c}{RV$_{2}$} & \mcol{1}{c}{$\sigma_{RV_{2}}$} & \mcol{1}{c}{RV$_{3}$} & \mcol{1}{c}{$\sigma_{RV_{3}}$} & \mcol{1}{c}{Source}\\
\mcol{1}{c}{\rm Nr} & \mcol{1}{c}{ } & \mcol{1}{c}{\kms} & \mcol{1}{c}{\kms} & \mcol{1}{c}{\kms} & \mcol{1}{c}{\kms} & \mcol{1}{c}{\kms} & \mcol{1}{c}{\kms} & \mcol{1}{c}{[1:\textit{H}, 2:\textit{A}]}\\
\hline
\small 5219533  & \small 2457298.48087  & \small -27.963  & \small  0.314  & \small 36.905  & \small  0.301  & \small  12.078  & \small  19.435  & \small 1 \\
\small 5219533  & \small 2457323.37432  & \small -26.188  & \small  0.328  & \small 35.589  & \small  0.347  & \small  11.412  & \small  20.592  & \small 1 \\
\small 5219533  & \small 2457466.74899  & \small  53.706  & \small  0.450  & \small-45.677  & \small  0.336  & \small   4.845  & \small  22.887  & \small 1 \\
\small 5219533  & \small 2457467.76323  & \small  61.562  & \small  0.479  & \small-53.721  & \small  0.386  & \small   6.956  & \small  16.694  & \small 1 \\
\small 5219533  & \small 2457507.70192  & \small   9.256  & \small  0.424  & \small -3.374  & \small  0.950  & \small   9.220  & \small  24.098  & \small 1 \\
\small 5219533  & \small 2457508.72411  & \small   6.755  & \small  1.773  & \small -0.724  & \small  3.810  & \small  29.269  & \small  28.208  & \small 1 \\
\hline                                                                                                              
\end{tabular}
\label{tab:radvel8}
\end{table*}

\clearpage
\setcounter{table}{1}
\begin{table*}
\centering
\setlength{\tabcolsep}{1.5mm}
\caption{List of radial velocity data obtained in this work (cont'ed).}
\begin{tabular}{ccrrrrrrc}
\hline
\hline
\mcol{1}{c}{\rm KIC} & \mcol{1}{c}{\rm HJD} & \mcol{1}{c}{\rm RV$_{1}$} & \mcol{1}{c}{$\sigma_{RV_{1}}$} & \mcol{1}{c}{RV$_{2}$} & \mcol{1}{c}{$\sigma_{RV_{2}}$} & \mcol{1}{c}{RV$_{3}$} & \mcol{1}{c}{$\sigma_{RV_{3}}$} & \mcol{1}{c}{Source}\\
\mcol{1}{c}{\rm Nr} & \mcol{1}{c}{ } & \mcol{1}{c}{\kms} & \mcol{1}{c}{\kms} & \mcol{1}{c}{\kms} & \mcol{1}{c}{\kms} & \mcol{1}{c}{\kms} & \mcol{1}{c}{\kms} & \mcol{1}{c}{[1:\textit{H}, 2:\textit{A}]}\\
\hline
\small 8975515  & \small 2456781.58945  & \small -16.248  & \small  9.076  & \small-20.870  & \small  0.549 & \small  0.000  & \small 0.000   & \small 1 \\
\small 8975515  & \small 2456804.69168  & \small -16.947  & \small  8.817  & \small-21.259  & \small  0.644 & \small  0.000  & \small 0.000   & \small 1 \\
\small 8975515  & \small 2456818.59219  & \small -16.238  & \small  8.482  & \small-20.980  & \small  0.618 & \small  0.000  & \small 0.000   & \small 1 \\
\small 8975515  & \small 2456966.34638  & \small -20.604  & \small  9.282  & \small-17.674  & \small  0.633 & \small  0.000  & \small 0.000   & \small 1 \\
\small 8975515  & \small 2456973.44729  & \small -19.369  & \small 10.309  & \small-17.666  & \small  0.597 & \small  0.000  & \small 0.000   & \small 1 \\
\small 8975515  & \small 2457159.58194  & \small -21.556  & \small  9.393  & \small-15.794  & \small  0.611 & \small  0.000  & \small 0.000   & \small 1 \\
\hline                                                                                                              
\end{tabular}
\label{tab:radvel9}
\end{table*}


\section{Tables which are also available in electronic version.}
\label{sect:Tables}

\setcounter{table}{1}
\begin{sidewaystable*}
\centering
\caption{Comparison of atmospheric stellar parameters obtained in this work with published values (assuming other parameters are equivalent).}
\begin{tabular}{llccrcccccccccc}
\hline\hline
\mcol{1}{c}{\rm KIC} & \mcol{1}{l}{\rm Sp T} & \mcol{1}{c}{$T_{\mathrm{eff}}$} & \mcol{1}{c}{$\mathrm{log}\,g$} & \mcol{1}{r}{Kp} 
& Category &  \mcol{1}{c}{Mean $T_{\mathrm{eff}}$} & \mcol{1}{c}{Mean $\mathrm{log}\,g$} & \mcol{1}{c}{GIR $T_{\mathrm{eff}}$} 
& \mcol{1}{c}{GIR $\mathrm{log}\,g$} & \mcol{1}{c}{GIR $v\,\mathrm{sin}\,i$} & \mcol{4}{c}{ Literature values}\\
\mcol{1}{c}{Nr} & \mcol{1}{l}{KIC} & \mcol{1}{c}{KIC} & \mcol{1}{c}{KIC} & \mcol{1}{r}{mag} &&& &&& & $T_{\mathrm{eff}}$ & $\mathrm{log}\,g$ & $v\,\mathrm{sin}\,i$ & Source\\
\hline
\setlength{\tabcolsep}{1.5mm}
\small 3429637 & \small (h)A9m &  7200 &  4 &  7.7 &  P+VAR  &   7350 &  <=4.0 &  7256 &  2.66 &  49 &  7200 &  4 &  50 &  (11)\\
\small 3437940 &  Fm &  7430 &  3.86 &  8.5 &  P   &    7700 &  <=3.9   &  7702 &  3.51 &  111 &  7700 &  4.1 &  120 &  (11)\\ 
\small 3851151 &  A2 &  8190 &  4.1  &  9.8 &  S?  &    8600 &    3.8   &  8571 &  3.71 &  124 &  8300 &  4   &  119 &  (15)\\  
\small 4044353 &  A2 &  8300 &  3.56 &  9.8 &  S   &    8300 &  4.1-4.5 &  8309 &  4.3  &  107 &  8000 &  3.8 &  110 &  (15)\\
\end{tabular}
\tablefoot{(11) \citet{Catanzaro2011MNRAS.411.1167C}, (13) \citet{Tkachenko2013A&A...556A..52T}, (15) \citet{Niemczura2015MNRAS.450.2764N}
\tablefoottext{c}{Range [510-520]~nm} }
\label{tab:complit}
\end{sidewaystable*}

\setcounter{table}{2}
\begin{sidewaystable*}
\centering
\caption{Comparison of atmospheric stellar parameters obtained in this work with the {\sc Lamost} results based on the {\sc Rotfit} procedure (22 targets in common).}
\begin{tabular}{llccrcccccccccc}
\hline\hline
\mcol{1}{c}{\rm KIC} & \mcol{1}{c}{\rm Sp T} & \mcol{1}{c}{$T_{\mathrm{eff}}$} & \mcol{1}{c}{$\mathrm{log}\,g$} & \mcol{1}{c}{Kp} & Category & 
\mcol{3}{c}{{\sc Girfit} values} & \mcol{1}{c}{\rm Sp T} & \mcol{5}{c}{{\sc Rotfit} values}\\
\mcol{1}{c}{\rm Nr} & \mcol{1}{c}{KIC} & \mcol{1}{c}{KIC} & \mcol{1}{c}{KIC} & \mcol{1}{c}{mag} & \mcol{1}{c}{} & \mcol{1}{c}{$T_{\mathrm{eff}}$} & 
\mcol{1}{c}{$\mathrm{log}\,g$} & \mcol{1}{c}{$v\,\mathrm{sin}\,i$} & \mcol{1}{c}{\rm {\sc Lamost}} & \mcol{1}{c}{$T_{\mathrm{eff}}$} & \mcol{1}{c}{$\mathrm{log}\,g$} 
& \mcol{1}{c}{metals} & \mcol{1}{c}{RV} & \mcol{1}{c}{$v\,\mathrm{sin}\,i$}\\
\hline
\setlength{\tabcolsep}{1.5mm}
\small 3097912 &\small     A5 & \small 7880 & \small 3.56 & \small 9.40 & \small S   & \small 8147 & \small 4.17 & \small 118 & \small A5IV& \small    7891 $\pm$ 140& \small  3.90 $\pm$ 0.12& \small -0.13 $\pm$ 0.11& \small -28.8 $\pm$ 24.1&\\ 
\small 3453494 &\small     A5 & \small 7810 & \small 3.84 & \small 9.59 & \small S   & \small 7941 & \small 4.69 & \small 230 & \small A0III/IV& \small7720 $\pm$ 160& \small  3.86 $\pm$ 0.15& \small -0.13 $\pm$ 0.12& \small  -3.5 $\pm$ 26.9&\small 183 $\pm$ 63\\
\small 4044353 &\small     A2 & \small 8300 & \small 3.56 & \small 9.85 & \small S   & \small 8309 & \small 4.3  & \small 107 & \small A3V& \small     8436 $\pm$ 543& \small  3.81 $\pm$ 0.14& \small -0.18 $\pm$ 0.12& \small -18.1 $\pm$ 33.1&\\ 
\small 4281581 &\small     A2 & \small 8140 & \small 3.84 & \small 9.40 & \small S   & \small 8406 & \small 3.55 & \small 104 & \small A2III& \small   8675 $\pm$ 162& \small  3.64 $\pm$ 0.19& \small -0.18 $\pm$ 0.11& \small   1.0 $\pm$ 23.0&\\ 
\small 4480321 &\small     A3 & \small 7147 & \small 3.87 & \small10.29 & \small SB3 & \small 7900  & \small 4$^*$ & \small 160$^*$ & \small A9V& \small     7540 $\pm$ 67& \small   3.85 $\pm$ 0.10& \small  0.01 $\pm$ 0.13& \small -11.1 $\pm$ 27.0&\\   
\small 4671225 &\small     A7 & \small 8230 & \small 3.27 & \small 9.97  & \small S   & \small 8008 & \small 3.76 & \small 152 & \small A4V& \small     8201 $\pm$ 373& \small  3.87 $\pm$ 0.16& \small -0.16 $\pm$ 0.12& \small  21.9 $\pm$ 33.2&\\  
\small 5219533 &\small     Am & \small 7410 & \small 3.94 & \small 9.20 & \small SB3  & \small 8300 & \small 4$^*$  & \small 10$^*$ & \small A6m& \small     7835 $\pm$ 327& \small  3.88 $\pm$ 0.11& \small -0.09 $\pm$ 0.13& \small  10.1 $\pm$ 22.3&\\ 
\small 5219533 &\small     Am & \small 7410 & \small 3.94 & \small 9.20 & \small SB3  & \small 8300 & \small 4$^*$  & \small 10$^*$ & \small A6m& \small     7920 $\pm$ 194& \small  3.88 $\pm$ 0.12& \small -0.11 $\pm$ 0.12& \small  45.9 $\pm$ 29.5&\\ 
\small 5965837 &\small     F2 & \small 6520 & \small 4.15 & \small 9.18 & \small P   & \small 6800 & \small 3.3\tablefootmark{c} & \small15 & \small F6II& \small    7079 $\pm$ 172& \small  3.71 $\pm$ 0.14& \small  0.24 $\pm$ 0.13& \small -55.8 $\pm$ 19.7&\\ 
\small 6289468 &\small     A2 & \small 8270 & \small 3.74 & \small 9.40 & \small S   & \small 8208 & \small 3.58 & \small 164 & \small A4III& \small   8278 $\pm$ 99& \small   3.85 $\pm$ 0.12& \small -0.20 $\pm$ 0.10& \small   3.3 $\pm$ 34.7&\\   
\small 6289468 &\small     A2 & \small 8270 & \small 3.74 & \small 9.40 & \small S   & \small 8208 & \small 3.58 & \small 164 & \small A2V& \small 8947 $\pm$ 545& \small  3.89 $\pm$ 0.13& \small -0.15 $\pm$ 0.13& \small -13.4 $\pm$ 22.0&\small 191 $\pm$ 74\\   
\small 7668791 &\small     A2 & \small 8150 & \small 3.89 & \small 9.81 & \small S   & \small 8146 & \small 3.85 & \small  49 & \small A5IV& \small 7625 $\pm$ 245& \small  3.90 $\pm$ 0.12& \small -0.07 $\pm$ 0.12& \small -34.7 $\pm$ 17.4&\\ 
\small 7748238 &\small     A5 & \small 7230 & \small 3.47 & \small 9.53 & \small S   & \small 7380 & \small 4.65 & \small 123 & \small F0IV& \small    7235 $\pm$ 177& \small  3.91 $\pm$ 0.13& \small  0.05 $\pm$ 0.14& \small -27.5 $\pm$ 20.3&\\ 
\small 7770282 &\small     F0 & \small 7450 & \small 3.5  & \small 9.73 & \small P?  & \small 7921 & \small 4.19 & \small  52 & \small A9V  & \small   7496 $\pm$ 200& \small  3.90 $\pm$ 0.13& \small -0.02 $\pm$ 0.12& \small  -2.1 $\pm$ 27.1&\\ 
\small 7770282 &\small     F0 & \small 7450 & \small 3.5  & \small 9.73 & \small P?  & \small 7921 & \small 4.19 & \small  52 & \small A5IV & \small   7873 $\pm$ 130& \small  3.93 $\pm$ 0.12& \small -0.14 $\pm$ 0.12& \small  25.8 $\pm$ 18.1&\\ 
\small 7770282 &\small     F0 & \small 7450 & \small 3.5  & \small 9.73 & \small P?  & \small 7921 & \small 4.19 & \small  52 & \small A7III& \small   7737 $\pm$ 192& \small  3.91 $\pm$ 0.12& \small -0.08 $\pm$ 0.12& \small  -1.4 $\pm$ 15.1&\\ 
\small 8975515 &\small     A2 & \small 7180 & \small 3.9  & \small 9.52 & \small SB2 & \small 7440 & \small 4$^*$ & \small 164 & \small A7III& \small   7359 $\pm$ 172& \small  3.90 $\pm$ 0.13& \small -0.01 $\pm$ 0.13& \small -24.1 $\pm$ 21.0&\\ 
\small 9351622 &\small     F0 & \small 7450 & \small 3.53 & \small 9.11 & \small S   & \small 7711 & \small 3.53 & \small 77  & \small A9V& \small     7512 $\pm$ 177& \small  3.82 $\pm$ 0.14& \small  0.02 $\pm$ 0.15& \small -16.2 $\pm$ 18.3&\\  
\small 9413057 &\small     A2 & \small 8470 & \small 3.87 & \small 9.64 & \small S   & \small 8698 & \small 3.79 & \small 190 & \small A1V& \small     8547 $\pm$ 212& \small  3.90 $\pm$ 0.12& \small -0.16 $\pm$ 0.13& \small  25.0 $\pm$ 37.4& \small 196 $\pm$ 61\\
\small 9650390 &\small     A0  & \small 8390 & \small 3.68 & \small 9.37 & \small S   & \small 8297 & \small 3.2  & \small 272 & \small A1V& \small     8828 $\pm$ 471& \small  3.84 $\pm$ 0.11& \small -0.20 $\pm$ 0.11& \small   3.5 $\pm$ 31.3& \small 284 $\pm$ 69\\
\small 9775454 &\small     F1IV & \small --   & \small --   & \small 8.20 & \small SB1 & \small  7287 & \small 4.25 & \small 65 & \small F0V& \small     7255 $\pm$ 87& \small   3.92 $\pm$ 0.14& \small  0.14 $\pm$ 0.13& \small -13.2 $\pm$ 27.7&\\  
\small 9790479 &\small     A2   & \small 7840 & \small 4.04 & \small 9.91 & \small SB1 & \small 8239 & \small 4.28 & \small 40 & \small A5IV& \small    8068 $\pm$ 384& \small  3.87 $\pm$ 0.12& \small -0.11 $\pm$ 0.13& \small  -0.6 $\pm$ 16.7&\\ 
\small 9970568 &\small     A2   & \small 7790 & \small 3.64 & \small 9.58 & \small S   & \small 7843 & \small 4.27 & \small 252 & \small A9V& \small     8234 $\pm$ 676& \small  3.94 $\pm$ 0.12& \small -0.08 $\pm$ 0.13& \small  14.1 $\pm$ 23.5& \small 237 $\pm$ 70\\
\small 10537907 &\small    F0   & \small 7500 & \small 3.45 & \small 9.91 & \small P+VAR? & \small 7660 & \small 4.26 & \small 118 & \small A7III& \small   7417 $\pm$ 124& \small  3.91 $\pm$ 0.13& \small -0.04 $\pm$ 0.12& \small  -5.4 $\pm$ 19.4&\\ 
\small 11193046 &\small    A2   & \small 8170 & \small 3.7  & \small 9.56 & \small S   & \small 7920 & \small 3.9  & \small 208 & \small A9V& \small     7891 $\pm$ 579& \small  3.88 $\pm$ 0.13& \small -0.09 $\pm$ 0.13& \small  23.7 $\pm$ 24.5& \small 205 $\pm$ 59\\
\small 11572666 &\small     --  & \small 7040 & \small 3.49 & \small 9.70 & \small SB2 & \small 7900 & \small 4$^*$ & \small 253 & \small A8V& \small     7335 $\pm$ 224& \small  3.93 $\pm$ 0.13& \small  0.01 $\pm$ 0.13& \small  -4.7 $\pm$ 26.7& \small 200 $\pm$ 56\\
\end{tabular}
\tablefoot{$^*$ kept fixed while determining the other parameters
\tablefoottext{c}{Range [510-520]~nm} }
\label{tab:rotfit}
\end{sidewaystable*}


\end{document}